\documentclass[acmtocl]{acmtrans2m}
\usepackage{isabelle}
\usepackage{amsmath}
\usepackage{amssymb}
\usepackage{isabellesym}
\usepackage{mathpartir}
\usepackage{paralist}
\usepackage{xspace}

\newcommand{\Var}{\mathit{Var}}
\newcommand{\Id}{\mathit{Id}}
\newtheorem{theorem}{Theorem}
\newtheorem{lemma}{Lemma}
\newtheorem{corollary}{Corollary}

\newcommand{\labelSec}[1]{\label{sec:#1}}
\newcommand{\refSec}[1]{Sec.~\ref{sec:#1}}
\newcommand{\labelFig}[1]{\label{fig:#1}}
\newcommand{\refFig}[1]{Fig.~\ref{fig:#1}}

\newcommand{\labelThm}[1]{\label{thm:#1}}
\newcommand{\refThm}[1]{Thm.~\ref{thm:#1}}
\newcommand{\labelLem}[1]{\label{lem:#1}}
\newcommand{\refLem}[1]{Lem.~\ref{lem:#1}}

\newcommand{\labelCor}[1]{\label{cor:#1}}
\newcommand{\refCor}[1]{Cor.~\ref{cor:#1}}

\newcommand{\HP}{\text{HP05}\xspace}

\isabellestyle{it}

\newcommand{\isasymextends}{\isamath{\extends}}
\newcommand{\extends}{\geq}
\newcommand{\isasymtranslates}{\isamath{\translates}}
\newcommand{\translates}{\leftrightsquigarrow}

\renewcommand{\isabeginpar}{\par\relax}
\renewcommand{\isaendpar}{\par\relax}

\renewcommand{\isamarkupsection}[1]{\section{#1}}
\renewcommand{\isamarkupsubsection}[1]{\subsection{#1}}

\renewcommand{\isasymsharp}{\isamath{\#}}
\def\dn{\stackrel{\mbox{\scriptsize def}}{=}}
\renewcommand{\Gamma}{\varGamma}
\renewcommand{\Sigma}{\varSigma}
\renewcommand{\emptyset}{\varnothing}
\renewcommand{\isasymtheta}{\isamath{\theta}}
\renewcommand{\isacharat}{\!@\!}

\DeclareRobustCommand{\isactrlbsub}{\emph\bgroup\isascriptstyle\math{}\sb\bgroup}
\DeclareRobustCommand{\isactrlesub}{\egroup\endmath\egroup}

\acmVolume{V} 
\acmNumber{N}
\acmYear{YY}
\acmMonth{Month}
\begin{document}

\title{Mechanizing the Metatheory of LF}

\author{CHRISTIAN URBAN\\
TU Munich
\and
JAMES CHENEY\\
University of Edinburgh
\and
STEFAN BERGHOFER\\
TU Munich}
            
\begin{abstract}
  LF is a dependent type theory in which many other formal systems can
  be conveniently embedded.  However, correct use of LF relies on
  nontrivial metatheoretic developments such as proofs of correctness
  of decision procedures for LF's judgments.  Although detailed
  informal proofs of these properties have been published, they have
  not been formally verified in a theorem prover. We have formalized 
  these properties within Isabelle/HOL using
  the Nominal Datatype Package, closely following a recent article by
  Harper and Pfenning.  In the process, we identified and resolved a
  gap in one of the proofs and a small number of minor lacunae
  in others.  We also formally derive a version of the type checking 
  algorithm from which Isabelle/HOL can generate executable code. 
  Besides its intrinsic interest, our formalization
  provides a foundation for studying the adequacy of LF encodings, the
  correctness of Twelf-style metatheoretic reasoning, and the
  metatheory of extensions to LF.
\end{abstract}

            
            

\begin{bottomstuff}
 This is a revised and expanded version of a conference
 paper~\cite{UrbanCheneyBerghofer08}.  

 Cheney was supported by a Royal Society University Research
 Fellowship and by EPSRC grant GR/S63205/01.
 Urban was supported by an Emmy Noether Grant from the DFG.

 Corresponding author: J. Cheney, Informatics Forum, 10 Crichton
 Street, Edinburgh EH8 9AB, Scotland, email:
 \texttt{jcheney@inf.ed.ac.uk}.
\end{bottomstuff}
\maketitle

% LaTeXSugar
\begin{isabellebody}%
\def\isabellecontext{LaTeXsugar}%
\isadelimtheory
\isanewline
\endisadelimtheory
\isatagtheory
\endisatagtheory
{\isafoldtheory}%
\isadelimtheory
\endisadelimtheory
\isadelimtheory
\endisadelimtheory
\isatagtheory
\endisatagtheory
{\isafoldtheory}%
\isadelimtheory
\endisadelimtheory
\end{isabellebody}%

% LaTeXSugar

% generated text of all theories
%
\begin{isabellebody}%
\def\isabellecontext{Paper}%
\isadelimtheory
\endisadelimtheory
\isatagtheory
\endisatagtheory
{\isafoldtheory}%
\isadelimtheory
\endisadelimtheory
\isamarkupsection{Introduction%
}
\isamarkuptrue%
\begin{isamarkuptext}%
The (Edinburgh) Logical Framework (LF) is a dependent type theory
  introduced by Harper, Honsell and
  Plotkin~\citeyear{HarperHonsellPlotkin93} as a framework for
  specifying and reasoning about formal systems. It has found many
  applications, including proof-carrying code~\cite{Necula97}.  The
  Twelf system \cite{PfenningSchuermann99} has been used to mechanize
  reasoning about LF specifications.

  The cornerstone of LF is the idea of encoding \emph{judgments-as-types} and
  \emph{proofs-as-terms} whereby judgments of a specified formal system are
  represented as LF-types and the LF-terms inhabiting these LF-types
  correspond to valid deductions for these judgments.  Hence, the validity of
  a deduction in a specified system is equivalent to a type checking problem
  in LF. Therefore correct use of LF to encode other logics depends on the
  proofs of correctness of type checking algorithms for LF.
 
  Type checking in LF is decidable, but proving decidability is
  nontrivial because types may contain expressions with computational
  behavior.  This means that typechecking depends on equality-tests
  for LF-terms and LF-types. Several algorithms for such
  equality-tests have been proposed in the
  literature~\cite{Coquand91,Goguen05POPL,HarperPfenning05}. Harper
  and Pfenning \citeyear{HarperPfenning05} present a type-driven
  algorithm that is practical and also has been extended to a
  variety of richer languages.  The correctness of this algorithm is
  proved by establishing soundness and completeness with respect to
  the definitional equality rules of LF. These proofs are involved:
  Harper and Pfenning's detailed pencil-and-paper proof spans more
  than 30 pages, yet still omits many cases and lemmas.

  We present a formalization of the main results of Harper and
  Pfenning's article.  To our knowledge this is the first
  formalization of these or comparable results.  While most of the
  formal proofs go through as described by \citeN{HarperPfenning05},
  we found a few do \emph{not} go through as described, and there is a
  \emph{gap} in the proof of soundness.  Although the problem can be
  avoided easily by adding to or changing the rules of
  \citeN{HarperPfenning05}, we found that it was still possible to prove the
  original results, though the argument was nontrivial.  Our
  formalization was essential not only to find this gap in Harper and
  Pfenning's argument, but also to find and validate the possible
  repairs relatively quickly.
  
  We used Isabelle/HOL~\cite{IsabelleTutorial} and the Nominal
  Datatype
  Package~\cite{UrbanBerghoferNorrish07,UrbanTasson05,Urban08JAR} for
  our formalization. The latter provides an infrastructure for
  reasoning conveniently about datatypes with a built-in notion of
  alpha-equivalence: it allows to specify such datatypes, provides
  appropriate recursion combinators and derives strong induction
  principles that have the usual variable convention already built-in.
  The Nominal Datatype Package has already been used to formalize
  logical relation arguments similar to (but much simpler than) those
  in Harper and Pfenning's completeness proof~\cite{NarbouxUrban07};
  it is worth noting that logical relations proofs are currently not
  easy to formalize in Twelf itself, despite the recent breakthrough
  by~\citeN{Schuermann}.

  Besides proving the correctness of their equivalence
  algorithm, Harper and Pfenning also sketched a proof of
  decidability.  Unfortunately, since Isabelle/HOL is based on
  classical logic, proving decidability results of this kind is not
  straightforward.  We have formalized the essential parts of the
  decidability proof by providing inductive definitions of the
  complements of the relations we wish to decide.  It is clear by
  inspection that these relations define recursively enumerable sets,
  which implies decidability, but we have not formalized this part of
  the proof. A complete proof of decidability would require first
  developing a substantial amount of computability theory within
  Isabelle/HOL, a problem of independent interest we leave for future
  work.

  We followed the arguments in Harper and Pfenning's
  article very closely using the Nominal Datatype Package for our
  formalisation, but the current system does not allow us to generate
  executable code directly from definitions involving nominal
  datatypes.  We therefore also implemented a type-checking algorithm
  based on the locally nameless approach for representing binders
  \cite{McKinnaP99,AydemirEtAl08}.  We proved that the nominal
  datatype formalization of Harper and Pfenning's algorithm is
  equivalent to the locally nameless formulation.  Moreover, by making
  the choice of fresh names explicit, we can generate a working ML
  implementation directly from the verified formalization.%
\end{isamarkuptext}%
\isamarkuptrue%
\begin{isamarkuptext}%
\paragraph*{Outline} 
  We first briefly review LF and its representation in the Nominal
  Datatype Package (\refSec{background}).  In \refSec{formalization},
  we report on our formalization.  To ease comparison,
  \refSec{formalization} follows the structure
  of~\citeN{HarperPfenning05} closely, although this article is
  self-contained.
  Sections~\ref{sec:syntactic}--\ref{sec:typechecking} summarize our
  formalization of the basic syntactic properties of LF and soundness
  and completeness of the equivalence and typechecking algorithms.  We
  discuss additional lemmas, proof details, and other complications
  arising during the formalization, and discuss the gap in the
  soundness proof and its solutions in detail.  The remainder of
  \refSec{formalization} reports upon formalizations of additional
  results whose proofs were only sketched by~\citeN{HarperPfenning05}.
  These include
  \begin{compactenum}
  \item the admissibility of strengthening and strong extensionality rules
    (\refSec{strengthening}), 
  \item a partial formalization of decidability of algorithmic
    typechecking for LF, and a discussion of the current limitations
    of Isabelle/HOL in formalizing proofs about decidability
    (\refSec{decidability}),
  \item the existence and uniqueness of quasicanonical forms
    (\refSec{quasicanonical}), and
  \item a partial formalization of an example proof of adequacy
    (\refSec{adequacy}), and a discussion of complications in the
    proof sketched in~\cite{HarperPfenning05}.
  \end{compactenum}
  In \refSec{locally-nameless} we define and verify the correctness of
  a type checking algorithm based on the locally nameless representation 
  of binders, from which
  Isabelle/HOL can generate executable code.  This amounts to a
  verified typechecker for LF, an original contribution of this
  article.  \refSec{discussion} summarizes the authors' experience
  with the formalization, \refSec{related} discusses related and
  future work and \refSec{concl} concludes.

  \paragraph*{Contributions}
  The metatheory of LF is well-understood: it had been studied for
  many years before the definitive presentation
  in~\citeN{HarperPfenning05}.  Their main results were not in serious
  doubt, and formalizing such work might strike some readers as
  perverse or pedantic.  Nevertheless, our formalization is an
  original and significant contribution to the study of logical
  frameworks and mechanized metatheory, because:
  \begin{compactenum}
  \item it tests the capabilities of the Nominal Datatype Package for
    formalizing a large and complex metatheoretical development,
  \item it provides high confidence in algorithms that are widely
    trusted but have never been mechanically verified,
  \item it elucidates a few subtle issues in the basic metatheory of
    LF, and
  \item it constitutes a re-usable library of formalized results about
    LF, providing a foundation for verification of Twelf-style
    meta-reasoning about LF specifications, extensions to LF, or
    related type theories that are not as well-understood.
  \end{compactenum}

  This article is a revised and extended version of a previous
  conference paper presenting our initial formalization of the
  metatheory of LF~\cite{UrbanCheneyBerghofer08}.  The formal
  development described by this article can be obtained by request
  from the authors, and is available at
  \texttt{http://isabelle.in.tum.de/nominal/LF/}.%
\end{isamarkuptext}%
\isamarkuptrue%
\isamarkupsection{Background%
}
\isamarkuptrue%
\begin{isamarkuptext}%
\labelSec{background}%
\end{isamarkuptext}%
\isamarkuptrue%
\begin{isamarkuptext}%
\noindent
  This article assumes some familiarity with formalization in
  Isabelle/HOL and its ML-like notation for functions and definitions.
  We used the Nominal Datatype Package in
  Isabelle/HOL~\cite{UrbanBerghoferNorrish07,UrbanTasson05,Urban08JAR}
  to formalize the syntax and judgments of LF. The key features we
  rely upon are
  \begin{enumerate}
  \item support for \emph{nominal datatypes} with a built-in notion of
    binding (i.e.~$\alpha$-equivalence classes), 
  \item facilities for defining functions over nominal datatypes (such
    as substitution) by \emph{(nominal) primitive recursion}, and
  \item \emph{strong induction principles} for datatypes and inductive
    definitions that build in Barendregt-style renaming
    conventions.
  \end{enumerate}
  Together, these features make it possible to formalize most of the
  definitions and proofs following their paper versions closely.  We
  will not review the features of this system in this article, but
  will discuss details of the formalization only when they introduce
  complications.  The interested reader is referred to previous work
  on nominal techniques and the Nominal Datatype Package for further
  details~\cite{GabbayPitts02,Pitts06,UrbanBerghoferNorrish07,UrbanTasson05,Urban08JAR}.%
\end{isamarkuptext}%
\isamarkuptrue%
\isamarkupsubsection{Syntax of LF%
}
\isamarkuptrue%
\begin{isamarkuptext}%
The logical framework LF~\cite{HarperHonsellPlotkin93} is a
  dependently-typed lambda-calculus. We present it here following
  closely the article by Harper and
  Pfenning~\citeyear{HarperPfenning05}, to which we refer from now on
  as \HP for brevity.  The syntax of LF includes \emph{kinds},
  \emph{type families} and \emph{objects} defined by the grammar:

  \begin{center}
  \begin{tabular}{lr@ {\hspace{2mm}}c@ {\hspace{2mm}}l}
    \textit{Kinds} & \isa{K}, \isa{L} &::=& \isa{type} $\mid$ \isa{{\isasymPi}x{\isacharcolon}A{\isachardot}\ K}\\
    \textit{Type families} & \isa{A}, \isa{B} &::=& \isa{a} 
  $\mid$ \isa{{\isasymPi}x{\isacharcolon}A\isactrlisub {\isadigit{1}}{\isachardot}\ A\isactrlisub {\isadigit{2}}} $\mid$ \isa{A\ M}\\
    \textit{Objects} & \isa{M}, \isa{N} &::=& \isa{c} $\mid$ \isa{x} $\mid$ 
                      \isa{{\isasymlambda}x{\isacharcolon}A{\isachardot}\ M} $\mid$ \isa{M\isactrlisub {\isadigit{1}}\ M\isactrlisub {\isadigit{2}}}
  \end{tabular}
  \end{center}
  
  \noindent
  where variables $x$ and constants $c$ and $a$ are drawn from countably
  infinite, disjoint sets $\Var$ and $\Id$ of \emph{variables} and
  \emph{identifiers}, respectively.  Traditionally, LF has included
  $\lambda$-abstraction at the level of both types and objects.  However,
  \citeN{GeuversBarendsen99} established that type-level
  $\lambda$-abstraction is superfluous in LF.  Accordingly, \HP omits
  type-level $\lambda$-abstraction, and so do we.

  We formalize the syntax of LF using nominal datatypes since the
  constructors $\lambda$ and $\Pi$ bind variables.  Substitutions are
  represented as lists of variable-term pairs and we define capture
  avoiding substitution in the standard way as

  \begin{center}
  \begin{tabular}{r@ {\hspace{2mm}}c@ {\hspace{2mm}}ll}
  \isa{x{\isacharbrackleft}{\isasymsigma}{\isacharbrackright}} & $=$ & \isa{lookup\ {\isasymsigma}\ x}\\
  \isa{c{\isacharbrackleft}{\isasymsigma}{\isacharbrackright}} & $=$ & \isa{c}\\
  \isa{{\isacharparenleft}M\ N{\isacharparenright}{\isacharbrackleft}{\isasymsigma}{\isacharbrackright}} & $=$ & \isa{M{\isacharbrackleft}{\isasymsigma}{\isacharbrackright}\ N{\isacharbrackleft}{\isasymsigma}{\isacharbrackright}}\\
  \isa{{\isacharparenleft}{\isasymlambda}y{\isacharcolon}A{\isachardot}\ M{\isacharparenright}{\isacharbrackleft}{\isasymsigma}{\isacharbrackright}} & $=$ & \isa{{\isasymlambda}y{\isacharcolon}A{\isacharbrackleft}{\isasymsigma}{\isacharbrackright}{\isachardot}\ M{\isacharbrackleft}{\isasymsigma}{\isacharbrackright}}
  & provided \isa{y\ {\isasymsharp}\ {\isasymsigma}}\smallskip\\

  \isa{a{\isacharbrackleft}{\isasymsigma}{\isacharbrackright}} & $=$ & \isa{a}\\
  \isa{{\isacharparenleft}A\ M{\isacharparenright}{\isacharbrackleft}{\isasymsigma}{\isacharbrackright}} & $=$ & \isa{A{\isacharbrackleft}{\isasymsigma}{\isacharbrackright}\ M{\isacharbrackleft}{\isasymsigma}{\isacharbrackright}}\\
  \isa{{\isacharparenleft}{\isasymPi}y{\isacharcolon}A{\isachardot}\ B{\isacharparenright}{\isacharbrackleft}{\isasymsigma}{\isacharbrackright}} & $=$ & \isa{{\isasymPi}y{\isacharcolon}A{\isacharbrackleft}{\isasymsigma}{\isacharbrackright}{\isachardot}\ B{\isacharbrackleft}{\isasymsigma}{\isacharbrackright}}
  & provided \isa{y\ {\isasymsharp}\ {\isasymsigma}}\smallskip\\

  \isa{type{\isacharbrackleft}{\isasymsigma}{\isacharbrackright}} & $=$ & \isa{type}\\
  \isa{{\isacharparenleft}{\isasymPi}y{\isacharcolon}A{\isachardot}\ K{\isacharparenright}{\isacharbrackleft}{\isasymsigma}{\isacharbrackright}} & $=$ & \isa{{\isasymPi}y{\isacharcolon}A{\isacharbrackleft}{\isasymsigma}{\isacharbrackright}{\isachardot}\ K{\isacharbrackleft}{\isasymsigma}{\isacharbrackright}}
  & provided \isa{y\ {\isasymsharp}\ {\isasymsigma}}\\
  \end{tabular}
  \end{center}

  \noindent
  where the variable case is defined in terms of the auxiliary function 
  \isa{lookup}:
  
  \begin{center}
  \begin{tabular}{@ {}l@ {}}
  \isa{lookup\ {\isacharbrackleft}{\isacharbrackright}\ x\ {\isacharequal}\ x}\\
  \isa{lookup\ {\isacharparenleft}{\isacharparenleft}y{\isacharcomma}\ M{\isacharparenright}{\isacharcolon}{\isacharcolon}{\isasymsigma}{\isacharparenright}\ x\ {\isacharequal}\ {\isacharparenleft}\textrm{if}\ x\ {\isacharequal}\ y\ \textrm{then}\ M\ \textrm{else}\ lookup\ {\isasymsigma}\ x{\isacharparenright}}\\
  \end{tabular}  
  \end{center}

  \noindent
  The side-conditions \isa{y\ {\isasymsharp}\ {\isasymsigma}}
  in the above definition are freshness constraints provided
  automatically by the Nominal Datatype Package and stand for \isa{y} not occurring freely in the substitution \isa{{\isasymsigma}}. Substitution for a single variable is defined as a
  special case: \isa{{\isacharparenleft}$-${\isacharparenright}{\isacharbrackleft}x{\isacharcolon}{\isacharequal}M{\isacharbrackright}} $\dn$ \isa{{\isacharparenleft}{\isacharminus}{\isacharparenright}{\isacharbrackleft}{\isacharparenleft}x{\isacharcomma}M{\isacharparenright}{\isacharbrackright}}.

  We use ML-like notation $[]$ for the empty list and $x::L$ for list
  construction.  LF includes \emph{signatures} \isa{{\isasymSigma}}
  and \emph{contexts} \isa{{\isasymGamma}}, both of which we represent
  as lists of pairs.  The former consist of pairs of the form \isa{{\isacharparenleft}c{\isacharcomma}\ A{\isacharparenright}} or \isa{{\isacharparenleft}a{\isacharcomma}\ K{\isacharparenright}} associating the constant $c$ with type
  $A$ and the constant $a$ with kind $K$ respectively, and the latter
  consists of pairs \isa{{\isacharparenleft}x{\isacharcomma}\ A{\isacharparenright}} associating the variable \isa{x} with type $A$.  Accordingly, we write \mbox{\isa{{\isacharparenleft}x{\isacharcomma}\ A{\isacharparenright}{\isacharcolon}{\isacharcolon}{\isasymGamma}}} for context construction (rather than
  $\Gamma,x{:}A$), \isa{{\isasymGamma}\ {\isacharat}\ {\isasymGamma}{\isacharprime}} for context concatenation
  and \mbox{\isa{{\isacharparenleft}x{\isacharcomma}\ A{\isacharparenright}\ {\isasymin}\ {\isasymGamma}}} for context membership
  (similarly for \isa{{\isasymSigma}}). Context inclusion is
  defined as follows:
  \begin{center}
  \isa{{\isasymGamma}\isactrlisub {\isadigit{1}}\ {\isasymsubseteq}\ {\isasymGamma}\isactrlisub {\isadigit{2}}} $\dn$ \isa{{\isasymforall}x\ A{\isachardot}\ {\isacharparenleft}x{\isacharcomma}\ A{\isacharparenright}\ {\isasymin}\ {\isasymGamma}\isactrlisub {\isadigit{1}}} implies \isa{{\isacharparenleft}x{\isacharcomma}\ A{\isacharparenright}\ {\isasymin}\ {\isasymGamma}\isactrlisub {\isadigit{2}}}
  \end{center}%
\end{isamarkuptext}%
\isamarkuptrue%
\isamarkupsubsection{Validity and Definitional Equivalence%
}
\isamarkuptrue%
\begin{isamarkuptext}%
\HP defines two judgments for identifying valid signatures and contexts,
  which we formalize in \refFig{valid-sig-ctx}. In contrast with \HP, we make
  explicit that the new bindings do not occur previously in \isa{{\isasymSigma}} or
  \isa{{\isasymGamma}}, using freshness constraints such as \mbox{\isa{x\ {\isasymsharp}\ {\isasymGamma}}}.  We also
  make the dependence of all judgments on \isa{{\isasymSigma}} explicit.

  Central in \HP are the definitions of the validity and definitional
  equivalence judgments for LF, and of algorithmic judgments for
  checking equivalence.  The validity and definitional equivalence
  rules are shown in \refFig{lf-wf} and \ref{fig:lf-equiv}.  There are
  three judgments for validity and three for equivalence corresponding
  to objects, type families and kinds respectively:
  \begin{center}
\begin{tabular}{cccc}
 & 
Objects
& 
Type families
& 
Kinds
\\
Validity 
&  
\isa{{\isasymGamma}\ {\isasymturnstile}\isactrlisub {\isasymSigma}\ M\ {\isacharcolon}\ A}
&
\isa{{\isasymGamma}\ {\isasymturnstile}\isactrlisub {\isasymSigma}\ A\ {\isacharcolon}\ K}
&
\isa{{\isasymGamma}\ {\isasymturnstile}\isactrlisub {\isasymSigma}\ K\ {\isacharcolon}\ kind}
\\
Equivalence
&
   \isa{{\isasymGamma}\ {\isasymturnstile}\isactrlisub {\isasymSigma}\ M\ {\isacharequal}\ N\ {\isacharcolon}\ A}
& 
  \isa{{\isasymGamma}\ {\isasymturnstile}\isactrlisub {\isasymSigma}\ A\ {\isacharequal}\ B\ {\isacharcolon}\ K}
&
   \isa{{\isasymGamma}\ {\isasymturnstile}\isactrlisub {\isasymSigma}\ K\ {\isacharequal}\ L\ {\isacharcolon}\ kind}
\end{tabular}
\end{center}
  These six judgments are defined simultaneously with signature
  validity (\isa{{\isasymturnstile}\ {\isasymSigma}\ sig} ) and context validity
  (\isa{{\isasymturnstile}\isactrlisub {\isasymSigma}\ {\isasymGamma}\ ctx} ) by induction.  We
  added explicit validity hypotheses to some of the rules; these are
  left implicit in \HP.  We also added some (redundant) freshness
  constraints to some rules in order to be able to use strong
  induction principles~\cite{UrbanBerghoferNorrish07}.

  %%%%%%%%%%%%%%%%%%%%%%%%%%%%%%%%%%%%%%%%%%%%%%%%%%%%%%%%%%%%%%%%%%
  \begin{figure}
   \fbox{\isa{{\isasymturnstile}\ {\isasymSigma}\ sig}}
   \begin{center}
      \begin{tabular}{@ {}c@ {}}
        \isa{\mbox{}\inferrule{\mbox{}}{\mbox{{\isasymturnstile}\ {\isacharbrackleft}{\isacharbrackright}\ sig}}}$\qquad$
        \isa{\mbox{}\inferrule{\mbox{{\isasymturnstile}\ {\isasymSigma}\ sig}\\\ \mbox{{\isacharbrackleft}{\isacharbrackright}\ {\isasymturnstile}\isactrlisub {\isasymSigma}\ K\ {\isacharcolon}\ kind}\\\ \mbox{a\ {\isasymsharp}\ {\isasymSigma}}}{\mbox{{\isasymturnstile}\ {\isacharparenleft}a{\isacharcomma}\ K{\isacharparenright}{\isacharcolon}{\isacharcolon}{\isasymSigma}\ sig}}}$\qquad$
        \isa{\mbox{}\inferrule{\mbox{{\isasymturnstile}\ {\isasymSigma}\ sig}\\\ \mbox{{\isacharbrackleft}{\isacharbrackright}\ {\isasymturnstile}\isactrlisub {\isasymSigma}\ A\ {\isacharcolon}\ type}\\\ \mbox{c\ {\isasymsharp}\ {\isasymSigma}}}{\mbox{{\isasymturnstile}\ {\isacharparenleft}c{\isacharcomma}\ A{\isacharparenright}{\isacharcolon}{\isacharcolon}{\isasymSigma}\ sig}}}
      \end{tabular}
    \end{center}
    \fbox{\isa{{\isasymturnstile}\isactrlisub {\isasymSigma}\ {\isasymGamma}\ ctx}}
    \begin{center}
      \begin{tabular}{@ {}c@ {}}
        \isa{\mbox{}\inferrule{\mbox{{\isasymturnstile}\ {\isasymSigma}\ sig}}{\mbox{{\isasymturnstile}\isactrlisub {\isasymSigma}\ {\isacharbrackleft}{\isacharbrackright}\ ctx}}}$\qquad$
        \isa{\mbox{}\inferrule{\mbox{{\isasymturnstile}\isactrlisub {\isasymSigma}\ {\isasymGamma}\ ctx}\\\ \mbox{{\isasymGamma}\ {\isasymturnstile}\isactrlisub {\isasymSigma}\ A\ {\isacharcolon}\ type}\\\ \mbox{x\ {\isasymsharp}\ {\isasymGamma}}}{\mbox{{\isasymturnstile}\isactrlisub {\isasymSigma}\ {\isacharparenleft}x{\isacharcomma}\ A{\isacharparenright}{\isacharcolon}{\isacharcolon}{\isasymGamma}\ ctx}}}
      \end{tabular}
    \end{center}
    \caption{Validity rules for signatures and contexts}
\label{fig:valid-sig-ctx}
\fbox{\isa{{\isasymGamma}\ {\isasymturnstile}\isactrlisub {\isasymSigma}\ M\ {\isacharcolon}\ A}}
\begin{center}
      \begin{tabular}{@ {}c@ {}}
    \isa{\mbox{}\inferrule{\mbox{{\isasymturnstile}\isactrlisub {\isasymSigma}\ {\isasymGamma}\ ctx}\\\ \mbox{{\isacharparenleft}x{\isacharcomma}\ A{\isacharparenright}\ {\isasymin}\ {\isasymGamma}}}{\mbox{{\isasymGamma}\ {\isasymturnstile}\isactrlisub {\isasymSigma}\ x\ {\isacharcolon}\ A}}}$\;\;\;$ 
    \isa{\mbox{}\inferrule{\mbox{{\isasymturnstile}\isactrlisub {\isasymSigma}\ {\isasymGamma}\ ctx}\\\ \mbox{{\isacharparenleft}c{\isacharcomma}\ A{\isacharparenright}\ {\isasymin}\ {\isasymSigma}}}{\mbox{{\isasymGamma}\ {\isasymturnstile}\isactrlisub {\isasymSigma}\ c\ {\isacharcolon}\ A}}}\medskip\\%$\;\;\;$
    % \begin{minipage}[b]{8cm}
    \isa{\mbox{}\inferrule{\mbox{{\isasymGamma}\ {\isasymturnstile}\isactrlisub {\isasymSigma}\ M\isactrlisub {\isadigit{1}}\ {\isacharcolon}\ {\isasymPi}x{\isacharcolon}A\isactrlisub {\isadigit{2}}{\isachardot}\ A\isactrlisub {\isadigit{1}}}\\\ \mbox{{\isasymGamma}\ {\isasymturnstile}\isactrlisub {\isasymSigma}\ M\isactrlisub {\isadigit{2}}\ {\isacharcolon}\ A\isactrlisub {\isadigit{2}}}\\\ \mbox{x\ {\isasymsharp}\ {\isasymGamma}}}{\mbox{{\isasymGamma}\ {\isasymturnstile}\isactrlisub {\isasymSigma}\ M\isactrlisub {\isadigit{1}}\ M\isactrlisub {\isadigit{2}}\ {\isacharcolon}\ A\isactrlisub {\isadigit{1}}{\isacharbrackleft}x{\isacharcolon}{\isacharequal}M\isactrlisub {\isadigit{2}}{\isacharbrackright}}}}
    % \end{minipage}
    \medskip\\

    % \begin{minipage}[b]{6cm}
    \isa{\mbox{}\inferrule{\mbox{{\isasymGamma}\ {\isasymturnstile}\isactrlisub {\isasymSigma}\ A\isactrlisub {\isadigit{1}}\ {\isacharcolon}\ type}\\\ \mbox{{\isacharparenleft}x{\isacharcomma}\ A\isactrlisub {\isadigit{1}}{\isacharparenright}{\isacharcolon}{\isacharcolon}{\isasymGamma}\ {\isasymturnstile}\isactrlisub {\isasymSigma}\ M\isactrlisub {\isadigit{2}}\ {\isacharcolon}\ A\isactrlisub {\isadigit{2}}}\\\ \mbox{x\ {\isasymsharp}\ {\isacharparenleft}{\isasymGamma}{\isacharcomma}\ A\isactrlisub {\isadigit{1}}{\isacharparenright}}}{\mbox{{\isasymGamma}\ {\isasymturnstile}\isactrlisub {\isasymSigma}\ {\isasymlambda}x{\isacharcolon}A\isactrlisub {\isadigit{1}}{\isachardot}\ M\isactrlisub {\isadigit{2}}\ {\isacharcolon}\ {\isasymPi}x{\isacharcolon}A\isactrlisub {\isadigit{1}}{\isachardot}\ A\isactrlisub {\isadigit{2}}}}}
    % \end{minipage}
    % $\;\;\;$
    \medskip\\
    \isa{\mbox{}\inferrule{\mbox{{\isasymGamma}\ {\isasymturnstile}\isactrlisub {\isasymSigma}\ M\ {\isacharcolon}\ A}\\\ \mbox{{\isasymGamma}\ {\isasymturnstile}\isactrlisub {\isasymSigma}\ A\ {\isacharequal}\ B\ {\isacharcolon}\ type}}{\mbox{{\isasymGamma}\ {\isasymturnstile}\isactrlisub {\isasymSigma}\ M\ {\isacharcolon}\ B}}}
 \end{tabular}
\end{center}
\fbox{\isa{{\isasymGamma}\ {\isasymturnstile}\isactrlisub {\isasymSigma}\ A\ {\isacharcolon}\ K}}
\begin{center}
      \begin{tabular}{@ {}c@ {}}
    \isa{\mbox{}\inferrule{\mbox{{\isasymturnstile}\isactrlisub {\isasymSigma}\ {\isasymGamma}\ ctx}\\\ \mbox{{\isacharparenleft}a{\isacharcomma}\ K{\isacharparenright}\ {\isasymin}\ {\isasymSigma}}}{\mbox{{\isasymGamma}\ {\isasymturnstile}\isactrlisub {\isasymSigma}\ a\ {\isacharcolon}\ K}}}\medskip\\%$\;\;\;$
    \isa{\mbox{}\inferrule{\mbox{{\isasymGamma}\ {\isasymturnstile}\isactrlisub {\isasymSigma}\ A\ {\isacharcolon}\ {\isasymPi}x{\isacharcolon}B{\isachardot}\ K}\\\ \mbox{{\isasymGamma}\ {\isasymturnstile}\isactrlisub {\isasymSigma}\ M\ {\isacharcolon}\ B}\\\ \mbox{x\ {\isasymsharp}\ {\isasymGamma}}}{\mbox{{\isasymGamma}\ {\isasymturnstile}\isactrlisub {\isasymSigma}\ A\ M\ {\isacharcolon}\ K{\isacharbrackleft}x{\isacharcolon}{\isacharequal}M{\isacharbrackright}}}}\medskip\\

    % \begin{minipage}[b]{6cm}
    \isa{\mbox{}\inferrule{\mbox{{\isasymGamma}\ {\isasymturnstile}\isactrlisub {\isasymSigma}\ A\isactrlisub {\isadigit{1}}\ {\isacharcolon}\ type}\\\ \mbox{{\isacharparenleft}x{\isacharcomma}\ A\isactrlisub {\isadigit{1}}{\isacharparenright}{\isacharcolon}{\isacharcolon}{\isasymGamma}\ {\isasymturnstile}\isactrlisub {\isasymSigma}\ A\isactrlisub {\isadigit{2}}\ {\isacharcolon}\ type}\\\ \mbox{x\ {\isasymsharp}\ {\isacharparenleft}{\isasymGamma}{\isacharcomma}\ A\isactrlisub {\isadigit{1}}{\isacharparenright}}}{\mbox{{\isasymGamma}\ {\isasymturnstile}\isactrlisub {\isasymSigma}\ {\isasymPi}x{\isacharcolon}A\isactrlisub {\isadigit{1}}{\isachardot}\ A\isactrlisub {\isadigit{2}}\ {\isacharcolon}\ type}}}
    % \end{minipage}
    % $\;\;\;$
    \medskip\\
    \isa{\mbox{}\inferrule{\mbox{{\isasymGamma}\ {\isasymturnstile}\isactrlisub {\isasymSigma}\ A\ {\isacharcolon}\ K}\\\ \mbox{{\isasymGamma}\ {\isasymturnstile}\isactrlisub {\isasymSigma}\ K\ {\isacharequal}\ L\ {\isacharcolon}\ kind}}{\mbox{{\isasymGamma}\ {\isasymturnstile}\isactrlisub {\isasymSigma}\ A\ {\isacharcolon}\ L}}}
  \end{tabular}
\end{center}
\fbox{\isa{{\isasymGamma}\ {\isasymturnstile}\isactrlisub {\isasymSigma}\ K\ {\isacharcolon}\ kind}}
\begin{center}
  \begin{tabular}{@ {}c@ {}}
    \isa{\mbox{}\inferrule{\mbox{{\isasymturnstile}\isactrlisub {\isasymSigma}\ {\isasymGamma}\ ctx}}{\mbox{{\isasymGamma}\ {\isasymturnstile}\isactrlisub {\isasymSigma}\ type\ {\isacharcolon}\ kind}}}\medskip\\
    \isa{\mbox{}\inferrule{\mbox{{\isasymGamma}\ {\isasymturnstile}\isactrlisub {\isasymSigma}\ A\ {\isacharcolon}\ type}\\\ \mbox{{\isacharparenleft}x{\isacharcomma}\ A{\isacharparenright}{\isacharcolon}{\isacharcolon}{\isasymGamma}\ {\isasymturnstile}\isactrlisub {\isasymSigma}\ K\ {\isacharcolon}\ kind}\\\ \mbox{x\ {\isasymsharp}\ {\isacharparenleft}{\isasymGamma}{\isacharcomma}\ A{\isacharparenright}}}{\mbox{{\isasymGamma}\ {\isasymturnstile}\isactrlisub {\isasymSigma}\ {\isasymPi}x{\isacharcolon}A{\isachardot}\ K\ {\isacharcolon}\ kind}}}\smallskip\\
  \end{tabular}
\end{center}
  \caption{Validity rules for kinds, type
    families and objects.}\labelFig{lf-wf}
  \end{figure}
  %%%%%%%%%%%%%%%%%%%%%%%%%%%%%%%%%%%%%%%%%%%%%%%%%%%%%%%%%%%%%%%%%%

  %%%%%%%%%%%%%%%%%%%%%%%%%%%%%%%%%%%%%%%%%%%%%%%%%%%%%%%%%%%%%%%%%%
  \begin{figure}

  \fbox{\isa{{\isasymGamma}\ {\isasymturnstile}\isactrlisub {\isasymSigma}\ M\ {\isacharequal}\ N\ {\isacharcolon}\ A}}
  \begin{center}\begin{tabular}{@ {}c@ {}}
      \isa{\mbox{}\inferrule{\mbox{{\isasymturnstile}\isactrlisub {\isasymSigma}\ {\isasymGamma}\ ctx}\\\ \mbox{{\isacharparenleft}x{\isacharcomma}\ A{\isacharparenright}\ {\isasymin}\ {\isasymGamma}}}{\mbox{{\isasymGamma}\ {\isasymturnstile}\isactrlisub {\isasymSigma}\ x\ {\isacharequal}\ x\ {\isacharcolon}\ A}}}$\;\;\;$ 
      \isa{\mbox{}\inferrule{\mbox{{\isasymturnstile}\isactrlisub {\isasymSigma}\ {\isasymGamma}\ ctx}\\\ \mbox{{\isacharparenleft}c{\isacharcomma}\ A{\isacharparenright}\ {\isasymin}\ {\isasymSigma}}}{\mbox{{\isasymGamma}\ {\isasymturnstile}\isactrlisub {\isasymSigma}\ c\ {\isacharequal}\ c\ {\isacharcolon}\ A}}}\medskip\\
      
 %     \begin{minipage}[b]{5.0cm}
        \isa{\mbox{}\inferrule{\mbox{{\isasymGamma}\ {\isasymturnstile}\isactrlisub {\isasymSigma}\ M\isactrlisub {\isadigit{1}}\ {\isacharequal}\ N\isactrlisub {\isadigit{1}}\ {\isacharcolon}\ {\isasymPi}x{\isacharcolon}A\isactrlisub {\isadigit{2}}{\isachardot}\ A\isactrlisub {\isadigit{1}}}\\\ \mbox{{\isasymGamma}\ {\isasymturnstile}\isactrlisub {\isasymSigma}\ M\isactrlisub {\isadigit{2}}\ {\isacharequal}\ N\isactrlisub {\isadigit{2}}\ {\isacharcolon}\ A\isactrlisub {\isadigit{2}}}\\\ \mbox{x\ {\isasymsharp}\ {\isasymGamma}}}{\mbox{{\isasymGamma}\ {\isasymturnstile}\isactrlisub {\isasymSigma}\ M\isactrlisub {\isadigit{1}}\ M\isactrlisub {\isadigit{2}}\ {\isacharequal}\ N\isactrlisub {\isadigit{1}}\ N\isactrlisub {\isadigit{2}}\ {\isacharcolon}\ A\isactrlisub {\isadigit{1}}{\isacharbrackleft}x{\isacharcolon}{\isacharequal}M\isactrlisub {\isadigit{2}}{\isacharbrackright}}}}
 %     \end{minipage}$\;\;\;$
\medskip\\
%      \begin{minipage}[b]{6.5cm}
        \isa{\mbox{}\inferrule{\mbox{{\isasymGamma}\ {\isasymturnstile}\isactrlisub {\isasymSigma}\ A\isactrlisub {\isadigit{1}}{\isacharprime}\ {\isacharequal}\ A\isactrlisub {\isadigit{1}}\ {\isacharcolon}\ type}\\\ \mbox{{\isasymGamma}\ {\isasymturnstile}\isactrlisub {\isasymSigma}\ A\isactrlisub {\isadigit{1}}{\isacharprime}{\isacharprime}\ {\isacharequal}\ A\isactrlisub {\isadigit{1}}\ {\isacharcolon}\ type}\\\ \mbox{{\isasymGamma}\ {\isasymturnstile}\isactrlisub {\isasymSigma}\ A\isactrlisub {\isadigit{1}}\ {\isacharcolon}\ type}\\\ \mbox{{\isacharparenleft}x{\isacharcomma}\ A\isactrlisub {\isadigit{1}}{\isacharparenright}{\isacharcolon}{\isacharcolon}{\isasymGamma}\ {\isasymturnstile}\isactrlisub {\isasymSigma}\ M\isactrlisub {\isadigit{2}}\ {\isacharequal}\ N\isactrlisub {\isadigit{2}}\ {\isacharcolon}\ A\isactrlisub {\isadigit{2}}}\\\ \mbox{x\ {\isasymsharp}\ {\isasymGamma}}}{\mbox{{\isasymGamma}\ {\isasymturnstile}\isactrlisub {\isasymSigma}\ {\isasymlambda}x{\isacharcolon}A\isactrlisub {\isadigit{1}}{\isacharprime}{\isachardot}\ M\isactrlisub {\isadigit{2}}\ {\isacharequal}\ {\isasymlambda}x{\isacharcolon}A\isactrlisub {\isadigit{1}}{\isacharprime}{\isacharprime}{\isachardot}\ N\isactrlisub {\isadigit{2}}\ {\isacharcolon}\ {\isasymPi}x{\isacharcolon}A\isactrlisub {\isadigit{1}}{\isachardot}\ A\isactrlisub {\isadigit{2}}}}}
%      \end{minipage}
        \medskip\\
      
%      \begin{minipage}[b]{5.8cm}
        \isa{\mbox{}\inferrule{\mbox{{\isasymGamma}\ {\isasymturnstile}\isactrlisub {\isasymSigma}\ M\ {\isacharcolon}\ {\isasymPi}x{\isacharcolon}A\isactrlisub {\isadigit{1}}{\isachardot}\ A\isactrlisub {\isadigit{2}}}\\\ \mbox{{\isasymGamma}\ {\isasymturnstile}\isactrlisub {\isasymSigma}\ N\ {\isacharcolon}\ {\isasymPi}x{\isacharcolon}A\isactrlisub {\isadigit{1}}{\isachardot}\ A\isactrlisub {\isadigit{2}}}\\\ \mbox{{\isasymGamma}\ {\isasymturnstile}\isactrlisub {\isasymSigma}\ A\isactrlisub {\isadigit{1}}\ {\isacharcolon}\ type}\\\ \mbox{{\isacharparenleft}x{\isacharcomma}\ A\isactrlisub {\isadigit{1}}{\isacharparenright}{\isacharcolon}{\isacharcolon}{\isasymGamma}\ {\isasymturnstile}\isactrlisub {\isasymSigma}\ M\ x\ {\isacharequal}\ N\ x\ {\isacharcolon}\ A\isactrlisub {\isadigit{2}}}\\\ \mbox{x\ {\isasymsharp}\ {\isasymGamma}}}{\mbox{{\isasymGamma}\ {\isasymturnstile}\isactrlisub {\isasymSigma}\ M\ {\isacharequal}\ N\ {\isacharcolon}\ {\isasymPi}x{\isacharcolon}A\isactrlisub {\isadigit{1}}{\isachardot}\ A\isactrlisub {\isadigit{2}}}}}
%      \end{minipage}
        \medskip\\
%      \begin{minipage}[b]{7cm}
        \isa{\mbox{}\inferrule{\mbox{{\isasymGamma}\ {\isasymturnstile}\isactrlisub {\isasymSigma}\ A\isactrlisub {\isadigit{1}}\ {\isacharcolon}\ type}\\\ \mbox{{\isacharparenleft}x{\isacharcomma}\ A\isactrlisub {\isadigit{1}}{\isacharparenright}{\isacharcolon}{\isacharcolon}{\isasymGamma}\ {\isasymturnstile}\isactrlisub {\isasymSigma}\ M\isactrlisub {\isadigit{2}}\ {\isacharequal}\ N\isactrlisub {\isadigit{2}}\ {\isacharcolon}\ A\isactrlisub {\isadigit{2}}}\\\ \mbox{{\isasymGamma}\ {\isasymturnstile}\isactrlisub {\isasymSigma}\ M\isactrlisub {\isadigit{1}}\ {\isacharequal}\ N\isactrlisub {\isadigit{1}}\ {\isacharcolon}\ A\isactrlisub {\isadigit{1}}}\\\ \mbox{x\ {\isasymsharp}\ {\isasymGamma}}}{\mbox{{\isasymGamma}\ {\isasymturnstile}\isactrlisub {\isasymSigma}\ {\isacharparenleft}{\isasymlambda}x{\isacharcolon}A\isactrlisub {\isadigit{1}}{\isachardot}\ M\isactrlisub {\isadigit{2}}{\isacharparenright}\ M\isactrlisub {\isadigit{1}}\ {\isacharequal}\ N\isactrlisub {\isadigit{2}}{\isacharbrackleft}x{\isacharcolon}{\isacharequal}N\isactrlisub {\isadigit{1}}{\isacharbrackright}\ {\isacharcolon}\ A\isactrlisub {\isadigit{2}}{\isacharbrackleft}x{\isacharcolon}{\isacharequal}M\isactrlisub {\isadigit{1}}{\isacharbrackright}}}}
%      \end{minipage}
        \medskip\\
      
      \isa{\mbox{}\inferrule{\mbox{{\isasymGamma}\ {\isasymturnstile}\isactrlisub {\isasymSigma}\ M\ {\isacharequal}\ N\ {\isacharcolon}\ A}}{\mbox{{\isasymGamma}\ {\isasymturnstile}\isactrlisub {\isasymSigma}\ N\ {\isacharequal}\ M\ {\isacharcolon}\ A}}}$\qquad$
      \isa{\mbox{}\inferrule{\mbox{{\isasymGamma}\ {\isasymturnstile}\isactrlisub {\isasymSigma}\ M\ {\isacharequal}\ N\ {\isacharcolon}\ A}\\\ \mbox{{\isasymGamma}\ {\isasymturnstile}\isactrlisub {\isasymSigma}\ N\ {\isacharequal}\ P\ {\isacharcolon}\ A}}{\mbox{{\isasymGamma}\ {\isasymturnstile}\isactrlisub {\isasymSigma}\ M\ {\isacharequal}\ P\ {\isacharcolon}\ A}}}\medskip\\
      \isa{\mbox{}\inferrule{\mbox{{\isasymGamma}\ {\isasymturnstile}\isactrlisub {\isasymSigma}\ M\ {\isacharequal}\ N\ {\isacharcolon}\ A}\\\ \mbox{{\isasymGamma}\ {\isasymturnstile}\isactrlisub {\isasymSigma}\ A\ {\isacharequal}\ B\ {\isacharcolon}\ type}}{\mbox{{\isasymGamma}\ {\isasymturnstile}\isactrlisub {\isasymSigma}\ M\ {\isacharequal}\ N\ {\isacharcolon}\ B}}}
    \end{tabular}\end{center}

  \fbox{\isa{{\isasymGamma}\ {\isasymturnstile}\isactrlisub {\isasymSigma}\ A\ {\isacharequal}\ B\ {\isacharcolon}\ K}}
  \begin{center}\begin{tabular}{@ {}c@ {}}
  \isa{\mbox{}\inferrule{\mbox{{\isasymturnstile}\isactrlisub {\isasymSigma}\ {\isasymGamma}\ ctx}\\\ \mbox{{\isacharparenleft}a{\isacharcomma}\ K{\isacharparenright}\ {\isasymin}\ {\isasymSigma}}}{\mbox{{\isasymGamma}\ {\isasymturnstile}\isactrlisub {\isasymSigma}\ a\ {\isacharequal}\ a\ {\isacharcolon}\ K}}}\medskip\\
%  \begin{minipage}[b]{4cm}
  \isa{\mbox{}\inferrule{\mbox{{\isasymGamma}\ {\isasymturnstile}\isactrlisub {\isasymSigma}\ A\ {\isacharequal}\ B\ {\isacharcolon}\ {\isasymPi}x{\isacharcolon}C{\isachardot}\ K}\\\ \mbox{{\isasymGamma}\ {\isasymturnstile}\isactrlisub {\isasymSigma}\ M\ {\isacharequal}\ N\ {\isacharcolon}\ C}\\\ \mbox{x\ {\isasymsharp}\ {\isasymGamma}}}{\mbox{{\isasymGamma}\ {\isasymturnstile}\isactrlisub {\isasymSigma}\ A\ M\ {\isacharequal}\ B\ N\ {\isacharcolon}\ K{\isacharbrackleft}x{\isacharcolon}{\isacharequal}M{\isacharbrackright}}}}
%  \end{minipage}
  \medskip\\

  %\begin{minipage}[b]{5.5cm}
  \isa{\mbox{}\inferrule{\mbox{{\isasymGamma}\ {\isasymturnstile}\isactrlisub {\isasymSigma}\ A\isactrlisub {\isadigit{1}}\ {\isacharequal}\ B\isactrlisub {\isadigit{1}}\ {\isacharcolon}\ type}\\\ \mbox{{\isasymGamma}\ {\isasymturnstile}\isactrlisub {\isasymSigma}\ A\isactrlisub {\isadigit{1}}\ {\isacharcolon}\ type}\\\ \mbox{{\isacharparenleft}x{\isacharcomma}\ A\isactrlisub {\isadigit{1}}{\isacharparenright}{\isacharcolon}{\isacharcolon}{\isasymGamma}\ {\isasymturnstile}\isactrlisub {\isasymSigma}\ A\isactrlisub {\isadigit{2}}\ {\isacharequal}\ B\isactrlisub {\isadigit{2}}\ {\isacharcolon}\ type}\\\ \mbox{x\ {\isasymsharp}\ {\isasymGamma}}}{\mbox{{\isasymGamma}\ {\isasymturnstile}\isactrlisub {\isasymSigma}\ {\isasymPi}x{\isacharcolon}A\isactrlisub {\isadigit{1}}{\isachardot}\ A\isactrlisub {\isadigit{2}}\ {\isacharequal}\ {\isasymPi}x{\isacharcolon}B\isactrlisub {\isadigit{1}}{\isachardot}\ B\isactrlisub {\isadigit{2}}\ {\isacharcolon}\ type}}}
  %\end{minipage}
  \medskip\\
  \isa{\mbox{}\inferrule{\mbox{{\isasymGamma}\ {\isasymturnstile}\isactrlisub {\isasymSigma}\ A\ {\isacharequal}\ B\ {\isacharcolon}\ K}}{\mbox{{\isasymGamma}\ {\isasymturnstile}\isactrlisub {\isasymSigma}\ B\ {\isacharequal}\ A\ {\isacharcolon}\ K}}}$\qquad$

  \isa{\mbox{}\inferrule{\mbox{{\isasymGamma}\ {\isasymturnstile}\isactrlisub {\isasymSigma}\ A\ {\isacharequal}\ B\ {\isacharcolon}\ K}\\\ \mbox{{\isasymGamma}\ {\isasymturnstile}\isactrlisub {\isasymSigma}\ B\ {\isacharequal}\ C\ {\isacharcolon}\ K}}{\mbox{{\isasymGamma}\ {\isasymturnstile}\isactrlisub {\isasymSigma}\ A\ {\isacharequal}\ C\ {\isacharcolon}\ K}}}
  \medskip\\
  \isa{\mbox{}\inferrule{\mbox{{\isasymGamma}\ {\isasymturnstile}\isactrlisub {\isasymSigma}\ A\ {\isacharequal}\ B\ {\isacharcolon}\ K}\\\ \mbox{{\isasymGamma}\ {\isasymturnstile}\isactrlisub {\isasymSigma}\ K\ {\isacharequal}\ L\ {\isacharcolon}\ kind}}{\mbox{{\isasymGamma}\ {\isasymturnstile}\isactrlisub {\isasymSigma}\ A\ {\isacharequal}\ B\ {\isacharcolon}\ L}}}
     \end{tabular}\end{center}

  {\fbox{\isa{{\isasymGamma}\ {\isasymturnstile}\isactrlisub {\isasymSigma}\ K\ {\isacharequal}\ L\ {\isacharcolon}\ kind}}}
  \begin{center}\begin{tabular}{@ {}c@ {}}

  \isa{\mbox{}\inferrule{\mbox{{\isasymturnstile}\isactrlisub {\isasymSigma}\ {\isasymGamma}\ ctx}}{\mbox{{\isasymGamma}\ {\isasymturnstile}\isactrlisub {\isasymSigma}\ type\ {\isacharequal}\ type\ {\isacharcolon}\ kind}}}\medskip\\
  %\begin{minipage}[b]{5.5cm}
  \isa{\mbox{}\inferrule{\mbox{{\isasymGamma}\ {\isasymturnstile}\isactrlisub {\isasymSigma}\ A\ {\isacharequal}\ B\ {\isacharcolon}\ type}\\\ \mbox{{\isasymGamma}\ {\isasymturnstile}\isactrlisub {\isasymSigma}\ A\ {\isacharcolon}\ type}\\\ \mbox{{\isacharparenleft}x{\isacharcomma}\ A{\isacharparenright}{\isacharcolon}{\isacharcolon}{\isasymGamma}\ {\isasymturnstile}\isactrlisub {\isasymSigma}\ K\ {\isacharequal}\ L\ {\isacharcolon}\ kind}\\\ \mbox{x\ {\isasymsharp}\ {\isasymGamma}}}{\mbox{{\isasymGamma}\ {\isasymturnstile}\isactrlisub {\isasymSigma}\ {\isasymPi}x{\isacharcolon}A{\isachardot}\ K\ {\isacharequal}\ {\isasymPi}x{\isacharcolon}B{\isachardot}\ L\ {\isacharcolon}\ kind}}}
  %\end{minipage}
  \medskip\\

  \isa{\mbox{}\inferrule{\mbox{{\isasymGamma}\ {\isasymturnstile}\isactrlisub {\isasymSigma}\ K\ {\isacharequal}\ L\ {\isacharcolon}\ kind}}{\mbox{{\isasymGamma}\ {\isasymturnstile}\isactrlisub {\isasymSigma}\ L\ {\isacharequal}\ K\ {\isacharcolon}\ kind}}}$\qquad$
  \isa{\mbox{}\inferrule{\mbox{{\isasymGamma}\ {\isasymturnstile}\isactrlisub {\isasymSigma}\ K\ {\isacharequal}\ L\ {\isacharcolon}\ kind}\\\ \mbox{{\isasymGamma}\ {\isasymturnstile}\isactrlisub {\isasymSigma}\ L\ {\isacharequal}\ L{\isacharprime}\ {\isacharcolon}\ kind}}{\mbox{{\isasymGamma}\ {\isasymturnstile}\isactrlisub {\isasymSigma}\ K\ {\isacharequal}\ L{\isacharprime}\ {\isacharcolon}\ kind}}}
     \end{tabular}\end{center}
   
  \caption{Definitional equivalence rules for kinds, type
    families and objects.}\labelFig{lf-equiv}
  \end{figure}%
\end{isamarkuptext}%
\isamarkuptrue%
\isamarkupsubsection{Algorithmic Equivalence%
}
\isamarkuptrue%
\begin{isamarkuptext}%
The definitional equivalence judgment captures equivalence between
  LF terms, types and kinds declaratively, but it is highly
  nondeterministic due to the symmetry, transitivity and conversion
  rules.  Accordingly, \HP introduces algorithmic equivalence
  judgments that are type- and syntax-directed, and the main
  contribution of that article is the proof that the algorithmic and
  declarative systems coincide.

  A crucial point of the algorithm in \HP is that it does not analyze
  the precise types of objects or kinds of types during equivalence checking; rather
  it only uses approximate \emph{simple types} \isa{{\isasymtau}} and
  \emph{simple kinds} \isa{{\isasymkappa}} defined as follows:

  \begin{center}
  \isa{{\isasymtau}}  ::= \isa{a}$^-$ $\mid$ \isa{{\isasymtau}\ {\isasymrightarrow}\ {\isasymtau}{\isacharprime}} $\quad$
  \isa{{\isasymkappa}} ::= \isa{type\isactrlisup {\isacharminus}} $\mid$ \isa{{\isasymtau}\ {\isasymrightarrow}\ {\isasymkappa}}
  \end{center}
  
  \noindent
  This simplification is sufficient for obtaining a sound and complete
  equivalence checking algorithm, and also simplifies the proof
  development in a number of places.

  Similarly, \emph{simple contexts} \isa{{\isasymDelta}}, \isa{{\isasymTheta}} consist of lists of pairs \isa{{\isacharparenleft}x{\isacharcomma}\ {\isasymtau}{\isacharparenright}} of
  variables and simple types.  We write \isa{{\isasymturnstile}\ {\isasymDelta}\ sctx}
  to indicate that \isa{{\isasymDelta}} is valid, i.e.~has no repeated
  variables, and write \isa{{\isasymDelta}\ {\isasymextends}\ {\isasymDelta}{\isacharprime}} to
  indicate that \isa{{\isasymDelta}} contains all of the bindings of
  \isa{{\isasymDelta}{\isacharprime}} and \isa{{\isasymDelta}} is a valid simple
  context.

  Finally, we also introduce \emph{simple signatures}, also written
  \isa{{\isasymSigma}}, consisting of lists of pairs \isa{{\isacharparenleft}c{\isacharcomma}\ {\isasymtau}{\isacharparenright}} or \isa{{\isacharparenleft}a{\isacharcomma}\ {\isasymkappa}{\isacharparenright}} of constants and simple
  kinds or types.  We write \isa{{\isasymturnstile}\ {\isasymSigma}\ ssig} to indicate
  that \isa{{\isasymSigma}} is a well-formed simple signature with no
  repeated type or kind assignments.

  The \emph{erasure} function translates families and kinds to simple
  types and simple kinds:

  \begin{center}
  \begin{tabular}{@ {}cc@ {}}
  \begin{tabular}{@ {}r@ {\hspace{1mm}}c@ {\hspace{1mm}}l@ {}}
  \isa{{\isacharparenleft}a{\isacharparenright}\isactrlisup {\isacharminus}} & $=$ &
  \isa{a\isactrlisup {\isacharminus}}\\
  \isa{{\isacharparenleft}A\ M{\isacharparenright}\isactrlisup {\isacharminus}} & $=$ & 
  \isa{A\isactrlisup {\isacharminus}}\\
  \isa{{\isacharparenleft}{\isasymPi}x{\isacharcolon}A\isactrlisub {\isadigit{1}}{\isachardot}\ A\isactrlisub {\isadigit{2}}{\isacharparenright}\isactrlisup {\isacharminus}} & $=$ &
  \isa{A\isactrlisub {\isadigit{1}}\isactrlisup {\isacharminus}\ {\isasymrightarrow}\ A\isactrlisub {\isadigit{2}}\isactrlisup {\isacharminus}}\\
  \end{tabular} &
  \begin{tabular}{@ {}r@ {\hspace{1mm}}c@ {\hspace{1mm}}l@ {}}
  \isa{{\isacharparenleft}type{\isacharparenright}\isactrlisup {\isacharminus}} & $=$ &
  \isa{type\isactrlisup {\isacharminus}}\\
  \isa{{\isacharparenleft}{\isasymPi}x{\isacharcolon}A{\isachardot}\ K{\isacharparenright}\isactrlisup {\isacharminus}} & $=$ &
  \isa{A\isactrlisup {\isacharminus}\ {\isasymrightarrow}\ K\isactrlisup {\isacharminus}}\\
  \end{tabular}
  \end{tabular}
  \end{center}

  \noindent
  Similarly, we write \isa{{\isasymGamma}\isactrlisup {\isacharminus}} for the
  simple context resulting from replacing each binding \isa{{\isacharparenleft}x{\isacharcomma}\ A{\isacharparenright}}
  in \isa{{\isasymGamma}} with \isa{{\isacharparenleft}x{\isacharcomma}\ A\isactrlisup {\isacharminus}{\isacharparenright}}.
  Likewise, we extend the erasure function to map signatures  \isa{{\isasymSigma}}  to simple signatures \isa{{\isasymSigma}\isactrlisup {\isacharminus}} in the natural way.

  The rules for the algorithm also employ a \emph{weak head reduction}
  relation \mbox{\isa{{\isacharparenleft}$-${\isacharparenright}\ $\stackrel{\mathrm{whr}}{\longrightarrow}$\ {\isacharparenleft}$-${\isacharparenright}}} which performs beta-reductions only at the head of the
  top-level application of a term. It is defined as

  \begin{center}
  \isa{\mbox{}\inferrule{\mbox{x\ {\isasymsharp}\ {\isacharparenleft}A\isactrlisub {\isadigit{1}}{\isacharcomma}\ M\isactrlisub {\isadigit{1}}{\isacharparenright}}}{\mbox{{\isacharparenleft}{\isasymlambda}x{\isacharcolon}A\isactrlisub {\isadigit{1}}{\isachardot}\ M\isactrlisub {\isadigit{2}}{\isacharparenright}\ M\isactrlisub {\isadigit{1}}\ $\stackrel{\mathrm{whr}}{\longrightarrow}$\ M\isactrlisub {\isadigit{2}}{\isacharbrackleft}x{\isacharcolon}{\isacharequal}M\isactrlisub {\isadigit{1}}{\isacharbrackright}}}}$\;\;\;$
  \isa{\mbox{}\inferrule{\mbox{M\isactrlisub {\isadigit{1}}\ $\stackrel{\mathrm{whr}}{\longrightarrow}$\ M\isactrlisub {\isadigit{1}}{\isacharprime}}}{\mbox{M\isactrlisub {\isadigit{1}}\ M\isactrlisub {\isadigit{2}}\ $\stackrel{\mathrm{whr}}{\longrightarrow}$\ M\isactrlisub {\isadigit{1}}{\isacharprime}\ M\isactrlisub {\isadigit{2}}}}}
  \end{center}

  The rules for the equivalence checking algorithm are given in
  \refFig{alg-equiv}. There are five algorithmic equivalence judgments: 
  \begin{center}
\begin{tabular}{cccc}
 & 
Objects
& 
Type families
& 
Kinds
\\
Algorithmic 
&  
\isa{{\isasymDelta}\ {\isasymturnstile}\isactrlbsub {\isasymSigma}\isactrlesub \ M\ {\isasymLeftrightarrow}\ N\ {\isacharcolon}\ {\isasymtau}}
&
\isa{{\isasymDelta}\ {\isasymturnstile}\isactrlbsub {\isasymSigma}\isactrlesub \ A\ {\isasymLeftrightarrow}\ B\ {\isacharcolon}\ {\isasymkappa}}
 &
\isa{{\isasymDelta}\ {\isasymturnstile}\isactrlbsub {\isasymSigma}\isactrlesub \ K\ {\isasymLeftrightarrow}\ L\ {\isacharcolon}\ kind\isactrlisup {\isacharminus}}
\\
Structural 
&
   \isa{{\isasymDelta}\ {\isasymturnstile}\isactrlbsub {\isasymSigma}\isactrlesub \ M\ {\isasymleftrightarrow}\ N\ {\isacharcolon}\ {\isasymtau}}
& 
 \isa{{\isasymDelta}\ {\isasymturnstile}\isactrlbsub {\isasymSigma}\isactrlesub \ A\ {\isasymleftrightarrow}\ B\ {\isacharcolon}\ {\isasymkappa}}
&
\end{tabular}
\end{center}
 Note that the algorithmic rules are type-
  (or kind-) directed while the structural rules are syntax-directed.

  %%%%%%%%%%%%%%%%%%%%%%%%%%%%%%%%%%%%%%%%%%%%%%%%%%%%%%%%%%%%%%%%%%
  \begin{figure}
    \fbox{\isa{{\isasymDelta}\ {\isasymturnstile}\isactrlbsub {\isasymSigma}\isactrlesub \ M\ {\isasymLeftrightarrow}\ N\ {\isacharcolon}\ {\isasymtau}}}
    \begin{center}  
      \begin{tabular}{@ {}c@ {}}
        \isa{\mbox{}\inferrule{\mbox{M\ $\stackrel{\mathrm{whr}}{\longrightarrow}$\ M{\isacharprime}}\\\ \mbox{{\isasymDelta}\ {\isasymturnstile}\isactrlbsub {\isasymSigma}\isactrlesub \ M{\isacharprime}\ {\isasymLeftrightarrow}\ N\ {\isacharcolon}\ a\isactrlisup {\isacharminus}}}{\mbox{{\isasymDelta}\ {\isasymturnstile}\isactrlbsub {\isasymSigma}\isactrlesub \ M\ {\isasymLeftrightarrow}\ N\ {\isacharcolon}\ a\isactrlisup {\isacharminus}}}}\medskip\\%$\;\;\;$ 
        \isa{\mbox{}\inferrule{\mbox{N\ $\stackrel{\mathrm{whr}}{\longrightarrow}$\ N{\isacharprime}}\\\ \mbox{{\isasymDelta}\ {\isasymturnstile}\isactrlbsub {\isasymSigma}\isactrlesub \ M\ {\isasymLeftrightarrow}\ N{\isacharprime}\ {\isacharcolon}\ a\isactrlisup {\isacharminus}}}{\mbox{{\isasymDelta}\ {\isasymturnstile}\isactrlbsub {\isasymSigma}\isactrlesub \ M\ {\isasymLeftrightarrow}\ N\ {\isacharcolon}\ a\isactrlisup {\isacharminus}}}}\medskip\\
        \isa{\mbox{}\inferrule{\mbox{{\isasymDelta}\ {\isasymturnstile}\isactrlbsub {\isasymSigma}\isactrlesub \ M\ {\isasymleftrightarrow}\ N\ {\isacharcolon}\ a\isactrlisup {\isacharminus}}}{\mbox{{\isasymDelta}\ {\isasymturnstile}\isactrlbsub {\isasymSigma}\isactrlesub \ M\ {\isasymLeftrightarrow}\ N\ {\isacharcolon}\ a\isactrlisup {\isacharminus}}}}\medskip\\%$\;\;\;$ 
        \isa{\mbox{}\inferrule{\mbox{{\isacharparenleft}x{\isacharcomma}\ {\isasymtau}\isactrlisub {\isadigit{1}}{\isacharparenright}{\isacharcolon}{\isacharcolon}{\isasymDelta}\ {\isasymturnstile}\isactrlbsub {\isasymSigma}\isactrlesub \ M\ x\ {\isasymLeftrightarrow}\ N\ x\ {\isacharcolon}\ {\isasymtau}\isactrlisub {\isadigit{2}}}\\\ \mbox{x\ {\isasymsharp}\ {\isacharparenleft}{\isasymDelta}{\isacharcomma}\ M{\isacharcomma}\ N{\isacharparenright}}}{\mbox{{\isasymDelta}\ {\isasymturnstile}\isactrlbsub {\isasymSigma}\isactrlesub \ M\ {\isasymLeftrightarrow}\ N\ {\isacharcolon}\ {\isasymtau}\isactrlisub {\isadigit{1}}\ {\isasymrightarrow}\ {\isasymtau}\isactrlisub {\isadigit{2}}}}}\bigskip\\  
      \end{tabular}
    \end{center}
    
    \fbox{\isa{{\isasymDelta}\ {\isasymturnstile}\isactrlbsub {\isasymSigma}\isactrlesub \ M\ {\isasymleftrightarrow}\ N\ {\isacharcolon}\ {\isasymtau}}}
    
    \begin{center}
      \begin{tabular}{@ {}c@ {}}
        \isa{\mbox{}\inferrule{\mbox{{\isacharparenleft}x{\isacharcomma}\ {\isasymtau}{\isacharparenright}\ {\isasymin}\ {\isasymDelta}}\\\ \mbox{{\isasymturnstile}\ {\isasymDelta}\ sctx}\\\ \mbox{{\isasymturnstile}\ {\isasymSigma}\ ssig}}{\mbox{{\isasymDelta}\ {\isasymturnstile}\isactrlbsub {\isasymSigma}\isactrlesub \ x\ {\isasymleftrightarrow}\ x\ {\isacharcolon}\ {\isasymtau}}}}\medskip\\%$\;\;\;$ 
        \isa{\mbox{}\inferrule{\mbox{{\isacharparenleft}c{\isacharcomma}\ {\isasymtau}{\isacharparenright}\ {\isasymin}\ {\isasymSigma}}\\\ \mbox{{\isasymturnstile}\ {\isasymDelta}\ sctx}\\\ \mbox{{\isasymturnstile}\ {\isasymSigma}\ ssig}}{\mbox{{\isasymDelta}\ {\isasymturnstile}\isactrlbsub {\isasymSigma}\isactrlesub \ c\ {\isasymleftrightarrow}\ c\ {\isacharcolon}\ {\isasymtau}}}}\medskip\\%$\;\;\;$ 
        % \begin{minipage}[b]{5cm}
        \isa{\mbox{}\inferrule{\mbox{{\isasymDelta}\ {\isasymturnstile}\isactrlbsub {\isasymSigma}\isactrlesub \ M\isactrlisub {\isadigit{1}}\ {\isasymleftrightarrow}\ N\isactrlisub {\isadigit{1}}\ {\isacharcolon}\ {\isasymtau}\isactrlisub {\isadigit{2}}\ {\isasymrightarrow}\ {\isasymtau}\isactrlisub {\isadigit{1}}}\\\ \mbox{{\isasymDelta}\ {\isasymturnstile}\isactrlbsub {\isasymSigma}\isactrlesub \ M\isactrlisub {\isadigit{2}}\ {\isasymLeftrightarrow}\ N\isactrlisub {\isadigit{2}}\ {\isacharcolon}\ {\isasymtau}\isactrlisub {\isadigit{2}}}}{\mbox{{\isasymDelta}\ {\isasymturnstile}\isactrlbsub {\isasymSigma}\isactrlesub \ M\isactrlisub {\isadigit{1}}\ M\isactrlisub {\isadigit{2}}\ {\isasymleftrightarrow}\ N\isactrlisub {\isadigit{1}}\ N\isactrlisub {\isadigit{2}}\ {\isacharcolon}\ {\isasymtau}\isactrlisub {\isadigit{1}}}}}
        % \end{minipage}
      \end{tabular}
    \end{center}
      
      \fbox{\isa{{\isasymDelta}\ {\isasymturnstile}\isactrlbsub {\isasymSigma}\isactrlesub \ A\ {\isasymLeftrightarrow}\ B\ {\isacharcolon}\ {\isasymkappa}}}
      
      \begin{center}
        \begin{tabular}{@ {}c@ {}}
          \isa{\mbox{}\inferrule{\mbox{{\isasymDelta}\ {\isasymturnstile}\isactrlbsub {\isasymSigma}\isactrlesub \ A\ {\isasymleftrightarrow}\ B\ {\isacharcolon}\ type\isactrlisup {\isacharminus}}}{\mbox{{\isasymDelta}\ {\isasymturnstile}\isactrlbsub {\isasymSigma}\isactrlesub \ A\ {\isasymLeftrightarrow}\ B\ {\isacharcolon}\ type\isactrlisup {\isacharminus}}}}\medskip\\%$\;\;\;$
          \isa{\mbox{}\inferrule{\mbox{{\isacharparenleft}x{\isacharcomma}\ {\isasymtau}{\isacharparenright}{\isacharcolon}{\isacharcolon}{\isasymDelta}\ {\isasymturnstile}\isactrlbsub {\isasymSigma}\isactrlesub \ A\ x\ {\isasymLeftrightarrow}\ B\ x\ {\isacharcolon}\ {\isasymkappa}}\\\ \mbox{x\ {\isasymsharp}\ {\isacharparenleft}{\isasymDelta}{\isacharcomma}\ A{\isacharcomma}\ B{\isacharparenright}}}{\mbox{{\isasymDelta}\ {\isasymturnstile}\isactrlbsub {\isasymSigma}\isactrlesub \ A\ {\isasymLeftrightarrow}\ B\ {\isacharcolon}\ {\isasymtau}\ {\isasymrightarrow}\ {\isasymkappa}}}}\medskip\\
          
          % \begin{minipage}[b]{8cm}
          \isa{\mbox{}\inferrule{\mbox{{\isasymDelta}\ {\isasymturnstile}\isactrlbsub {\isasymSigma}\isactrlesub \ A\isactrlisub {\isadigit{1}}\ {\isasymLeftrightarrow}\ B\isactrlisub {\isadigit{1}}\ {\isacharcolon}\ type\isactrlisup {\isacharminus}}\\\ \mbox{{\isacharparenleft}x{\isacharcomma}\ A\isactrlisub {\isadigit{1}}\isactrlisup {\isacharminus}{\isacharparenright}{\isacharcolon}{\isacharcolon}{\isasymDelta}\ {\isasymturnstile}\isactrlbsub {\isasymSigma}\isactrlesub \ A\isactrlisub {\isadigit{2}}\ {\isasymLeftrightarrow}\ B\isactrlisub {\isadigit{2}}\ {\isacharcolon}\ type\isactrlisup {\isacharminus}}\\\ \mbox{x\ {\isasymsharp}\ {\isacharparenleft}{\isasymDelta}{\isacharcomma}\ A\isactrlisub {\isadigit{1}}{\isacharcomma}\ B\isactrlisub {\isadigit{1}}{\isacharparenright}}}{\mbox{{\isasymDelta}\ {\isasymturnstile}\isactrlbsub {\isasymSigma}\isactrlesub \ {\isasymPi}x{\isacharcolon}A\isactrlisub {\isadigit{1}}{\isachardot}\ A\isactrlisub {\isadigit{2}}\ {\isasymLeftrightarrow}\ {\isasymPi}x{\isacharcolon}B\isactrlisub {\isadigit{1}}{\isachardot}\ B\isactrlisub {\isadigit{2}}\ {\isacharcolon}\ type\isactrlisup {\isacharminus}}}}
          % \end{minipage}
        \end{tabular}
      \end{center}
      
      \fbox{\isa{{\isasymDelta}\ {\isasymturnstile}\isactrlbsub {\isasymSigma}\isactrlesub \ A\ {\isasymleftrightarrow}\ B\ {\isacharcolon}\ {\isasymkappa}}}
      
      \begin{center}
        \begin{tabular}{@ {}c@ {}}
          \isa{\mbox{}\inferrule{\mbox{{\isacharparenleft}a{\isacharcomma}\ {\isasymkappa}{\isacharparenright}\ {\isasymin}\ {\isasymSigma}}\\\ \mbox{{\isasymturnstile}\ {\isasymDelta}\ sctx}\\\ \mbox{{\isasymturnstile}\ {\isasymSigma}\ ssig}}{\mbox{{\isasymDelta}\ {\isasymturnstile}\isactrlbsub {\isasymSigma}\isactrlesub \ a\ {\isasymleftrightarrow}\ a\ {\isacharcolon}\ {\isasymkappa}}}}\medskip\\%$\;\;\;$ 
          \isa{\mbox{}\inferrule{\mbox{{\isasymDelta}\ {\isasymturnstile}\isactrlbsub {\isasymSigma}\isactrlesub \ A\ {\isasymleftrightarrow}\ B\ {\isacharcolon}\ {\isasymtau}\ {\isasymrightarrow}\ {\isasymkappa}}\\\ \mbox{{\isasymDelta}\ {\isasymturnstile}\isactrlbsub {\isasymSigma}\isactrlesub \ M\ {\isasymLeftrightarrow}\ N\ {\isacharcolon}\ {\isasymtau}}}{\mbox{{\isasymDelta}\ {\isasymturnstile}\isactrlbsub {\isasymSigma}\isactrlesub \ A\ M\ {\isasymleftrightarrow}\ B\ N\ {\isacharcolon}\ {\isasymkappa}}}}
        \end{tabular}
      \end{center}
      
      \fbox{\isa{{\isasymDelta}\ {\isasymturnstile}\isactrlbsub {\isasymSigma}\isactrlesub \ K\ {\isasymLeftrightarrow}\ L\ {\isacharcolon}\ kind\isactrlisup {\isacharminus}}}
      
      \begin{center}
        \begin{tabular}{@ {}c@ {}}
        \isa{\mbox{}\inferrule{\mbox{{\isasymturnstile}\ {\isasymDelta}\ sctx}\\\ \mbox{{\isasymturnstile}\ {\isasymSigma}\ ssig}}{\mbox{{\isasymDelta}\ {\isasymturnstile}\isactrlbsub {\isasymSigma}\isactrlesub \ type\ {\isasymLeftrightarrow}\ type\ {\isacharcolon}\ kind\isactrlisup {\isacharminus}}}}\medskip\\%$\;\;\;$ 
        %\begin{minipage}[b]{9cm}
          \isa{\mbox{}\inferrule{\mbox{{\isasymDelta}\ {\isasymturnstile}\isactrlbsub {\isasymSigma}\isactrlesub \ A\ {\isasymLeftrightarrow}\ B\ {\isacharcolon}\ type\isactrlisup {\isacharminus}}\\\ \mbox{{\isacharparenleft}x{\isacharcomma}\ A\isactrlisup {\isacharminus}{\isacharparenright}{\isacharcolon}{\isacharcolon}{\isasymDelta}\ {\isasymturnstile}\isactrlbsub {\isasymSigma}\isactrlesub \ K\ {\isasymLeftrightarrow}\ L\ {\isacharcolon}\ kind\isactrlisup {\isacharminus}}\\\ \mbox{x\ {\isasymsharp}\ {\isacharparenleft}{\isasymDelta}{\isacharcomma}\ A{\isacharcomma}\ B{\isacharparenright}}}{\mbox{{\isasymDelta}\ {\isasymturnstile}\isactrlbsub {\isasymSigma}\isactrlesub \ {\isasymPi}x{\isacharcolon}A{\isachardot}\ K\ {\isasymLeftrightarrow}\ {\isasymPi}x{\isacharcolon}B{\isachardot}\ L\ {\isacharcolon}\ kind\isactrlisup {\isacharminus}}}} 
        %\end{minipage}
      \end{tabular}
    \end{center}
    
    \caption{Algorithmic equivalence rules}\labelFig{alg-equiv}
  \end{figure}	
  %%%%%%%%%%%%%%%%%%%%%%%%%%%%%%%%%%%%%%%%%%%%%%%%%%%%%%%%%%%%%%%%%%%
\end{isamarkuptext}%
\isamarkuptrue%
\begin{isamarkuptext}%
The main results of \HP are soundness and completeness of the algorithmic
  judgments relative to the equivalence judgments:

  \begin{theorem}[(Completeness)]\label{thm:completeness}
  ~
  \begin{compactenum}
  \item \isa{{\normalsize{}If\,}\ {\isasymGamma}\ {\isasymturnstile}\isactrlisub {\isasymSigma}\ M\ {\isacharequal}\ N\ {\isacharcolon}\ A\ {\normalsize \,then\,}\ {\isasymGamma}\isactrlisup {\isacharminus}\ {\isasymturnstile}\isactrlbsub {\isasymSigma}\isactrlisup {\isacharminus}\isactrlesub \ M\ {\isasymLeftrightarrow}\ N\ {\isacharcolon}\ A\isactrlisup {\isacharminus}{\isachardot}}
  \item \isa{{\normalsize{}If\,}\ {\isasymGamma}\ {\isasymturnstile}\isactrlisub {\isasymSigma}\ A\ {\isacharequal}\ B\ {\isacharcolon}\ K\ {\normalsize \,then\,}\ {\isasymGamma}\isactrlisup {\isacharminus}\ {\isasymturnstile}\isactrlbsub {\isasymSigma}\isactrlisup {\isacharminus}\isactrlesub \ A\ {\isasymLeftrightarrow}\ B\ {\isacharcolon}\ K\isactrlisup {\isacharminus}{\isachardot}}
  \item \isa{{\normalsize{}If\,}\ {\isasymGamma}\ {\isasymturnstile}\isactrlisub {\isasymSigma}\ K\ {\isacharequal}\ L\ {\isacharcolon}\ kind\ {\normalsize \,then\,}\ {\isasymGamma}\isactrlisup {\isacharminus}\ {\isasymturnstile}\isactrlbsub {\isasymSigma}\isactrlisup {\isacharminus}\isactrlesub \ K\ {\isasymLeftrightarrow}\ L\ {\isacharcolon}\ kind\isactrlisup {\isacharminus}{\isachardot}}
  \end{compactenum}
  \end{theorem}%
\end{isamarkuptext}%
\isamarkuptrue%
\begin{isamarkuptext}%
\begin{theorem}[(Soundness)]\labelThm{soundness}\raggedright
  ~
  \begin{compactenum}
  \item \isa{{\normalsize{}If\,}\ \mbox{{\isasymGamma}\isactrlisup {\isacharminus}\ {\isasymturnstile}\isactrlbsub {\isasymSigma}\isactrlisup {\isacharminus}\isactrlesub \ M\ {\isasymLeftrightarrow}\ N\ {\isacharcolon}\ A\isactrlisup {\isacharminus}}\ {\normalsize \,and\,}\ \mbox{{\isasymGamma}\ {\isasymturnstile}\isactrlisub {\isasymSigma}\ M\ {\isacharcolon}\ A}\ {\normalsize \,and\,}\ \mbox{{\isasymGamma}\ {\isasymturnstile}\isactrlisub {\isasymSigma}\ N\ {\isacharcolon}\ A}\ {\normalsize\linebreak[0] \,then\,\linebreak[0]}\ \mbox{{\isasymGamma}\ {\isasymturnstile}\isactrlisub {\isasymSigma}\ M\ {\isacharequal}\ N\ {\isacharcolon}\ A{\isachardot}}}
  \item \isa{{\normalsize{}If\,}\ \mbox{{\isasymGamma}\isactrlisup {\isacharminus}\ {\isasymturnstile}\isactrlbsub {\isasymSigma}\isactrlisup {\isacharminus}\isactrlesub \ A\ {\isasymLeftrightarrow}\ B\ {\isacharcolon}\ K\isactrlisup {\isacharminus}}\ {\normalsize \,and\,}\ \mbox{{\isasymGamma}\ {\isasymturnstile}\isactrlisub {\isasymSigma}\ A\ {\isacharcolon}\ K}\ {\normalsize \,and\,}\ \mbox{{\isasymGamma}\ {\isasymturnstile}\isactrlisub {\isasymSigma}\ B\ {\isacharcolon}\ K}\ {\normalsize\linebreak[0] \,then\,\linebreak[0]}\ \mbox{{\isasymGamma}\ {\isasymturnstile}\isactrlisub {\isasymSigma}\ A\ {\isacharequal}\ B\ {\isacharcolon}\ K{\isachardot}}}
  \item \isa{{\normalsize{}If\,}\ \mbox{{\isasymGamma}\isactrlisup {\isacharminus}\ {\isasymturnstile}\isactrlbsub {\isasymSigma}\isactrlisup {\isacharminus}\isactrlesub \ K\ {\isasymLeftrightarrow}\ L\ {\isacharcolon}\ kind\isactrlisup {\isacharminus}}\ {\normalsize \,and\,}\ \mbox{{\isasymGamma}\ {\isasymturnstile}\isactrlisub {\isasymSigma}\ K\ {\isacharcolon}\ kind}\ {\normalsize \,and\,}\ \mbox{{\isasymGamma}\ {\isasymturnstile}\isactrlisub {\isasymSigma}\ L\ {\isacharcolon}\ kind}\ {\normalsize\linebreak[0] \,then\,\linebreak[0]}\ \mbox{{\isasymGamma}\ {\isasymturnstile}\isactrlisub {\isasymSigma}\ K\ {\isacharequal}\ L\ {\isacharcolon}\ kind{\isachardot}}}
  \end{compactenum}
  \end{theorem}

  \noindent
  In what follows, we outline the proofs of these results and discuss how we
  have formalized them, paying particular attention to places where additional
  lemmas or different proof techniques were needed. We also discuss the gap in
  the soundness proof of \HP, along with several solutions.%
\end{isamarkuptext}%
\isamarkuptrue%
\isamarkupsection{The formalization%
}
\isamarkuptrue%
\begin{isamarkuptext}%
\labelSec{formalization}%
\end{isamarkuptext}%
\isamarkuptrue%
\isamarkupsubsection{Syntactic properties%
}
\isamarkuptrue%
\begin{isamarkuptext}%
\labelSec{syntactic}%
\end{isamarkuptext}%
\isamarkuptrue%
\begin{isamarkuptext}%
\noindent
  The proof in \HP starts by developing of a number of useful
  metatheoretic properties for the validity and equality judgments
  (shown in \refFig{lf-wf}), such as weakening, substitution,
  generalizations of the conversion rules and inversion principles.
  Most of these properties have multiple parts corresponding to the
  eight different judgments in the definitional theory of LF.  We will
  list the main properties; however, to aid readability we will only
  show the statements of most of these properties for the object-level
  judgments, and we omit symmetric cases.  The full formal statements
  of the syntactic properties can be found in the electronic appendix.

  To prove the main syntactic properties we needed two technical
  lemmas having to do with the implicit freshness and validity
  assumptions that must be handled explicitly in our
  formalization. Both are straightforward by induction, and both are
  needed frequently.

  \begin{lemma}[(Freshness)]\labelLem{freshness}
    If \isa{x\ {\isasymsharp}\ {\isasymGamma}} and \isa{{\isasymGamma}\ {\isasymturnstile}\isactrlisub {\isasymSigma}\ M\ {\isacharcolon}\ A} then \mbox{\isa{x\ {\isasymsharp}\ M}} and
    \isa{x\ {\isasymsharp}\ A}.  Similarly, if \isa{x\ {\isasymsharp}\ {\isasymGamma}} and \isa{{\isasymGamma}\ {\isasymturnstile}\isactrlisub {\isasymSigma}\ M\ {\isacharequal}\ N\ {\isacharcolon}\ A} then \mbox{\isa{x\ {\isasymsharp}\ M}} and \mbox{\isa{x\ {\isasymsharp}\ N}} and
    \isa{x\ {\isasymsharp}\ A}.
  \end{lemma}
  
  \begin{lemma}[(Implicit Validity)]\labelLem{implicit-validity}
    If \isa{{\isasymGamma}\ {\isasymturnstile}\isactrlisub {\isasymSigma}\ M\ {\isacharcolon}\ A} or \isa{{\isasymGamma}\ {\isasymturnstile}\isactrlisub {\isasymSigma}\ M\ {\isacharequal}\ N\ {\isacharcolon}\ A} then
    \isa{{\isasymturnstile}\ {\isasymSigma}\ sig} and \isa{{\isasymturnstile}\isactrlisub {\isasymSigma}\ {\isasymGamma}\ ctx}.  
  \end{lemma}

  \begin{lemma}[(Weakening)] 
  Suppose \isa{{\isasymturnstile}\isactrlisub {\isasymSigma}\ {\isasymGamma}\isactrlisub {\isadigit{2}}\ ctx} and \isa{{\isasymGamma}\isactrlisub {\isadigit{1}}\ {\isasymsubseteq}\ {\isasymGamma}\isactrlisub {\isadigit{2}}}.
    ~
    \begin{compactenum} 
    \item If \isa{{\isasymGamma}\isactrlisub {\isadigit{1}}\ {\isasymturnstile}\isactrlisub {\isasymSigma}\ M\ {\isacharcolon}\ A} 
          then \isa{{\isasymGamma}\isactrlisub {\isadigit{2}}\ {\isasymturnstile}\isactrlisub {\isasymSigma}\ M\ {\isacharcolon}\ A}.
    \item If \isa{{\isasymGamma}\isactrlisub {\isadigit{1}}\ {\isasymturnstile}\isactrlisub {\isasymSigma}\ M\ {\isacharequal}\ N\ {\isacharcolon}\ A}
          then \isa{{\isasymGamma}\isactrlisub {\isadigit{2}}\ {\isasymturnstile}\isactrlisub {\isasymSigma}\ M\ {\isacharequal}\ N\ {\isacharcolon}\ A}.
    \end{compactenum}
  \end{lemma}
  
  \begin{lemma}[(Substitution)]  Suppose \isa{{\isasymGamma}\isactrlisub {\isadigit{2}}\ {\isasymturnstile}\isactrlisub {\isasymSigma}\ P\ {\isacharcolon}\ C} and let 
    \mbox{\isa{{\isasymGamma}\ {\isacharequal}\ {\isasymGamma}\isactrlisub {\isadigit{1}}\ {\isacharat}\ {\isacharbrackleft}{\isacharparenleft}y{\isacharcomma}\ C{\isacharparenright}{\isacharbrackright}\ {\isacharat}\ {\isasymGamma}\isactrlisub {\isadigit{2}}}}.
    \begin{compactenum} 
    \item If \isa{{\isasymturnstile}\isactrlisub {\isasymSigma}\ {\isasymGamma}\ ctx} 
          then \isa{{\isasymturnstile}\isactrlisub {\isasymSigma}\ {\isasymGamma}\isactrlisub {\isadigit{1}}{\isacharbrackleft}y{\isacharcolon}{\isacharequal}P{\isacharbrackright}\ {\isacharat}\ {\isasymGamma}\isactrlisub {\isadigit{2}}\ ctx}.
    \item If \isa{{\isasymGamma}\ {\isasymturnstile}\isactrlisub {\isasymSigma}\ M\ {\isacharcolon}\ B}
          then \isa{{\isasymGamma}\isactrlisub {\isadigit{1}}{\isacharbrackleft}y{\isacharcolon}{\isacharequal}P{\isacharbrackright}\ {\isacharat}\ {\isasymGamma}\isactrlisub {\isadigit{2}}\ {\isasymturnstile}\isactrlisub {\isasymSigma}\ M{\isacharbrackleft}y{\isacharcolon}{\isacharequal}P{\isacharbrackright}\ {\isacharcolon}\ B{\isacharbrackleft}y{\isacharcolon}{\isacharequal}P{\isacharbrackright}}.
    \item If \isa{{\isasymGamma}\ {\isasymturnstile}\isactrlisub {\isasymSigma}\ M\ {\isacharequal}\ N\ {\isacharcolon}\ A}
          then \isa{{\isasymGamma}\isactrlisub {\isadigit{1}}{\isacharbrackleft}y{\isacharcolon}{\isacharequal}P{\isacharbrackright}\ {\isacharat}\ {\isasymGamma}\isactrlisub {\isadigit{2}}\ {\isasymturnstile}\isactrlisub {\isasymSigma}\ M{\isacharbrackleft}y{\isacharcolon}{\isacharequal}P{\isacharbrackright}\ {\isacharequal}\ N{\isacharbrackleft}y{\isacharcolon}{\isacharequal}P{\isacharbrackright}\ {\isacharcolon}\ A{\isacharbrackleft}y{\isacharcolon}{\isacharequal}P{\isacharbrackright}}.
    \end{compactenum}
  \end{lemma}

  \begin{lemma}[(Context
    Conversion)] \labelLem{ctxconversion}\raggedright ~Assume that
    \isa{{\isasymGamma}\ {\isasymturnstile}\isactrlisub {\isasymSigma}\ B\ {\isacharcolon}\ type} and \isa{{\isasymGamma}\ {\isasymturnstile}\isactrlisub {\isasymSigma}\ A\ {\isacharequal}\ B\ {\isacharcolon}\ type}.  Then:
    \begin{compactenum}
    \item If \isa{{\isacharparenleft}x{\isacharcomma}\ A{\isacharparenright}{\isacharcolon}{\isacharcolon}{\isasymGamma}\ {\isasymturnstile}\isactrlisub {\isasymSigma}\ M\ {\isacharcolon}\ C} then \isa{{\isacharparenleft}x{\isacharcomma}\ B{\isacharparenright}{\isacharcolon}{\isacharcolon}{\isasymGamma}\ {\isasymturnstile}\isactrlisub {\isasymSigma}\ M\ {\isacharcolon}\ C}
    \item If \isa{{\isacharparenleft}x{\isacharcomma}\ A{\isacharparenright}{\isacharcolon}{\isacharcolon}{\isasymGamma}\ {\isasymturnstile}\isactrlisub {\isasymSigma}\ C\ {\isacharcolon}\ K} then \isa{{\isacharparenleft}x{\isacharcomma}\ B{\isacharparenright}{\isacharcolon}{\isacharcolon}{\isasymGamma}\ {\isasymturnstile}\isactrlisub {\isasymSigma}\ C\ {\isacharcolon}\ K}
    \end{compactenum}
  \end{lemma}

  \begin{lemma}[(Functionality for Typing)]\labelLem{functionality}\raggedright
    Assume that \isa{{\isasymGamma}\ {\isasymturnstile}\isactrlisub {\isasymSigma}\ M\ {\isacharcolon}\ C}
    and \mbox{\isa{{\isasymGamma}\ {\isasymturnstile}\isactrlisub {\isasymSigma}\ N\ {\isacharcolon}\ C}}
    and \isa{{\isasymGamma}\ {\isasymturnstile}\isactrlisub {\isasymSigma}\ M\ {\isacharequal}\ N\ {\isacharcolon}\ C}.
    Then if \isa{{\isasymGamma}{\isacharprime}\ {\isacharat}\ {\isacharbrackleft}{\isacharparenleft}y{\isacharcomma}\ C{\isacharparenright}{\isacharbrackright}\ {\isacharat}\ {\isasymGamma}\ {\isasymturnstile}\isactrlisub {\isasymSigma}\ P\ {\isacharcolon}\ B} then \isa{{\isasymGamma}{\isacharprime}{\isacharbrackleft}y{\isacharcolon}{\isacharequal}M{\isacharbrackright}\ {\isacharat}\ {\isasymGamma}\ {\isasymturnstile}\isactrlisub {\isasymSigma}\ P{\isacharbrackleft}y{\isacharcolon}{\isacharequal}M{\isacharbrackright}\ {\isacharequal}\ P{\isacharbrackleft}y{\isacharcolon}{\isacharequal}N{\isacharbrackright}\ {\isacharcolon}\ B{\isacharbrackleft}y{\isacharcolon}{\isacharequal}M{\isacharbrackright}}.
 \end{lemma}

  \noindent
  Since our judgements contain explicit validity hypotheses for 
  contexts, the proof of \refLem{functionality} relies on
  the fact that functionality holds also for contexts, namely

  \begin{lemma}[(Functionality for Contexts)]\raggedright
  \isa{{\normalsize{}If\,}\ \mbox{{\isasymturnstile}\isactrlisub {\isasymSigma}\ {\isasymGamma}{\isacharprime}\ {\isacharat}\ {\isacharbrackleft}{\isacharparenleft}x{\isacharcomma}\ C{\isacharparenright}{\isacharbrackright}\ {\isacharat}\ {\isasymGamma}\ ctx}\ {\normalsize \,and\,}\ \mbox{{\isasymGamma}\ {\isasymturnstile}\isactrlisub {\isasymSigma}\ M\ {\isacharcolon}\ C}\ {\normalsize\linebreak[0] \,then\,\linebreak[0]}\ \mbox{{\isasymturnstile}\isactrlisub {\isasymSigma}\ {\isasymGamma}{\isacharprime}{\isacharbrackleft}x{\isacharcolon}{\isacharequal}M{\isacharbrackright}\ {\isacharat}\ {\isasymGamma}\ ctx{\isachardot}}}
  \end{lemma}

  \noindent
  This fact can be established by induction on \isa{{\isasymGamma}{\isacharprime}}.

  \begin{lemma}[(Validity)] \labelLem{validity}
  Objects, types and kinds appearing in derivable judgments are valid, that is 
    \begin{compactenum}
    \item \isa{{\normalsize{}If\,}\ {\isasymGamma}\ {\isasymturnstile}\isactrlisub {\isasymSigma}\ M\ {\isacharcolon}\ A\ {\normalsize \,then\,}\ {\isasymGamma}\ {\isasymturnstile}\isactrlisub {\isasymSigma}\ A\ {\isacharcolon}\ type{\isachardot}}
   \item \isa{{\normalsize{}If\,}\ {\isasymGamma}\ {\isasymturnstile}\isactrlisub {\isasymSigma}\ M\ {\isacharequal}\ N\ {\isacharcolon}\ B\ {\normalsize \,then\,}\ {\isasymGamma}\ {\isasymturnstile}\isactrlisub {\isasymSigma}\ M\ {\isacharcolon}\ B\ \textrm{and\linebreak[1]}\ {\isasymGamma}\ {\isasymturnstile}\isactrlisub {\isasymSigma}\ N\ {\isacharcolon}\ B\ \textrm{and\linebreak[1]}\ {\isasymGamma}\ {\isasymturnstile}\isactrlisub {\isasymSigma}\ B\ {\isacharcolon}\ type{\isachardot}}
   \end{compactenum}
  \end{lemma}

  \begin{lemma}[(Typing inversion)]
    The validity rules are invertible, up to conversion of types and
    kinds.  
    \begin{compactenum}
    \item \isa{{\normalsize{}If\,}\ {\isasymGamma}\ {\isasymturnstile}\isactrlisub {\isasymSigma}\ x\ {\isacharcolon}\ A\ {\normalsize \,then\,}\ {\isasymexists}B{\isachardot}\ {\isacharparenleft}x{\isacharcomma}\ B{\isacharparenright}\ {\isasymin}\ {\isasymGamma}\ \textrm{and\linebreak[1]}\ {\isasymGamma}\ {\isasymturnstile}\isactrlisub {\isasymSigma}\ A\ {\isacharequal}\ B\ {\isacharcolon}\ type{\isachardot}}
    \item \isa{{\normalsize{}If\,}\ {\isasymGamma}\ {\isasymturnstile}\isactrlisub {\isasymSigma}\ c\ {\isacharcolon}\ A\ {\normalsize \,then\,}\ {\isasymexists}B{\isachardot}\ {\isacharparenleft}c{\isacharcomma}\ B{\isacharparenright}\ {\isasymin}\ {\isasymSigma}\ \textrm{and\linebreak[1]}\ {\isasymGamma}\ {\isasymturnstile}\isactrlisub {\isasymSigma}\ A\ {\isacharequal}\ B\ {\isacharcolon}\ type{\isachardot}}
    \item \isa{{\normalsize{}If\,}\ {\isasymGamma}\ {\isasymturnstile}\isactrlisub {\isasymSigma}\ M\isactrlisub {\isadigit{1}}\ M\isactrlisub {\isadigit{2}}\ {\isacharcolon}\ A\ {\normalsize \,then\,}\ {\isasymexists}x\ A\isactrlisub {\isadigit{1}}\ A\isactrlisub {\isadigit{2}}{\isachardot}\ {\isasymGamma}\ {\isasymturnstile}\isactrlisub {\isasymSigma}\ M\isactrlisub {\isadigit{1}}\ {\isacharcolon}\ {\isasymPi}x{\isacharcolon}A\isactrlisub {\isadigit{2}}{\isachardot}\ A\isactrlisub {\isadigit{1}}\ \textrm{and\linebreak[1]}\ {\isasymGamma}\ {\isasymturnstile}\isactrlisub {\isasymSigma}\ M\isactrlisub {\isadigit{2}}\ {\isacharcolon}\ A\isactrlisub {\isadigit{2}}\ \textrm{and\linebreak[1]}\ {\isasymGamma}\ {\isasymturnstile}\isactrlisub {\isasymSigma}\ A\ {\isacharequal}\ A\isactrlisub {\isadigit{1}}{\isacharbrackleft}x{\isacharcolon}{\isacharequal}M\isactrlisub {\isadigit{2}}{\isacharbrackright}\ {\isacharcolon}\ type{\isachardot}}
    \item \isa{{\normalsize{}If\,}\ {\isasymGamma}\ {\isasymturnstile}\isactrlisub {\isasymSigma}\ {\isasymlambda}x{\isacharcolon}A{\isachardot}\ M\ {\isacharcolon}\ B\ {\normalsize \,and\,}\ x\ {\isasymsharp}\ {\isasymGamma}\ {\normalsize \,then\,}\ {\isasymexists}A{\isacharprime}{\isachardot}\ {\isasymGamma}\ {\isasymturnstile}\isactrlisub {\isasymSigma}\ B\ {\isacharequal}\ {\isasymPi}x{\isacharcolon}A{\isachardot}\ A{\isacharprime}\ {\isacharcolon}\ type\ \textrm{and\linebreak[1]}\ {\isasymGamma}\ {\isasymturnstile}\isactrlisub {\isasymSigma}\ A\ {\isacharcolon}\ type\ \textrm{and\linebreak[1]}\ {\isacharparenleft}x{\isacharcomma}\ A{\isacharparenright}{\isacharcolon}{\isacharcolon}{\isasymGamma}\ {\isasymturnstile}\isactrlisub {\isasymSigma}\ M\ {\isacharcolon}\ A{\isacharprime}{\isachardot}}
   \end{compactenum}
  \end{lemma}

  Next \HP established some inversion and invertibility properties for
  definitional equality:
  \begin{lemma}[(Equality inversion)]\mbox{}    
    ~
    \begin{compactenum} 
    \item \isa{{\normalsize{}If\,}\ {\isasymGamma}\ {\isasymturnstile}\isactrlisub {\isasymSigma}\ type\ {\isacharequal}\ L\ {\isacharcolon}\ kind\ {\normalsize \,then\,}\ L\ {\isacharequal}\ type{\isachardot}}
   \item \isa{{\normalsize{}If\,}\ {\isasymGamma}\ {\isasymturnstile}\isactrlisub {\isasymSigma}\ A\ {\isacharequal}\ {\isasymPi}x{\isacharcolon}B\isactrlisub {\isadigit{1}}{\isachardot}\ B\isactrlisub {\isadigit{2}}\ {\isacharcolon}\ type\ {\normalsize \,and\,}\ x\ {\isasymsharp}\ {\isasymGamma}\ {\normalsize \,then\,}\ {\isasymexists}A\isactrlisub {\isadigit{1}}\ A\isactrlisub {\isadigit{2}}{\isachardot}\ A\ {\isacharequal}\ {\isasymPi}x{\isacharcolon}A\isactrlisub {\isadigit{1}}{\isachardot}\ A\isactrlisub {\isadigit{2}}\ \textrm{and\linebreak[1]}\ {\isasymGamma}\ {\isasymturnstile}\isactrlisub {\isasymSigma}\ A\isactrlisub {\isadigit{1}}\ {\isacharequal}\ B\isactrlisub {\isadigit{1}}\ {\isacharcolon}\ type\ \textrm{and\linebreak[1]}\ {\isacharparenleft}x{\isacharcomma}\ A\isactrlisub {\isadigit{1}}{\isacharparenright}{\isacharcolon}{\isacharcolon}{\isasymGamma}\ {\isasymturnstile}\isactrlisub {\isasymSigma}\ A\isactrlisub {\isadigit{2}}\ {\isacharequal}\ B\isactrlisub {\isadigit{2}}\ {\isacharcolon}\ type{\isachardot}}
    \item \isa{{\normalsize{}If\,}\ {\isasymGamma}\ {\isasymturnstile}\isactrlisub {\isasymSigma}\ K\ {\isacharequal}\ {\isasymPi}x{\isacharcolon}B\isactrlisub {\isadigit{1}}{\isachardot}\ L\isactrlisub {\isadigit{2}}\ {\isacharcolon}\ kind\ {\normalsize \,and\,}\ x\ {\isasymsharp}\ {\isasymGamma}\ {\normalsize \,then\,}\ {\isasymexists}A\isactrlisub {\isadigit{1}}\ K\isactrlisub {\isadigit{2}}{\isachardot}\ K\ {\isacharequal}\ {\isasymPi}x{\isacharcolon}A\isactrlisub {\isadigit{1}}{\isachardot}\ K\isactrlisub {\isadigit{2}}\ \textrm{and\linebreak[1]}\ {\isasymGamma}\ {\isasymturnstile}\isactrlisub {\isasymSigma}\ A\isactrlisub {\isadigit{1}}\ {\isacharequal}\ B\isactrlisub {\isadigit{1}}\ {\isacharcolon}\ type\ \textrm{and\linebreak[1]}\ {\isacharparenleft}x{\isacharcomma}\ A\isactrlisub {\isadigit{1}}{\isacharparenright}{\isacharcolon}{\isacharcolon}{\isasymGamma}\ {\isasymturnstile}\isactrlisub {\isasymSigma}\ K\isactrlisub {\isadigit{2}}\ {\isacharequal}\ L\isactrlisub {\isadigit{2}}\ {\isacharcolon}\ kind{\isachardot}}
    \end{compactenum}
  \end{lemma}

  Finally, we can prove that the product type constructor is
  injective up to definitional equality, which is needed for soundness:
  
  \begin{lemma}[(Product injectivity)]
  Suppose \isa{x\ {\isasymsharp}\ {\isasymGamma}}.
  \begin{compactenum}
  \item If \isa{{\isasymGamma}\ {\isasymturnstile}\isactrlisub {\isasymSigma}\ {\isasymPi}x{\isacharcolon}A\isactrlisub {\isadigit{1}}{\isachardot}\ A\isactrlisub {\isadigit{2}}\ {\isacharequal}\ {\isasymPi}x{\isacharcolon}B\isactrlisub {\isadigit{1}}{\isachardot}\ B\isactrlisub {\isadigit{2}}\ {\isacharcolon}\ type}
        then \isa{{\isasymGamma}\ {\isasymturnstile}\isactrlisub {\isasymSigma}\ A\isactrlisub {\isadigit{1}}\ {\isacharequal}\ B\isactrlisub {\isadigit{1}}\ {\isacharcolon}\ type}
             and\\ 
             \mbox{\isa{{\isacharparenleft}x{\isacharcomma}\ A\isactrlisub {\isadigit{1}}{\isacharparenright}{\isacharcolon}{\isacharcolon}{\isasymGamma}\ {\isasymturnstile}\isactrlisub {\isasymSigma}\ A\isactrlisub {\isadigit{2}}\ {\isacharequal}\ B\isactrlisub {\isadigit{2}}\ {\isacharcolon}\ type}}.
  \item If \isa{{\isasymGamma}\ {\isasymturnstile}\isactrlisub {\isasymSigma}\ {\isasymPi}x{\isacharcolon}A{\isachardot}\ K\ {\isacharequal}\ {\isasymPi}x{\isacharcolon}B{\isachardot}\ L\ {\isacharcolon}\ kind}
        then \isa{{\isasymGamma}\ {\isasymturnstile}\isactrlisub {\isasymSigma}\ A\ {\isacharequal}\ B\ {\isacharcolon}\ type}
             and\\ 
             \mbox{\isa{{\isacharparenleft}x{\isacharcomma}\ A{\isacharparenright}{\isacharcolon}{\isacharcolon}{\isasymGamma}\ {\isasymturnstile}\isactrlisub {\isasymSigma}\ K\ {\isacharequal}\ L\ {\isacharcolon}\ kind}}.
  \end{compactenum}
  \end{lemma}

  All the metatheoretic properties given above can be proved as stated
  in \HP (appealing to Lem.~\ref{lem:freshness}
  and~\ref{lem:implicit-validity} as necessary); however, since all of
  the definitional judgments of LF are interdependent, each inductive
  proof must consider all 35 cases, making each proof nontrivial as a
  practical matter (it is one of the biggest parts of our
  formalization).

  \HP organize the proofs of these metatheoretic properties very
  neatly.  For example as shown in \refLem{validity} the validity
  judgment of terms implies the validity of the type.  However, in
  order to establish this a number of auxiliary facts have to be
  proved first which depend on this property. In order to get the
  proof through, some of \HP's rules given in \refFig{lf-wf} are
  formulated to explicitly include validity constraints such as \isa{{\isasymGamma}\ {\isasymturnstile}\isactrlisub {\isasymSigma}\ A\ {\isacharcolon}\ type} and \mbox{\isa{{\isasymGamma}\ {\isasymturnstile}\isactrlisub {\isasymSigma}\ K\ {\isacharcolon}\ kind}}.  After proving the
  above properties, however, we can show that these extra hypotheses
  are not needed, by establishing stronger forms of the rules:

  \begin{lemma}[(Strong versions of rules)]
    The following rules are admissible:
    \begin{compactenum}
    \item \isa{\mbox{}\inferrule{\mbox{{\isasymGamma}\ {\isasymturnstile}\isactrlisub {\isasymSigma}\ M\isactrlisub {\isadigit{1}}\ {\isacharcolon}\ {\isasymPi}x{\isacharcolon}A\isactrlisub {\isadigit{2}}{\isachardot}\ A\isactrlisub {\isadigit{1}}}\\\ \mbox{{\isasymGamma}\ {\isasymturnstile}\isactrlisub {\isasymSigma}\ M\isactrlisub {\isadigit{2}}\ {\isacharcolon}\ A\isactrlisub {\isadigit{2}}}}{\mbox{{\isasymGamma}\ {\isasymturnstile}\isactrlisub {\isasymSigma}\ M\isactrlisub {\isadigit{1}}\ M\isactrlisub {\isadigit{2}}\ {\isacharcolon}\ A\isactrlisub {\isadigit{1}}{\isacharbrackleft}x{\isacharcolon}{\isacharequal}M\isactrlisub {\isadigit{2}}{\isacharbrackright}}}}
    \item \isa{\mbox{}\inferrule{\mbox{{\isasymGamma}\ {\isasymturnstile}\isactrlisub {\isasymSigma}\ A\ {\isacharcolon}\ {\isasymPi}x{\isacharcolon}B{\isachardot}\ K}\\\ \mbox{{\isasymGamma}\ {\isasymturnstile}\isactrlisub {\isasymSigma}\ M\ {\isacharcolon}\ B}}{\mbox{{\isasymGamma}\ {\isasymturnstile}\isactrlisub {\isasymSigma}\ A\ M\ {\isacharcolon}\ K{\isacharbrackleft}x{\isacharcolon}{\isacharequal}M{\isacharbrackright}}}}
    \item \isa{\mbox{}\inferrule{\mbox{{\isacharparenleft}x{\isacharcomma}\ A\isactrlisub {\isadigit{1}}{\isacharparenright}{\isacharcolon}{\isacharcolon}{\isasymGamma}\ {\isasymturnstile}\isactrlisub {\isasymSigma}\ M\isactrlisub {\isadigit{2}}\ {\isacharequal}\ N\isactrlisub {\isadigit{2}}\ {\isacharcolon}\ A\isactrlisub {\isadigit{2}}}\\\ \mbox{{\isasymGamma}\ {\isasymturnstile}\isactrlisub {\isasymSigma}\ M\isactrlisub {\isadigit{1}}\ {\isacharequal}\ N\isactrlisub {\isadigit{1}}\ {\isacharcolon}\ A\isactrlisub {\isadigit{1}}}\\\ \mbox{x\ {\isasymsharp}\ {\isasymGamma}}}{\mbox{{\isasymGamma}\ {\isasymturnstile}\isactrlisub {\isasymSigma}\ {\isacharparenleft}{\isasymlambda}x{\isacharcolon}A\isactrlisub {\isadigit{1}}{\isachardot}\ M\isactrlisub {\isadigit{2}}{\isacharparenright}\ M\isactrlisub {\isadigit{1}}\ {\isacharequal}\ N\isactrlisub {\isadigit{2}}{\isacharbrackleft}x{\isacharcolon}{\isacharequal}N\isactrlisub {\isadigit{1}}{\isacharbrackright}\ {\isacharcolon}\ A\isactrlisub {\isadigit{2}}{\isacharbrackleft}x{\isacharcolon}{\isacharequal}M\isactrlisub {\isadigit{1}}{\isacharbrackright}}}}
    \end{compactenum}
  \end{lemma}%
\end{isamarkuptext}%
\isamarkuptrue%
\isamarkupsubsection{Algorithmic equivalence%
}
\isamarkuptrue%
\begin{isamarkuptext}%
\labelSec{algorithm}%
\end{isamarkuptext}%
\isamarkuptrue%
\begin{isamarkuptext}%
\noindent
  The main metatheoretic properties of algorithmic equivalence proved
  in Sec. 3 of \HP are symmetry and transitivity. Several properties
  of weak head reduction and erasure needed later in \HP are also
  proved.  Most of the proofs were straightforward to formalize, given
  the details in \HP (where provided).  However, there were a few
  missing lemmas and other complications.  The algorithmic system is
  less well-behaved than the definitional system because derivable
  judgments may have ill-formed arguments; for example, the judgment
  \mbox{\isa{{\isacharbrackleft}{\isacharbrackright}\ {\isasymturnstile}\isactrlbsub {\isasymSigma}\isactrlesub \ {\isacharparenleft}{\isasymlambda}x{\isacharcolon}a{\isachardot}\ c{\isacharparenright}\ y\ {\isasymLeftrightarrow}\ c\ {\isacharcolon}\ b\isactrlisup {\isacharminus}}} is derivable, for any object term \isa{y}, provided that \isa{{\isacharparenleft}c{\isacharcomma}\ b{\isacharparenright}\ {\isasymin}\ {\isasymSigma}} since
  \isa{{\isacharparenleft}{\isasymlambda}x{\isacharcolon}a{\isachardot}\ c{\isacharparenright}\ y\ $\stackrel{\mathrm{whr}}{\longrightarrow}$\ c}.
  Thus, analogues of \refLem{freshness}
  and~\ref{lem:implicit-validity} do not hold for the algorithmic
  system, and in rules involving binding we need to impose additional
  freshness constraints.  Moreover, proof search in the algorithmic
  system is not necessarily terminating because \mbox{\isa{{\isacharparenleft}$-${\isacharparenright}\ $\stackrel{\mathrm{whr}}{\longrightarrow}$\ {\isacharparenleft}$-${\isacharparenright}}} may diverge if called on ill-formed terms such
  as \isa{{\isacharparenleft}{\isasymlambda}x{\isacharcolon}a{\isachardot}\ x\ x{\isacharparenright}\ {\isacharparenleft}{\isasymlambda}x{\isacharcolon}a{\isachardot}\ x\ x{\isacharparenright}}.

  The erasure preservation lemma establishes basic properties of erasure which
  are frequently needed in \HP:

  \begin{lemma}[(Erasure preservation)]
    ~
    \begin{compactenum}
    \item \isa{{\normalsize{}If\,}\ {\isasymGamma}\ {\isasymturnstile}\isactrlisub {\isasymSigma}\ A\ {\isacharequal}\ B\ {\isacharcolon}\ K\ {\normalsize \,then\,}\ A\isactrlisup {\isacharminus}\ {\isacharequal}\ B\isactrlisup {\isacharminus}{\isachardot}} 
    \item \isa{{\normalsize{}If\,}\ {\isasymGamma}\ {\isasymturnstile}\isactrlisub {\isasymSigma}\ K\ {\isacharequal}\ L\ {\isacharcolon}\ kind\ {\normalsize \,then\,}\ K\isactrlisup {\isacharminus}\ {\isacharequal}\ L\isactrlisup {\isacharminus}{\isachardot}} 
    \item If \isa{{\isacharparenleft}x{\isacharcomma}\ A{\isacharparenright}{\isacharcolon}{\isacharcolon}{\isasymGamma}\ {\isasymturnstile}\isactrlisub {\isasymSigma}\ B\ {\isacharcolon}\ type} then \isa{B\isactrlisup {\isacharminus}\ {\isacharequal}\ B{\isacharbrackleft}x{\isacharcolon}{\isacharequal}M{\isacharbrackright}\isactrlisup {\isacharminus}} 
    \item If \isa{{\isacharparenleft}x{\isacharcomma}\ A{\isacharparenright}{\isacharcolon}{\isacharcolon}{\isasymGamma}\ {\isasymturnstile}\isactrlisub {\isasymSigma}\ K\ {\isacharcolon}\ kind} then \isa{K\isactrlisup {\isacharminus}\ {\isacharequal}\ K{\isacharbrackleft}x{\isacharcolon}{\isacharequal}M{\isacharbrackright}\isactrlisup {\isacharminus}} 
    \end{compactenum}
  \end{lemma}

  \noindent 
  However, we found that the hypotheses of parts 3 and 4 are unnecessary. Indeed, we can easily prove:

  \begin{lemma}[(Erasure cancels substitution)]
    For any type family \isa{A}, kind \isa{K}, and
    substitution \isa{{\isasymsigma}}, we have
    \begin{compactenum}
    \item \isa{A{\isacharbrackleft}{\isasymsigma}{\isacharbrackright}\isactrlisup {\isacharminus}\ {\isacharequal}\ A\isactrlisup {\isacharminus}}
    \item \isa{K{\isacharbrackleft}{\isasymsigma}{\isacharbrackright}\isactrlisup {\isacharminus}\ {\isacharequal}\ K\isactrlisup {\isacharminus}}
    \end{compactenum}
  \end{lemma}
  
  \noindent
  In the proofs of symmetry and transitivity of the algorithmic
  judgments (\refThm{alg-symmetry} and \refThm{alg-transitivity}), we
  also needed the following algorithmic erasure preservation lemma (it
  is omitted from \HP, but straightforward by induction):
  
  \begin{lemma}[(Algorithmic erasure preservation)]
    ~
    \begin{compactenum}
    \item \isa{{\normalsize{}If\,}\ {\isasymDelta}\ {\isasymturnstile}\isactrlbsub {\isasymSigma}\isactrlesub \ A\ {\isasymLeftrightarrow}\ B\ {\isacharcolon}\ {\isasymkappa}\ {\normalsize \,then\,}\ A\isactrlisup {\isacharminus}\ {\isacharequal}\ B\isactrlisup {\isacharminus}{\isachardot}} 
    \item \isa{{\normalsize{}If\,}\ {\isasymDelta}\ {\isasymturnstile}\isactrlbsub {\isasymSigma}\isactrlesub \ A\ {\isasymleftrightarrow}\ B\ {\isacharcolon}\ {\isasymkappa}\ {\normalsize \,then\,}\ A\isactrlisup {\isacharminus}\ {\isacharequal}\ B\isactrlisup {\isacharminus}{\isachardot}} 
    \item \isa{{\normalsize{}If\,}\ {\isasymDelta}\ {\isasymturnstile}\isactrlbsub {\isasymSigma}\isactrlesub \ K\ {\isasymLeftrightarrow}\ L\ {\isacharcolon}\ kind\isactrlisup {\isacharminus}\ {\normalsize \,then\,}\ K\isactrlisup {\isacharminus}\ {\isacharequal}\ L\isactrlisup {\isacharminus}{\isachardot}} 
    \end{compactenum}
  \end{lemma}
  
  The determinacy lemma establishes several important properties of weak head
  reduction and algorithmic equivalence.
  
  \begin{lemma}[(Determinacy)]
  Suppose that \isa{{\isasymturnstile}\ {\isasymSigma}\ ssig} and \mbox{\isa{{\isasymturnstile}\ {\isasymDelta}\ sctx}}. 
  \begin{compactenum}
  \item If \isa{M\ $\stackrel{\mathrm{whr}}{\longrightarrow}$\ M{\isacharprime}} and \isa{M\ $\stackrel{\mathrm{whr}}{\longrightarrow}$\ M{\isacharprime}{\isacharprime}} then \isa{M{\isacharprime}\ {\isacharequal}\ M{\isacharprime}{\isacharprime}}.
  \item \isa{{\normalsize{}If\,}\ {\isasymDelta}\ {\isasymturnstile}\isactrlbsub {\isasymSigma}\isactrlesub \ M\ {\isasymleftrightarrow}\ N\ {\isacharcolon}\ {\isasymtau}\ {\normalsize \,then\,}\ {\isasymnexists}M{\isacharprime}{\isachardot}\ M\ $\stackrel{\mathrm{whr}}{\longrightarrow}$\ M{\isacharprime}{\isachardot}} 
  \item \isa{{\normalsize{}If\,}\ {\isasymDelta}\ {\isasymturnstile}\isactrlbsub {\isasymSigma}\isactrlesub \ M\ {\isasymleftrightarrow}\ N\ {\isacharcolon}\ {\isasymtau}\ {\normalsize \,then\,}\ {\isasymnexists}N{\isacharprime}{\isachardot}\ N\ $\stackrel{\mathrm{whr}}{\longrightarrow}$\ N{\isacharprime}{\isachardot}} 
  \item If \isa{{\isasymDelta}\ {\isasymturnstile}\isactrlbsub {\isasymSigma}\isactrlesub \ M\ {\isasymLeftrightarrow}\ N\ {\isacharcolon}\ {\isasymtau}} and  \isa{{\isasymDelta}\ {\isasymturnstile}\isactrlbsub {\isasymSigma}\isactrlesub \ M\ {\isasymLeftrightarrow}\ N\ {\isacharcolon}\ {\isasymtau}{\isacharprime}} then 
  \mbox{\isa{{\isasymtau}\ {\isacharequal}\ {\isasymtau}{\isacharprime}}}.
  \item If \isa{{\isasymDelta}\ {\isasymturnstile}\isactrlbsub {\isasymSigma}\isactrlesub \ A\ {\isasymLeftrightarrow}\ B\ {\isacharcolon}\ {\isasymkappa}} and  \isa{{\isasymDelta}\ {\isasymturnstile}\isactrlbsub {\isasymSigma}\isactrlesub \ A\ {\isasymLeftrightarrow}\ B\ {\isacharcolon}\ {\isasymkappa}{\isacharprime}} then 
  \mbox{\isa{{\isasymkappa}\ {\isacharequal}\ {\isasymkappa}{\isacharprime}}}.
  \end{compactenum}
  \end{lemma}

  \noindent
  However, we needed generalized forms of parts 4 and 5 in the proof
  of transitivity (\refThm{alg-transitivity}). These properties are
  also later used in \refThm{alg-eq-quasidecidable} in proving
  decidability of the algorithmic rules.

  \begin{lemma}[(Generalized determinacy)]\labelLem{generalized-determinacy}
  Suppose that \isa{{\isasymturnstile}\ {\isasymSigma}\ sig} and \mbox{\isa{{\isasymturnstile}\ {\isasymDelta}\ sctx}}. 
  \begin{compactenum}
  \item If \isa{{\isasymDelta}\ {\isasymturnstile}\isactrlbsub {\isasymSigma}\isactrlesub \ M\ {\isasymLeftrightarrow}\ N\ {\isacharcolon}\ {\isasymtau}} and  \isa{{\isasymDelta}\ {\isasymturnstile}\isactrlbsub {\isasymSigma}\isactrlesub \ N\ {\isasymLeftrightarrow}\ P\ {\isacharcolon}\ {\isasymtau}{\isacharprime}} then 
  \mbox{\isa{{\isasymtau}\ {\isacharequal}\ {\isasymtau}{\isacharprime}}}.
  \item If \isa{{\isasymDelta}\ {\isasymturnstile}\isactrlbsub {\isasymSigma}\isactrlesub \ A\ {\isasymLeftrightarrow}\ B\ {\isacharcolon}\ {\isasymkappa}} and  \isa{{\isasymDelta}\ {\isasymturnstile}\isactrlbsub {\isasymSigma}\isactrlesub \ B\ {\isasymLeftrightarrow}\ C\ {\isacharcolon}\ {\isasymkappa}{\isacharprime}} then 
  \mbox{\isa{{\isasymkappa}\ {\isacharequal}\ {\isasymkappa}{\isacharprime}}}.
  \end{compactenum}
  \end{lemma}

  \noindent
  Verifying symmetry of the algorithmic judgments is then straightforward,
  using properties established so far.

  \begin{theorem}[(Symmetry of algorithmic equivalence)]
  ~\labelThm{alg-symmetry}
 \begin{compactenum}
  \item[1.] \isa{{\normalsize{}If\,}\ {\isasymDelta}\ {\isasymturnstile}\isactrlbsub {\isasymSigma}\isactrlesub \ M\ {\isasymLeftrightarrow}\ N\ {\isacharcolon}\ {\isasymtau}\ {\normalsize \,then\,}\ {\isasymDelta}\ {\isasymturnstile}\isactrlbsub {\isasymSigma}\isactrlesub \ N\ {\isasymLeftrightarrow}\ M\ {\isacharcolon}\ {\isasymtau}{\isachardot}} 
  \item[2.] \isa{{\normalsize{}If\,}\ {\isasymDelta}\ {\isasymturnstile}\isactrlbsub {\isasymSigma}\isactrlesub \ M\ {\isasymleftrightarrow}\ N\ {\isacharcolon}\ {\isasymtau}\ {\normalsize \,then\,}\ {\isasymDelta}\ {\isasymturnstile}\isactrlbsub {\isasymSigma}\isactrlesub \ N\ {\isasymleftrightarrow}\ M\ {\isacharcolon}\ {\isasymtau}{\isachardot}} 
  \item[3.] \isa{{\normalsize{}If\,}\ {\isasymDelta}\ {\isasymturnstile}\isactrlbsub {\isasymSigma}\isactrlesub \ A\ {\isasymLeftrightarrow}\ B\ {\isacharcolon}\ {\isasymkappa}\ {\normalsize \,then\,}\ {\isasymDelta}\ {\isasymturnstile}\isactrlbsub {\isasymSigma}\isactrlesub \ B\ {\isasymLeftrightarrow}\ A\ {\isacharcolon}\ {\isasymkappa}{\isachardot}} 
  \item[4.] \isa{{\normalsize{}If\,}\ {\isasymDelta}\ {\isasymturnstile}\isactrlbsub {\isasymSigma}\isactrlesub \ A\ {\isasymleftrightarrow}\ B\ {\isacharcolon}\ {\isasymkappa}\ {\normalsize \,then\,}\ {\isasymDelta}\ {\isasymturnstile}\isactrlbsub {\isasymSigma}\isactrlesub \ B\ {\isasymleftrightarrow}\ A\ {\isacharcolon}\ {\isasymkappa}{\isachardot}} 
  \item[5.] \isa{{\normalsize{}If\,}\ {\isasymDelta}\ {\isasymturnstile}\isactrlbsub {\isasymSigma}\isactrlesub \ K\ {\isasymLeftrightarrow}\ L\ {\isacharcolon}\ kind\isactrlisup {\isacharminus}\ {\normalsize \,then\,}\ {\isasymDelta}\ {\isasymturnstile}\isactrlbsub {\isasymSigma}\isactrlesub \ L\ {\isasymLeftrightarrow}\ K\ {\isacharcolon}\ kind\isactrlisup {\isacharminus}{\isachardot}} 
  \end{compactenum}
  \end{theorem}

  \noindent
  However, verifying transitivity required more work.  

  \begin{theorem}[(Transitivity of algorithmic equivalence)]\labelThm{alg-transitivity}
  Suppose that\\ \mbox{\isa{{\isasymturnstile}\ {\isasymSigma}\ ssig}} and \mbox{\isa{{\isasymturnstile}\ {\isasymDelta}\ sctx}}. 
  \begin{compactenum}
  \item %@ {thm lemma_3_5_12(1)} 
  If \isa{{\isasymDelta}\ {\isasymturnstile}\isactrlbsub {\isasymSigma}\isactrlesub \ M\ {\isasymLeftrightarrow}\ N\ {\isacharcolon}\ {\isasymtau}} and \isa{{\isasymDelta}\ {\isasymturnstile}\isactrlbsub {\isasymSigma}\isactrlesub \ N\ {\isasymLeftrightarrow}\ P\ {\isacharcolon}\ {\isasymtau}} 
  then \mbox{\isa{{\isasymDelta}\ {\isasymturnstile}\isactrlbsub {\isasymSigma}\isactrlesub \ M\ {\isasymLeftrightarrow}\ P\ {\isacharcolon}\ {\isasymtau}}}. 
  \item If \isa{{\isasymDelta}\ {\isasymturnstile}\isactrlbsub {\isasymSigma}\isactrlesub \ M\ {\isasymleftrightarrow}\ N\ {\isacharcolon}\ {\isasymtau}} and \isa{{\isasymDelta}\ {\isasymturnstile}\isactrlbsub {\isasymSigma}\isactrlesub \ N\ {\isasymleftrightarrow}\ P\ {\isacharcolon}\ {\isasymtau}} 
  then \mbox{\isa{{\isasymDelta}\ {\isasymturnstile}\isactrlbsub {\isasymSigma}\isactrlesub \ M\ {\isasymleftrightarrow}\ P\ {\isacharcolon}\ {\isasymtau}}}. 
   \item If \isa{{\isasymDelta}\ {\isasymturnstile}\isactrlbsub {\isasymSigma}\isactrlesub \ A\ {\isasymLeftrightarrow}\ B\ {\isacharcolon}\ {\isasymkappa}} and \isa{{\isasymDelta}\ {\isasymturnstile}\isactrlbsub {\isasymSigma}\isactrlesub \ B\ {\isasymLeftrightarrow}\ C\ {\isacharcolon}\ {\isasymkappa}} 
   then \mbox{\isa{{\isasymDelta}\ {\isasymturnstile}\isactrlbsub {\isasymSigma}\isactrlesub \ A\ {\isasymLeftrightarrow}\ C\ {\isacharcolon}\ {\isasymkappa}}}. 
   \item If \isa{{\isasymDelta}\ {\isasymturnstile}\isactrlbsub {\isasymSigma}\isactrlesub \ A\ {\isasymleftrightarrow}\ B\ {\isacharcolon}\ {\isasymkappa}} and \isa{{\isasymDelta}\ {\isasymturnstile}\isactrlbsub {\isasymSigma}\isactrlesub \ B\ {\isasymleftrightarrow}\ C\ {\isacharcolon}\ {\isasymkappa}} 
   then \mbox{\isa{{\isasymDelta}\ {\isasymturnstile}\isactrlbsub {\isasymSigma}\isactrlesub \ A\ {\isasymleftrightarrow}\ C\ {\isacharcolon}\ {\isasymkappa}}}. 
   \item If \isa{{\isasymDelta}\ {\isasymturnstile}\isactrlbsub {\isasymSigma}\isactrlesub \ K\ {\isasymLeftrightarrow}\ L\ {\isacharcolon}\ kind\isactrlisup {\isacharminus}} and \isa{{\isasymDelta}\ {\isasymturnstile}\isactrlbsub {\isasymSigma}\isactrlesub \ L\ {\isasymLeftrightarrow}\ L{\isacharprime}\ {\isacharcolon}\ kind\isactrlisup {\isacharminus}} 
   then \mbox{\isa{{\isasymDelta}\ {\isasymturnstile}\isactrlbsub {\isasymSigma}\isactrlesub \ K\ {\isasymLeftrightarrow}\ L{\isacharprime}\ {\isacharcolon}\ kind\isactrlisup {\isacharminus}}}. 
  \end{compactenum}
  \end{theorem}

  \begin{proof}
    As described in \HP, the proof is by simultaneous induction on the
    two derivations.  For types and kinds, this simultaneous induction
    can be avoided by performing induction over one derivation and
    using inversion principles.  For the object-level judgments (cases
    1 and 2), we formalize this argument in Isabelle by defining
    object-level algorithmic judgments instrumented with a height
    argument, and prove parts 1 and 2 by well-founded induction on the
    sum of the heights of the derivations.

    Because we use induction over the height of the instrumented
    derivation, we cannot take advantage of the ``strong'' induction
    principles for algorithmic
    derivations~\cite{UrbanBerghoferNorrish07}.  As a result, there
    are several cases where we need to perform some explicit
    $\alpha$-conversion and renaming steps; these are places in an
    informal proof where one usually appeals to renaming principles
    ``without loss of generality''.  In the current version of the
    nominal datatype package offers strong inversion principles that
    ameliorate this difficulty~\cite{BerghoferUrban08}.

    The generalized determinacy property
    (\refLem{generalized-determinacy}) is needed here in the case of
    structural equivalence of applications.
  \end{proof}%
\end{isamarkuptext}%
\isamarkuptrue%
\begin{isamarkuptext}%
\paragraph*{Strengthening}
At this point in the development, we can also prove that the
algorithmic judgments satisfy \emph{strengthening}; that is, unused
variables can be removed from the context without harming derivability
of a conclusion.  Strengthening is not discussed in \HP until later in
the article, but we found it helpful in the proof of soundness.  We
first need an (easily established) freshness-preservation property of
weak head reduction.

  \begin{lemma}[(Weak head reduction preserves freshness)]\labelLem{whr-fresh}\mbox{}\\
    \isa{{\normalsize{}If\,}\ \mbox{M\ $\stackrel{\mathrm{whr}}{\longrightarrow}$\ N}\ {\normalsize \,and\,}\ \mbox{x\ {\isasymsharp}\ M}\ {\normalsize\linebreak[0] \,then\,\linebreak[0]}\ \mbox{x\ {\isasymsharp}\ N{\isachardot}}}
  \end{lemma}

  With this property in hand, strengthening for algorithmic and
  structural equivalence can be established by induction on the
  structure of judgments, making use of basic properties of freshness,
  valid contexts, and the previous lemma as necessary.

  \begin{lemma}[(Strengthening of algorithmic
    equivalence)]\labelLem{algorithmic-strengthening}
~Suppose that \isa{x\ {\isasymsharp}\ {\isacharparenleft}{\isasymDelta}{\isacharprime}{\isacharcomma}\ M{\isacharcomma}\ N{\isacharparenright}}.  Then:
  \begin{compactenum}
    \item If \isa{{\isasymDelta}{\isacharprime}\ {\isacharat}\ {\isacharbrackleft}{\isacharparenleft}x{\isacharcomma}\ {\isasymtau}{\isacharprime}{\isacharparenright}{\isacharbrackright}\ {\isacharat}\ {\isasymDelta}\ {\isasymturnstile}\isactrlbsub {\isasymSigma}\isactrlesub \ M\ {\isasymLeftrightarrow}\ N\ {\isacharcolon}\ {\isasymtau}}  then \isa{{\isasymDelta}{\isacharprime}\ {\isacharat}\ {\isasymDelta}\ {\isasymturnstile}\isactrlbsub {\isasymSigma}\isactrlesub \ M\ {\isasymLeftrightarrow}\ N\ {\isacharcolon}\ {\isasymtau}}.
    \item If \isa{{\isasymDelta}{\isacharprime}\ {\isacharat}\ {\isacharbrackleft}{\isacharparenleft}x{\isacharcomma}\ {\isasymtau}{\isacharprime}{\isacharparenright}{\isacharbrackright}\ {\isacharat}\ {\isasymDelta}\ {\isasymturnstile}\isactrlbsub {\isasymSigma}\isactrlesub \ M\ {\isasymleftrightarrow}\ N\ {\isacharcolon}\ {\isasymtau}}  then \isa{{\isasymDelta}{\isacharprime}\ {\isacharat}\ {\isasymDelta}\ {\isasymturnstile}\isactrlbsub {\isasymSigma}\isactrlesub \ M\ {\isasymleftrightarrow}\ N\ {\isacharcolon}\ {\isasymtau}}.
    \end{compactenum}
  \end{lemma}
  \begin{proof}
    Straightforward induction on derivations, using properties of
    freshness.  \refLem{whr-fresh} is needed in
    the cases involving weak head reduction to maintain the freshness
    constraints needed for the induction hypothesis.
  \end{proof}%
\end{isamarkuptext}%
\isamarkuptrue%
\isamarkupsubsection{Completeness%
}
\isamarkuptrue%
\begin{isamarkuptext}%
\labelSec{completeness}%
\end{isamarkuptext}%
\isamarkuptrue%
\begin{isamarkuptext}%
\begin{figure}
    %\small
  \begin{center}
  \begin{tabular}{r@ {\hspace{2mm}}c@ {\hspace{2mm}}l}
  \isa{{\isasymDelta}\ {\isasymturnstile}\isactrlbsub {\isasymSigma}\isactrlesub \ M\ {\isacharequal}\ N\ {\isasymin}\ {\isasymlbrakk}a\isactrlisup {\isacharminus}{\isasymrbrakk}} & $=$ &  \isa{{\isasymDelta}\ {\isasymturnstile}\isactrlbsub {\isasymSigma}\isactrlesub \ M\ {\isasymLeftrightarrow}\ N\ {\isacharcolon}\ a\isactrlisup {\isacharminus}}\\
  \isa{{\isasymDelta}\ {\isasymturnstile}\isactrlbsub {\isasymSigma}\isactrlesub \ M\ {\isacharequal}\ N\ {\isasymin}\ {\isasymlbrakk}{\isasymtau}\ {\isasymrightarrow}\ {\isasymtau}{\isacharprime}{\isasymrbrakk}} & $=$  &  
  \isa{{\isasymforall}}\isa{{\isasymDelta}{\isacharprime}\ {\isasymextends}\ {\isasymDelta}}\isa{{\isacharcomma}\ M{\isacharprime}{\isacharcomma}\ N{\isacharprime}{\isachardot}}
  \isa{{\isasymDelta}{\isacharprime}\ {\isasymturnstile}\isactrlbsub {\isasymSigma}\isactrlesub \ M{\isacharprime}\ {\isacharequal}\ N{\isacharprime}\ {\isasymin}\ {\isasymlbrakk}{\isasymtau}{\isasymrbrakk}} \\
&&{\it implies} \isa{{\isasymDelta}{\isacharprime}\ {\isasymturnstile}\isactrlbsub {\isasymSigma}\isactrlesub \ M\ M{\isacharprime}\ {\isacharequal}\ N\ N{\isacharprime}\ {\isasymin}\ {\isasymlbrakk}{\isasymtau}{\isacharprime}{\isasymrbrakk}}\\
  \isa{{\isasymDelta}\ {\isasymturnstile}\isactrlbsub {\isasymSigma}\isactrlesub \ A\ {\isacharequal}\ B\ {\isasymin}\ {\isasymlbrakk}type\isactrlisup {\isacharminus}{\isasymrbrakk}} & $=$ &  \isa{{\isasymDelta}\ {\isasymturnstile}\isactrlbsub {\isasymSigma}\isactrlesub \ A\ {\isasymLeftrightarrow}\ B\ {\isacharcolon}\ type\isactrlisup {\isacharminus}}\\
  \isa{{\isasymDelta}\ {\isasymturnstile}\isactrlbsub {\isasymSigma}\isactrlesub \ A\ {\isacharequal}\ B\ {\isasymin}\ {\isasymlbrakk}{\isasymtau}\ {\isasymrightarrow}\ {\isasymkappa}{\isasymrbrakk}} & $=$ &  
  \isa{{\isasymforall}}\isa{{\isasymDelta}{\isacharprime}\ {\isasymextends}\ {\isasymDelta}}\isa{{\isacharcomma}\ M{\isacharprime}{\isacharcomma}\ N{\isacharprime}{\isachardot}}
  \isa{{\isasymDelta}{\isacharprime}\ {\isasymturnstile}\isactrlbsub {\isasymSigma}\isactrlesub \ M{\isacharprime}\ {\isacharequal}\ N{\isacharprime}\ {\isasymin}\ {\isasymlbrakk}{\isasymtau}{\isasymrbrakk}} \\
&&{\it implies} \isa{{\isasymDelta}{\isacharprime}\ {\isasymturnstile}\isactrlbsub {\isasymSigma}\isactrlesub \ A\ M{\isacharprime}\ {\isacharequal}\ B\ N{\isacharprime}\ {\isasymin}\ {\isasymlbrakk}{\isasymkappa}{\isasymrbrakk}}\\
  \isa{{\isasymDelta}\ {\isasymturnstile}\isactrlbsub {\isasymSigma}\isactrlesub \ K\ {\isacharequal}\ L\ {\isasymin}\ {\isasymlbrakk}kind\isactrlisup {\isacharminus}{\isasymrbrakk}} & $=$ &  \isa{{\isasymDelta}\ {\isasymturnstile}\isactrlbsub {\isasymSigma}\isactrlesub \ K\ {\isasymLeftrightarrow}\ L\ {\isacharcolon}\ kind\isactrlisup {\isacharminus}}\\
  \isa{{\isasymDelta}\ {\isasymturnstile}\isactrlbsub {\isasymSigma}\isactrlesub \ {\isacharbrackleft}{\isacharbrackright}\ {\isacharequal}\ {\isacharbrackleft}{\isacharbrackright}\ {\isasymin}\ {\isasymlbrakk}{\isacharbrackleft}{\isacharbrackright}{\isasymrbrakk}} & $=$ & \isa{True}\\
  \isa{{\isasymDelta}\ {\isasymturnstile}\isactrlbsub {\isasymSigma}\isactrlesub \ {\isacharparenleft}x{\isacharcomma}\ M{\isacharparenright}{\isacharcolon}{\isacharcolon}{\isasymsigma}\ {\isacharequal}\ {\isacharparenleft}x{\isacharcomma}\ N{\isacharparenright}{\isacharcolon}{\isacharcolon}{\isasymtheta}\ {\isasymin}\ {\isasymlbrakk}{\isacharparenleft}x{\isacharcomma}\ {\isasymtau}{\isacharparenright}{\isacharcolon}{\isacharcolon}{\isasymTheta}{\isasymrbrakk}} & $=$ & \isa{{\isasymDelta}\ {\isasymturnstile}\isactrlbsub {\isasymSigma}\isactrlesub \ {\isasymsigma}\ {\isacharequal}\ {\isasymtheta}\ {\isasymin}\ {\isasymlbrakk}{\isasymTheta}{\isasymrbrakk}} \text{\it and } \isa{x\ {\isasymsharp}\ {\isasymTheta}} \\
&&\text{\it and } \isa{{\isasymDelta}\ {\isasymturnstile}\isactrlbsub {\isasymSigma}\isactrlesub \ M\ {\isacharequal}\ N\ {\isasymin}\ {\isasymlbrakk}{\isasymtau}{\isasymrbrakk}}
  \end{tabular}
  \end{center}
  \caption{Logical relation definition}\labelFig{logrel-def}
  \end{figure}

  \noindent
  The proof of completeness involves a Kripke-style logical relations
  argument.  We can define the logical relation for objects, types,
  and substitutions, by induction on the structure of simple types
  \isa{{\isasymtau}} and kinds \isa{{\isasymkappa}} and simple contexts
  \isa{{\isasymTheta}}, respectively, as shown in \refFig{logrel-def}.
  This kind of logical relation is called Kripke-style because the
  case for function types is modeled on Kripke's possible-worlds
  semantics for intuitionistic logic: it is
  indexed by a variable context $\Delta$ and in the case for function
  types and kinds, we quantify over all valid extensions to $\Delta$ when
  considering the argument terms $M',N'$.

  The key steps in proving completeness are showing that logically related
  terms are algorithmically equivalent (\refThm{logrel-implies-alg-equiv}) and
  that definitionally equivalent terms are logically related
  (\refThm{def-eq-implies-logrel}).  Many properties can be established by an
  induction on the structure of types, appealing to the properties of the
  algorithmic judgments established in section 3 of \HP and the definition of
  the logical relation.  

  \begin{lemma}[(Logical relation weakening)]
  Suppose \isa{{\isasymDelta}{\isacharprime}\ {\isasymextends}\ {\isasymDelta}}.  
  \begin{compactenum}
  \item If \isa{{\isasymDelta}\ {\isasymturnstile}\isactrlbsub {\isasymSigma}\isactrlesub \ M\ {\isacharequal}\ N\ {\isasymin}\ {\isasymlbrakk}{\isasymtau}{\isasymrbrakk}} then \isa{{\isasymDelta}{\isacharprime}\ {\isasymturnstile}\isactrlbsub {\isasymSigma}\isactrlesub \ M\ {\isacharequal}\ N\ {\isasymin}\ {\isasymlbrakk}{\isasymtau}{\isasymrbrakk}}.
  \item If \isa{{\isasymDelta}\ {\isasymturnstile}\isactrlbsub {\isasymSigma}\isactrlesub \ A\ {\isacharequal}\ B\ {\isasymin}\ {\isasymlbrakk}{\isasymkappa}{\isasymrbrakk}} then \isa{{\isasymDelta}{\isacharprime}\ {\isasymturnstile}\isactrlbsub {\isasymSigma}\isactrlesub \ A\ {\isacharequal}\ B\ {\isasymin}\ {\isasymlbrakk}{\isasymkappa}{\isasymrbrakk}}.
  \item If \isa{{\isasymDelta}\ {\isasymturnstile}\isactrlbsub {\isasymSigma}\isactrlesub \ {\isasymsigma}\ {\isacharequal}\ {\isasymtheta}\ {\isasymin}\ {\isasymlbrakk}{\isasymTheta}{\isasymrbrakk}} then \isa{{\isasymDelta}{\isacharprime}\ {\isasymturnstile}\isactrlbsub {\isasymSigma}\isactrlesub \ {\isasymsigma}\ {\isacharequal}\ {\isasymtheta}\ {\isasymin}\ {\isasymlbrakk}{\isasymTheta}{\isasymrbrakk}}.
  \end{compactenum}
  \end{lemma}

  \begin{theorem}[(Logically related terms are algorithmically equivalent)]\labelThm{logrel-implies-alg-equiv}
  \mbox{Suppose \isa{{\isasymturnstile}\ {\isasymDelta}\ sctx}}.   
  \begin{compactenum}
  \item If \isa{{\isasymDelta}\ {\isasymturnstile}\isactrlbsub {\isasymSigma}\isactrlesub \ M\ {\isacharequal}\ N\ {\isasymin}\ {\isasymlbrakk}{\isasymtau}{\isasymrbrakk}} then \isa{{\isasymDelta}\ {\isasymturnstile}\isactrlbsub {\isasymSigma}\isactrlesub \ M\ {\isasymLeftrightarrow}\ N\ {\isacharcolon}\ {\isasymtau}}.
  \item If \isa{{\isasymDelta}\ {\isasymturnstile}\isactrlbsub {\isasymSigma}\isactrlesub \ M\ {\isasymleftrightarrow}\ N\ {\isacharcolon}\ {\isasymtau}} then \isa{{\isasymDelta}\ {\isasymturnstile}\isactrlbsub {\isasymSigma}\isactrlesub \ M\ {\isacharequal}\ N\ {\isasymin}\ {\isasymlbrakk}{\isasymtau}{\isasymrbrakk}}.
  \item If \isa{{\isasymDelta}\ {\isasymturnstile}\isactrlbsub {\isasymSigma}\isactrlesub \ A\ {\isacharequal}\ B\ {\isasymin}\ {\isasymlbrakk}{\isasymkappa}{\isasymrbrakk}} then \isa{{\isasymDelta}\ {\isasymturnstile}\isactrlbsub {\isasymSigma}\isactrlesub \ A\ {\isasymLeftrightarrow}\ B\ {\isacharcolon}\ {\isasymkappa}}.
  \item If \isa{{\isasymDelta}\ {\isasymturnstile}\isactrlbsub {\isasymSigma}\isactrlesub \ A\ {\isasymleftrightarrow}\ B\ {\isacharcolon}\ {\isasymkappa}} then \isa{{\isasymDelta}\ {\isasymturnstile}\isactrlbsub {\isasymSigma}\isactrlesub \ A\ {\isacharequal}\ B\ {\isasymin}\ {\isasymlbrakk}{\isasymkappa}{\isasymrbrakk}}.
  \end{compactenum}
  \end{theorem}

  \begin{lemma}[(Closure under head expansion)]
    ~
    \begin{compactenum}
    \item If \isa{M\ $\stackrel{\mathrm{whr}}{\longrightarrow}$\ M{\isacharprime}} and \isa{{\isasymDelta}\ {\isasymturnstile}\isactrlbsub {\isasymSigma}\isactrlesub \ M{\isacharprime}\ {\isacharequal}\ N\ {\isasymin}\ {\isasymlbrakk}{\isasymtau}{\isasymrbrakk}} then \isa{{\isasymDelta}\ {\isasymturnstile}\isactrlbsub {\isasymSigma}\isactrlesub \ M\ {\isacharequal}\ N\ {\isasymin}\ {\isasymlbrakk}{\isasymtau}{\isasymrbrakk}}.
    \item If  \isa{N\ $\stackrel{\mathrm{whr}}{\longrightarrow}$\ N{\isacharprime}} and \isa{{\isasymDelta}\ {\isasymturnstile}\isactrlbsub {\isasymSigma}\isactrlesub \ M\ {\isacharequal}\ N{\isacharprime}\ {\isasymin}\ {\isasymlbrakk}{\isasymtau}{\isasymrbrakk}} then \isa{{\isasymDelta}\ {\isasymturnstile}\isactrlbsub {\isasymSigma}\isactrlesub \ M\ {\isacharequal}\ N\ {\isasymin}\ {\isasymlbrakk}{\isasymtau}{\isasymrbrakk}}.
    \end{compactenum}
  \end{lemma}

  \begin{lemma}[(Logical relation symmetry)]
  ~
  \begin{compactenum}
  \item \isa{{\normalsize{}If\,}\ {\isasymDelta}\ {\isasymturnstile}\isactrlbsub {\isasymSigma}\isactrlesub \ M\ {\isacharequal}\ N\ {\isasymin}\ {\isasymlbrakk}{\isasymtau}{\isasymrbrakk}\ {\normalsize \,then\,}\ {\isasymDelta}\ {\isasymturnstile}\isactrlbsub {\isasymSigma}\isactrlesub \ N\ {\isacharequal}\ M\ {\isasymin}\ {\isasymlbrakk}{\isasymtau}{\isasymrbrakk}{\isachardot}}
  \item \isa{{\normalsize{}If\,}\ {\isasymDelta}\ {\isasymturnstile}\isactrlbsub {\isasymSigma}\isactrlesub \ A\ {\isacharequal}\ B\ {\isasymin}\ {\isasymlbrakk}{\isasymkappa}{\isasymrbrakk}\ {\normalsize \,then\,}\ {\isasymDelta}\ {\isasymturnstile}\isactrlbsub {\isasymSigma}\isactrlesub \ B\ {\isacharequal}\ A\ {\isasymin}\ {\isasymlbrakk}{\isasymkappa}{\isasymrbrakk}{\isachardot}}
  \item \isa{{\normalsize{}If\,}\ {\isasymDelta}\ {\isasymturnstile}\isactrlbsub {\isasymSigma}\isactrlesub \ {\isasymsigma}\ {\isacharequal}\ {\isasymtheta}\ {\isasymin}\ {\isasymlbrakk}{\isasymTheta}{\isasymrbrakk}\ {\normalsize \,then\,}\ {\isasymDelta}\ {\isasymturnstile}\isactrlbsub {\isasymSigma}\isactrlesub \ {\isasymtheta}\ {\isacharequal}\ {\isasymsigma}\ {\isasymin}\ {\isasymlbrakk}{\isasymTheta}{\isasymrbrakk}{\isachardot}}
  \end{compactenum}
  \end{lemma}

  \begin{lemma}[(Logical relation transitivity)]\mbox{}\\
  Suppose that \isa{{\isasymturnstile}\ {\isasymSigma}\ sig} and \mbox{\isa{{\isasymturnstile}\ {\isasymDelta}\ sctx}}. 
  \begin{compactenum}
  \item If \isa{{\isasymDelta}\ {\isasymturnstile}\isactrlbsub {\isasymSigma}\isactrlesub \ M\ {\isacharequal}\ N\ {\isasymin}\ {\isasymlbrakk}{\isasymtau}{\isasymrbrakk}} and \isa{{\isasymDelta}\ {\isasymturnstile}\isactrlbsub {\isasymSigma}\isactrlesub \ N\ {\isacharequal}\ P\ {\isasymin}\ {\isasymlbrakk}{\isasymtau}{\isasymrbrakk}}
  then \mbox{\isa{{\isasymDelta}\ {\isasymturnstile}\isactrlbsub {\isasymSigma}\isactrlesub \ M\ {\isacharequal}\ P\ {\isasymin}\ {\isasymlbrakk}{\isasymtau}{\isasymrbrakk}}}.
  \item If \isa{{\isasymDelta}\ {\isasymturnstile}\isactrlbsub {\isasymSigma}\isactrlesub \ A\ {\isacharequal}\ B\ {\isasymin}\ {\isasymlbrakk}{\isasymkappa}{\isasymrbrakk}} and \isa{{\isasymDelta}\ {\isasymturnstile}\isactrlbsub {\isasymSigma}\isactrlesub \ B\ {\isacharequal}\ C\ {\isasymin}\ {\isasymlbrakk}{\isasymkappa}{\isasymrbrakk}}
  then \mbox{\isa{{\isasymDelta}\ {\isasymturnstile}\isactrlbsub {\isasymSigma}\isactrlesub \ A\ {\isacharequal}\ C\ {\isasymin}\ {\isasymlbrakk}{\isasymkappa}{\isasymrbrakk}}}.
  \item If \isa{{\isasymDelta}\ {\isasymturnstile}\isactrlbsub {\isasymSigma}\isactrlesub \ {\isasymsigma}\ {\isacharequal}\ {\isasymtheta}\ {\isasymin}\ {\isasymlbrakk}{\isasymTheta}{\isasymrbrakk}} and \isa{{\isasymDelta}\ {\isasymturnstile}\isactrlbsub {\isasymSigma}\isactrlesub \ {\isasymtheta}\ {\isacharequal}\ {\isasymdelta}\ {\isasymin}\ {\isasymlbrakk}{\isasymTheta}{\isasymrbrakk}}
  then \mbox{\isa{{\isasymDelta}\ {\isasymturnstile}\isactrlbsub {\isasymSigma}\isactrlesub \ {\isasymsigma}\ {\isacharequal}\ {\isasymdelta}\ {\isasymin}\ {\isasymlbrakk}{\isasymTheta}{\isasymrbrakk}}}.
  \end{compactenum}
  \end{lemma}

  \noindent
  The proof that definitionally equal terms are logically related
  required some care to formalize.  The key step is showing that
  applying logically related substitutions to definitionally equal
  terms yields logically related terms.  Establishing this (via the
  following lemma) required identifying and proving a number of
  standard properties of simultaneous substitutions.  In contrast,
  reasoning about single substitutions sufficed almost everywhere else
  in the formalization.

  \begin{lemma}\labelLem{def-eq-implies-subst-logrel}
  Suppose \isa{{\isasymturnstile}\ {\isasymDelta}\ sctx} and \isa{{\isasymDelta}\ {\isasymturnstile}\isactrlbsub {\isasymSigma}\isactrlesub \ {\isasymsigma}\ {\isacharequal}\ {\isasymtheta}\ {\isasymin}\ {\isasymlbrakk}{\isasymGamma}\isactrlisup {\isacharminus}{\isasymrbrakk}}.
  \begin{compactenum}
  \item If \isa{{\isasymGamma}\ {\isasymturnstile}\isactrlisub {\isasymSigma}\ M\ {\isacharequal}\ N\ {\isacharcolon}\ A} then \isa{{\isasymDelta}\ {\isasymturnstile}\isactrlbsub {\isasymSigma}\isactrlisup {\isacharminus}\isactrlesub \ M{\isacharbrackleft}{\isasymsigma}{\isacharbrackright}\ {\isacharequal}\ N{\isacharbrackleft}{\isasymtheta}{\isacharbrackright}\ {\isasymin}\ {\isasymlbrakk}A\isactrlisup {\isacharminus}{\isasymrbrakk}}.
  \item If \isa{{\isasymGamma}\ {\isasymturnstile}\isactrlisub {\isasymSigma}\ A\ {\isacharequal}\ B\ {\isacharcolon}\ K} then \isa{{\isasymDelta}\ {\isasymturnstile}\isactrlbsub {\isasymSigma}\isactrlisup {\isacharminus}\isactrlesub \ A{\isacharbrackleft}{\isasymsigma}{\isacharbrackright}\ {\isacharequal}\ B{\isacharbrackleft}{\isasymtheta}{\isacharbrackright}\ {\isasymin}\ {\isasymlbrakk}K\isactrlisup {\isacharminus}{\isasymrbrakk}}.
  \end{compactenum}
  \end{lemma}

  \noindent
  The last step needed to establish completeness is to show that the identity
  substitution over a given context (written \isa{id\isactrlisub {\isasymGamma}}) is related to
  itself:

  \begin{lemma}
  \isa{{\normalsize{}If\,}\ {\isasymturnstile}\isactrlisub {\isasymSigma}\ {\isasymGamma}\ ctx\ {\normalsize \,then\,}\ {\isasymGamma}\isactrlisup {\isacharminus}\ {\isasymturnstile}\isactrlbsub {\isasymSigma}\isactrlisup {\isacharminus}\isactrlesub \ id\isactrlisub {\isasymGamma}\ {\isacharequal}\ id\isactrlisub {\isasymGamma}\ {\isasymin}\ {\isasymlbrakk}{\isasymGamma}\isactrlisup {\isacharminus}{\isasymrbrakk}{\isachardot}}
  \end{lemma}

  \begin{theorem}[(Definitionally equal terms are logically related)]\labelThm{def-eq-implies-logrel}
  ~
  \begin{compactenum}
  \item \isa{{\normalsize{}If\,}\ {\isasymGamma}\ {\isasymturnstile}\isactrlisub {\isasymSigma}\ M\ {\isacharequal}\ N\ {\isacharcolon}\ A\ {\normalsize \,then\,}\ {\isasymGamma}\isactrlisup {\isacharminus}\ {\isasymturnstile}\isactrlbsub {\isasymSigma}\isactrlisup {\isacharminus}\isactrlesub \ M\ {\isacharequal}\ N\ {\isasymin}\ {\isasymlbrakk}A\isactrlisup {\isacharminus}{\isasymrbrakk}{\isachardot}}
  \item \isa{{\normalsize{}If\,}\ {\isasymGamma}\ {\isasymturnstile}\isactrlisub {\isasymSigma}\ A\ {\isacharequal}\ B\ {\isacharcolon}\ K\ {\normalsize \,then\,}\ {\isasymGamma}\isactrlisup {\isacharminus}\ {\isasymturnstile}\isactrlbsub {\isasymSigma}\isactrlisup {\isacharminus}\isactrlesub \ A\ {\isacharequal}\ B\ {\isasymin}\ {\isasymlbrakk}K\isactrlisup {\isacharminus}{\isasymrbrakk}{\isachardot}}
  \end{compactenum}
  \end{theorem}

  \begin{corollary}[(Completeness)]\labelCor{completeness}
  ~
  \begin{compactenum}
  \item \isa{{\normalsize{}If\,}\ {\isasymGamma}\ {\isasymturnstile}\isactrlisub {\isasymSigma}\ M\ {\isacharequal}\ N\ {\isacharcolon}\ A\ {\normalsize \,then\,}\ {\isasymGamma}\isactrlisup {\isacharminus}\ {\isasymturnstile}\isactrlbsub {\isasymSigma}\isactrlisup {\isacharminus}\isactrlesub \ M\ {\isasymLeftrightarrow}\ N\ {\isacharcolon}\ A\isactrlisup {\isacharminus}{\isachardot}}
  \item \isa{{\normalsize{}If\,}\ {\isasymGamma}\ {\isasymturnstile}\isactrlisub {\isasymSigma}\ A\ {\isacharequal}\ B\ {\isacharcolon}\ K\ {\normalsize \,then\,}\ {\isasymGamma}\isactrlisup {\isacharminus}\ {\isasymturnstile}\isactrlbsub {\isasymSigma}\isactrlisup {\isacharminus}\isactrlesub \ A\ {\isasymLeftrightarrow}\ B\ {\isacharcolon}\ K\isactrlisup {\isacharminus}{\isachardot}}
  \item \isa{{\normalsize{}If\,}\ {\isasymGamma}\ {\isasymturnstile}\isactrlisub {\isasymSigma}\ K\ {\isacharequal}\ L\ {\isacharcolon}\ kind\ {\normalsize \,then\,}\ {\isasymGamma}\isactrlisup {\isacharminus}\ {\isasymturnstile}\isactrlbsub {\isasymSigma}\isactrlisup {\isacharminus}\isactrlesub \ K\ {\isasymLeftrightarrow}\ L\ {\isacharcolon}\ kind\isactrlisup {\isacharminus}{\isachardot}}
  \end{compactenum}
  \end{corollary}

  \noindent
  Note that part 3 of \refCor{completeness} was omitted from \HP, but it is
  straightforward to prove by induction given parts 1 and 2, and algorithmic
  transitivity and symmetry.%
\end{isamarkuptext}%
\isamarkuptrue%
\isamarkupsubsection{Soundness%
}
\isamarkuptrue%
\begin{isamarkuptext}%
\labelSec{soundness}%
\end{isamarkuptext}%
\isamarkuptrue%
\begin{isamarkuptext}%
\noindent
  Soundness of algorithmic object (or type, or kind) equivalence means
  that if two well-formed objects (or type families, or kinds
  respectively) are algorithmically equivalent then they are also
  definitionally equivalent.  For example, for objects,
  \refThm{soundness}(1) states: 
  \begin{quote}
    If \isa{{\isasymGamma}\isactrlisup {\isacharminus}\ {\isasymturnstile}\isactrlbsub {\isasymSigma}\isactrlisup {\isacharminus}\isactrlesub \ M\ {\isasymLeftrightarrow}\ N\ {\isacharcolon}\ A\isactrlisup {\isacharminus}} and \isa{{\isasymGamma}\ {\isasymturnstile}\isactrlisub {\isasymSigma}\ M\ {\isacharcolon}\ A} and \mbox{\isa{{\isasymGamma}\ {\isasymturnstile}\isactrlisub {\isasymSigma}\ N\ {\isacharcolon}\ A}} \\
    then  \isa{{\isasymGamma}\ {\isasymturnstile}\isactrlisub {\isasymSigma}\ M\ {\isacharequal}\ N\ {\isacharcolon}\ A}.
  \end{quote}
  First, though, since the algorithmic judgments perform weak head
  reduction, we must show that weak head reduction preserves
  well-formedness:

  \begin{lemma}[(Subject reduction)] 
  Suppose \isa{M\ $\stackrel{\mathrm{whr}}{\longrightarrow}$\ M{\isacharprime}} and \mbox{\isa{{\isasymGamma}\ {\isasymturnstile}\isactrlisub {\isasymSigma}\ M\ {\isacharcolon}\ A}}.  Then
  \isa{{\isasymGamma}\ {\isasymturnstile}\isactrlisub {\isasymSigma}\ M{\isacharprime}\ {\isacharcolon}\ A} and \isa{{\isasymGamma}\ {\isasymturnstile}\isactrlisub {\isasymSigma}\ M\ {\isacharequal}\ M{\isacharprime}\ {\isacharcolon}\ A}.
  \end{lemma}

  \noindent
  Naturally, since algorithmic and structural equivalences for objects
  and types are defined by simultaneous induction, we must also prove
  a simultaneous soundness property for the structural equivalence
  judgments.  For example, to prove \refThm{soundness}(1), we also
  need to show by simultaneous induction that:
  \begin{quote}
    If \isa{{\isasymGamma}\isactrlisup {\isacharminus}\ {\isasymturnstile}\isactrlbsub {\isasymSigma}\isactrlisup {\isacharminus}\isactrlesub \ M\ {\isasymleftrightarrow}\ N\ {\isacharcolon}\ {\isasymtau}} and \isa{{\isasymGamma}\ {\isasymturnstile}\isactrlisub {\isasymSigma}\ M\ {\isacharcolon}\ A} and \isa{{\isasymGamma}\ {\isasymturnstile}\isactrlisub {\isasymSigma}\ N\ {\isacharcolon}\ B}\\
        then
        \isa{{\isasymGamma}\ {\isasymturnstile}\isactrlisub {\isasymSigma}\ M\ {\isacharequal}\ N\ {\isacharcolon}\ A\ \textrm{and\linebreak[1]}\ {\isasymGamma}\ {\isasymturnstile}\isactrlisub {\isasymSigma}\ A\ {\isacharequal}\ B\ {\isacharcolon}\ type\ \textrm{and\linebreak[1]}\ A\isactrlisup {\isacharminus}\ {\isacharequal}\ {\isasymtau}\ \textrm{and\linebreak[1]}\ B\isactrlisup {\isacharminus}\ {\isacharequal}\ {\isasymtau}}.
\end{quote}
In contrast to completeness, the proof of soundness in \HP proceeds by
entirely syntactic techniques, by induction over the structure of
algorithmic and structural derivations, using standard syntactic
properties and subject reduction.  Our initial formalization attempt
followed the proofs given by \HP.  However, we encountered two
difficulties which were not addressed in the article.  Both
difficulties have to do with algorithmic rules for checking
equivalence at function types (or function kinds) using
extensionality.  In the rest of this section, we first discuss and
address a minor difficulty involving extensionality in the proof of
\refThm{soundness}(1).  We then discuss a more serious complication in
proving soundness at the level of types, and show how to fix the
problem.  We conclude by summarizing the soundness results.

  \paragraph*{Soundness for algorithmic object equivalence}
  In proving the soundness of algorithmic extensionality for objects arising
  in part 1 of \refThm{soundness}, recall that we have a derivation of the
  form:

  \begin{center}
  \isa{\mbox{}\inferrule{\mbox{{\isacharparenleft}x{\isacharcomma}\ {\isasymtau}\isactrlisub {\isadigit{1}}{\isacharparenright}{\isacharcolon}{\isacharcolon}{\isasymGamma}\isactrlisup {\isacharminus}\ {\isasymturnstile}\isactrlbsub {\isasymSigma}\isactrlesub \ M\ x\ {\isasymLeftrightarrow}\ N\ x\ {\isacharcolon}\ {\isasymtau}\isactrlisub {\isadigit{2}}}\\\ \mbox{x\ {\isasymsharp}\ {\isacharparenleft}{\isasymGamma}\isactrlisup {\isacharminus}{\isacharcomma}\ M{\isacharcomma}\ N{\isacharparenright}}}{\mbox{{\isasymGamma}\isactrlisup {\isacharminus}\ {\isasymturnstile}\isactrlbsub {\isasymSigma}\isactrlesub \ M\ {\isasymLeftrightarrow}\ N\ {\isacharcolon}\ {\isasymtau}\isactrlisub {\isadigit{1}}\ {\isasymrightarrow}\ {\isasymtau}\isactrlisub {\isadigit{2}}}}}
  \end{center}

  \noindent
  and we also know that \isa{{\isasymGamma}\ {\isasymturnstile}\isactrlisub {\isasymSigma}\ M\ {\isacharcolon}\ A} and \isa{{\isasymGamma}\ {\isasymturnstile}\isactrlisub {\isasymSigma}\ N\ {\isacharcolon}\ A} for some \isa{A} with \isa{A\isactrlisup {\isacharminus}\ {\isacharequal}\ {\isasymtau}\isactrlisub {\isadigit{1}}\ {\isasymrightarrow}\ {\isasymtau}\isactrlisub {\isadigit{2}}}.  In order to apply the induction hypothesis, we need to know
  that \isa{M\ x} and \isa{N\ x} are well-formed in
  an extended context \isa{{\isacharparenleft}x{\isacharcomma}\ A\isactrlisub {\isadigit{1}}{\isacharparenright}{\isacharcolon}{\isacharcolon}{\isasymGamma}}.  \HP's proof begins by
  assuming that \isa{{\isasymGamma}\ {\isasymturnstile}\isactrlisub {\isasymSigma}\ M\ {\isacharcolon}\ {\isasymPi}x{\isacharcolon}A\isactrlisub {\isadigit{1}}{\isachardot}\ A\isactrlisub {\isadigit{2}}} and
  \isa{{\isasymGamma}\ {\isasymturnstile}\isactrlisub {\isasymSigma}\ N\ {\isacharcolon}\ {\isasymPi}x{\isacharcolon}A\isactrlisub {\isadigit{1}}{\isachardot}\ A\isactrlisub {\isadigit{2}}}, and proceeding using
  inversion properties.  However, it is \emph{not} immediately clear that
  \isa{A\isactrlisup {\isacharminus}\ {\isacharequal}\ {\isasymtau}\isactrlisub {\isadigit{1}}\ {\isasymrightarrow}\ {\isasymtau}\isactrlisub {\isadigit{2}}} implies that \isa{A\ {\isacharequal}\ {\isasymPi}x{\isacharcolon}A\isactrlisub {\isadigit{1}}{\isachardot}\ A\isactrlisub {\isadigit{2}}} for some \isa{A\isactrlisub {\isadigit{1}}} and \isa{A\isactrlisub {\isadigit{2}}}; indeed, this can fail to be the case if \isa{A} is not
  well-formed.  Instead, we first need the following inversion principles for
  erasure:

  \begin{lemma}[(Erasure inversion)]\labelLem{erasure-inversion}
    ~
  \begin{compactenum}
  \item \isa{{\normalsize{}If\,}\ {\isasymGamma}\ {\isasymturnstile}\isactrlisub {\isasymSigma}\ A\ {\isacharcolon}\ {\isasymPi}x{\isacharcolon}B{\isachardot}\ K\ {\normalsize \,then\,}\ {\isasymexists}c{\isachardot}\ A\isactrlisup {\isacharminus}\ {\isacharequal}\ c\isactrlisup {\isacharminus}{\isachardot}} 
    \label{bug_fix1}
  \item If \isa{{\isasymtau}\isactrlisub {\isadigit{1}}\ {\isasymrightarrow}\ {\isasymtau}\isactrlisub {\isadigit{2}}\ {\isacharequal}\ A\isactrlisup {\isacharminus}} and
      \isa{{\isasymGamma}\ {\isasymturnstile}\isactrlisub {\isasymSigma}\ A\ {\isacharcolon}\ type} and
      \isa{x\ {\isasymsharp}\ A} then\\
      \isa{{\isasymexists}A\isactrlisub {\isadigit{1}}\ A\isactrlisub {\isadigit{2}}{\isachardot}\ A\ {\isacharequal}\ {\isasymPi}x{\isacharcolon}A\isactrlisub {\isadigit{1}}{\isachardot}\ A\isactrlisub {\isadigit{2}}}.
    \label{bug_fix2}
  \item \isa{{\normalsize{}If\,}\ {\isasymtau}\ {\isasymrightarrow}\ {\isasymkappa}\ {\isacharequal}\ K\isactrlisup {\isacharminus}\ {\normalsize \,and\,}\ x\ {\isasymsharp}\ K\ {\normalsize \,then\,}\ {\isasymexists}A\ L{\isachardot}\ K\ {\isacharequal}\ {\isasymPi}x{\isacharcolon}A{\isachardot}\ L{\isachardot}} 
    \label{bug_fix3}
  \end{compactenum}
  \end{lemma}
  \begin{proof}
    Part \ref{bug_fix1} follows by induction on the derivation.  Parts \ref{bug_fix2} and \ref{bug_fix3}
    follow by induction on the structure of \isa{A} and \isa{K} respectively.  In the case for type applications \isa{A\ M}, clearly \isa{A} has a $\Pi$-kind, but by part \ref{bug_fix1},
    \isa{A} erases to a constant, contradicting the assumption
    that \isa{A\isactrlisup {\isacharminus}\ {\isacharequal}\ {\isasymtau}\isactrlisub {\isadigit{1}}\ {\isasymrightarrow}\ {\isasymtau}\isactrlisub {\isadigit{2}}}. So the case is vacuous.  The remaining cases
    of part \ref{bug_fix2} are straightforward, as are the cases for part \ref{bug_fix3}.
  \end{proof}

  Using \refLem{erasure-inversion}, we can complete the proof of the
  first part of \refThm{soundness} as described in \HP:
  
  \begin{lemma}[(Soundness of algorithmic object equivalence)]
    \labelLem{alg-object-equiv-soundness}
    ~
    \begin{compactenum}
    \item If \isa{{\isasymGamma}\isactrlisup {\isacharminus}\ {\isasymturnstile}\isactrlbsub {\isasymSigma}\isactrlisup {\isacharminus}\isactrlesub \ M\ {\isasymLeftrightarrow}\ N\ {\isacharcolon}\ A\isactrlisup {\isacharminus}} and
             \isa{{\isasymGamma}\ {\isasymturnstile}\isactrlisub {\isasymSigma}\ M\ {\isacharcolon}\ A} and
             \mbox{\isa{{\isasymGamma}\ {\isasymturnstile}\isactrlisub {\isasymSigma}\ N\ {\isacharcolon}\ A}}
          then \\\isa{{\isasymGamma}\ {\isasymturnstile}\isactrlisub {\isasymSigma}\ M\ {\isacharequal}\ N\ {\isacharcolon}\ A}.
    \item If \isa{{\isasymGamma}\isactrlisup {\isacharminus}\ {\isasymturnstile}\isactrlbsub {\isasymSigma}\isactrlisup {\isacharminus}\isactrlesub \ M\ {\isasymleftrightarrow}\ N\ {\isacharcolon}\ {\isasymtau}} and
             \isa{{\isasymGamma}\ {\isasymturnstile}\isactrlisub {\isasymSigma}\ M\ {\isacharcolon}\ A} and
             \mbox{\isa{{\isasymGamma}\ {\isasymturnstile}\isactrlisub {\isasymSigma}\ N\ {\isacharcolon}\ B}}
          then \\
          \isa{{\isasymGamma}\ {\isasymturnstile}\isactrlisub {\isasymSigma}\ M\ {\isacharequal}\ N\ {\isacharcolon}\ A\ \textrm{and\linebreak[1]}\ {\isasymGamma}\ {\isasymturnstile}\isactrlisub {\isasymSigma}\ A\ {\isacharequal}\ B\ {\isacharcolon}\ type\ \textrm{and\linebreak[1]}\ A\isactrlisup {\isacharminus}\ {\isacharequal}\ {\isasymtau}\ \textrm{and\linebreak[1]}\ B\isactrlisup {\isacharminus}\ {\isacharequal}\ {\isasymtau}}.
     \end{compactenum}
  \end{lemma}%
\end{isamarkuptext}%
\isamarkuptrue%
\begin{isamarkuptext}%
\paragraph*{Soundness for algorithmic type equivalence} The second
  problem we encountered arises in the proof of soundness for the
  extensionality rule in the algorithmic type equivalence judgment
  (part~3 of \refThm{soundness}).  In this case, we have a derivation of the form:

  \begin{center}
  \isa{\mbox{}\inferrule{\mbox{{\isacharparenleft}x{\isacharcomma}\ {\isasymtau}{\isacharparenright}{\isacharcolon}{\isacharcolon}{\isasymGamma}\isactrlisup {\isacharminus}\ {\isasymturnstile}\isactrlbsub {\isasymSigma}\isactrlesub \ A\ x\ {\isasymLeftrightarrow}\ B\ x\ {\isacharcolon}\ {\isasymkappa}}\\\ \mbox{x\ {\isasymsharp}\ {\isacharparenleft}{\isasymGamma}\isactrlisup {\isacharminus}{\isacharcomma}\ A{\isacharcomma}\ B{\isacharparenright}}}{\mbox{{\isasymGamma}\isactrlisup {\isacharminus}\ {\isasymturnstile}\isactrlbsub {\isasymSigma}\isactrlesub \ A\ {\isasymLeftrightarrow}\ B\ {\isacharcolon}\ {\isasymtau}\ {\isasymrightarrow}\ {\isasymkappa}}}}
  \end{center}

  \noindent
  We can easily show that the induction hypothesis applies, using the same
  technique as above, ultimately deriving \mbox{\isa{{\isacharparenleft}x{\isacharcomma}\ A{\isacharprime}{\isacharparenright}{\isacharcolon}{\isacharcolon}{\isasymGamma}\ {\isasymturnstile}\isactrlisub {\isasymSigma}\ A\ x\ {\isacharequal}\ B\ x\ {\isacharcolon}\ K}} for some \isa{A{\isacharprime}} and \isa{K}.
  However, we cannot complete the proof of this case in the same way as for
  object extensionality, because \HP's variant of LF does \emph{not} include a
  type-level extensionality rule
  \begin{center}
    \isa{\mbox{}\inferrule{\mbox{{\isasymGamma}\ {\isasymturnstile}\isactrlisub {\isasymSigma}\ A\ {\isacharcolon}\ {\isasymPi}x{\isacharcolon}C{\isachardot}\ K}\\\ \mbox{{\isasymGamma}\ {\isasymturnstile}\isactrlisub {\isasymSigma}\ B\ {\isacharcolon}\ {\isasymPi}x{\isacharcolon}C{\isachardot}\ K}\\\ \mbox{{\isasymGamma}\ {\isasymturnstile}\isactrlisub {\isasymSigma}\ C\ {\isacharcolon}\ type}\\\ \mbox{{\isacharparenleft}x{\isacharcomma}\ C{\isacharparenright}{\isacharcolon}{\isacharcolon}{\isasymGamma}\ {\isasymturnstile}\isactrlisub {\isasymSigma}\ A\ x\ {\isacharequal}\ B\ x\ {\isacharcolon}\ K}\\\ \mbox{x\ {\isasymsharp}\ {\isasymGamma}}}{\mbox{{\isasymGamma}\ {\isasymturnstile}\isactrlisub {\isasymSigma}\ A\ {\isacharequal}\ B\ {\isacharcolon}\ {\isasymPi}x{\isacharcolon}C{\isachardot}\ K}}}
  \end{center}
  that would permit us to conclude that \isa{{\isasymGamma}\ {\isasymturnstile}\isactrlisub {\isasymSigma}\ A\ {\isacharequal}\ B\ {\isacharcolon}\ {\isasymPi}x{\isacharcolon}A{\isacharprime}{\isachardot}\ K}.
  
  It was not immediately clear to us whether the original proof could
  be repaired.  There appear to be several ways to fix this problem by
  changing the definitional or algorithmic rules.  One way is simply
  to add the above extensionality rule for types to the definitional
  system.  Using our formalization, we were easily able to verify that
  this solves the problem and does not introduce any new complications.
  For this we had to make sure that every proof done earlier is
  either not affected by this additional rule or can be extended to
  include it.

  \begin{figure}[tb]
    %\small
\fbox{\isa{{\isasymDelta}\ {\isasymturnstile}\isactrlbsub {\isasymSigma}\isactrlesub \ A\ {\isasymrightleftharpoons}\ B\ {\isacharcolon}\ {\isasymkappa}}}
  \begin{center}
    \begin{tabular}{c}
  \isa{\mbox{}\inferrule{\mbox{{\isacharparenleft}a{\isacharcomma}\ {\isasymkappa}{\isacharparenright}\ {\isasymin}\ {\isasymSigma}}\\\ \mbox{{\isasymturnstile}\ {\isasymSigma}\ ssig}\\\ \mbox{{\isasymturnstile}\ {\isasymDelta}\ sctx}}{\mbox{{\isasymDelta}\ {\isasymturnstile}\isactrlbsub {\isasymSigma}\isactrlesub \ a\ {\isasymrightleftharpoons}\ a\ {\isacharcolon}\ {\isasymkappa}}}} \hspace{5mm}
  \isa{\mbox{}\inferrule{\mbox{{\isasymDelta}\ {\isasymturnstile}\isactrlbsub {\isasymSigma}\isactrlesub \ A\ {\isasymrightleftharpoons}\ B\ {\isacharcolon}\ {\isasymtau}\ {\isasymrightarrow}\ {\isasymkappa}}\\\ \mbox{{\isasymDelta}\ {\isasymturnstile}\isactrlbsub {\isasymSigma}\isactrlesub \ M\ {\isasymLeftrightarrow}\ N\ {\isacharcolon}\ {\isasymtau}}}{\mbox{{\isasymDelta}\ {\isasymturnstile}\isactrlbsub {\isasymSigma}\isactrlesub \ A\ M\ {\isasymrightleftharpoons}\ B\ N\ {\isacharcolon}\ {\isasymkappa}}}}\smallskip\\
  \isa{\mbox{}\inferrule{\mbox{{\isasymDelta}\ {\isasymturnstile}\isactrlbsub {\isasymSigma}\isactrlesub \ A\isactrlisub {\isadigit{1}}\ {\isasymrightleftharpoons}\ B\isactrlisub {\isadigit{1}}\ {\isacharcolon}\ type\isactrlisup {\isacharminus}}\\\ \mbox{{\isacharparenleft}x{\isacharcomma}\ A\isactrlisub {\isadigit{1}}\isactrlisup {\isacharminus}{\isacharparenright}{\isacharcolon}{\isacharcolon}{\isasymDelta}\ {\isasymturnstile}\isactrlbsub {\isasymSigma}\isactrlesub \ A\isactrlisub {\isadigit{2}}\ {\isasymrightleftharpoons}\ B\isactrlisub {\isadigit{2}}\ {\isacharcolon}\ type\isactrlisup {\isacharminus}}\\\ \mbox{x\ {\isasymsharp}\ {\isacharparenleft}{\isasymDelta}{\isacharcomma}\ A\isactrlisub {\isadigit{1}}{\isacharcomma}\ B\isactrlisub {\isadigit{1}}{\isacharparenright}}}{\mbox{{\isasymDelta}\ {\isasymturnstile}\isactrlbsub {\isasymSigma}\isactrlesub \ {\isasymPi}x{\isacharcolon}A\isactrlisub {\isadigit{1}}{\isachardot}\ A\isactrlisub {\isadigit{2}}\ {\isasymrightleftharpoons}\ {\isasymPi}x{\isacharcolon}B\isactrlisub {\isadigit{1}}{\isachardot}\ B\isactrlisub {\isadigit{2}}\ {\isacharcolon}\ type\isactrlisup {\isacharminus}}}} 
  \end{tabular} 
  \end{center}
  \caption{Weak algorithmic type equivalence judgment}\labelFig{weak-type-equiv}
  \end{figure}

  A second solution, suggested by Harper\footnote{personal
    communication}, is to observe that the original algorithmic rules
  were unnecessarily general. In the absence of type-level
  $\lambda$-abstraction, the weaker, syntax-directed type equivalence
  rules shown in \refFig{weak-type-equiv} suffice.  We can easily
  prove that these rules are sound with respect to definitional type
  equivalence:
  \begin{lemma}[(Soundness of weak type equivalence)]\labelLem{weak-soundness}\mbox{}\\
    If \isa{{\isasymGamma}\isactrlisup {\isacharminus}\ {\isasymturnstile}\isactrlbsub {\isasymSigma}\isactrlisup {\isacharminus}\isactrlesub \ A\ {\isasymrightleftharpoons}\ B\ {\isacharcolon}\ {\isasymkappa}} and
    \isa{{\isasymGamma}\ {\isasymturnstile}\isactrlisub {\isasymSigma}\ A\ {\isacharcolon}\ K} and
    \isa{{\isasymGamma}\ {\isasymturnstile}\isactrlisub {\isasymSigma}\ B\ {\isacharcolon}\ L}
    then \isa{{\isasymGamma}\ {\isasymturnstile}\isactrlisub {\isasymSigma}\ A\ {\isacharequal}\ B\ {\isacharcolon}\ K},
    \isa{{\isasymGamma}\ {\isasymturnstile}\isactrlisub {\isasymSigma}\ K\ {\isacharequal}\ L\ {\isacharcolon}\ kind},
    \isa{K\isactrlisup {\isacharminus}\ {\isacharequal}\ {\isasymkappa}} and
    \isa{L\isactrlisup {\isacharminus}\ {\isacharequal}\ {\isasymkappa}}.
  \end{lemma}
  \begin{proof}
    Similar to the proof of soundness of algorithmic and structural type
    equivalence from \HP.  Requires soundness of object equivalence
    (\refLem{alg-object-equiv-soundness}).
  \end{proof}

  With this change, we can prove completeness using a slightly modified
  logical relation: the type-level logical relation needs to be
  redefined as
  
  \begin{center}
  \isa{{\isasymDelta}\ {\isasymturnstile}\isactrlbsub {\isasymSigma}\isactrlesub \ A\ {\isacharequal}\ B\ {\isasymin}\ {\isasymlbrakk}{\isasymkappa}{\isasymrbrakk}} $\;=\;$ \isa{{\isasymDelta}\ {\isasymturnstile}\isactrlbsub {\isasymSigma}\isactrlesub \ A\ {\isasymrightleftharpoons}\ B\ {\isacharcolon}\ {\isasymkappa}}\;.
  \end{center}%
\end{isamarkuptext}%
\isamarkuptrue%
\begin{isamarkuptext}%
The first two solutions however establish soundness only for
  variants of the definitions in \HP.  In particular, the first shows
  that the original algorithmic rules are sound with respect to a
  stronger notion of definitional equality, while the second gives a
  correct modified algorithm for the original definitional rules.
  Either solution appears reasonable, but neither tells us whether the
  \emph{original} equivalence algorithm is sound with respect to the
  \emph{original} definitional system in \HP.  We felt it was
  important to determine whether or not a change to the definitions is
  truly necessary to recover soundness.  In the rest of this section
  we show that the original results hold as stated.
  
  Since we already established that weak type equivalence implies
  definitional equivalence (for well-formed terms), it suffices to
  show that the original algorithmic type equivalence judgments imply
  weak type equivalence.  To do so, we need to show that weak type
  equivalence admits extensionality (\refLem{weak-extensionality}
  below).  This is nontrivial: we first need to develop some syntactic
  properties of algorithmic equivalence for objects, in particular
  that if \isa{{\isasymDelta}\ {\isasymturnstile}\isactrlbsub {\isasymSigma}\isactrlesub \ x\ {\isasymLeftrightarrow}\ x\ {\isacharcolon}\ {\isasymtau}} then \isa{{\isacharparenleft}x{\isacharcomma}\ {\isasymtau}{\isacharparenright}\ {\isasymin}\ {\isasymDelta}}.  This seems obvious, but the proof
  is slightly subtle because the algorithmic equivalence judgment is
  type-directed, not syntax-directed.  Indeed, if we try to prove this
  directly by induction, then in the case where $x$ has function type,
  the inductive hypothesis does not apply.  

  Instead, we need to show something more general: for any term
  $M_0$ of the form $x~y_1\cdots y_n$, if $M_0$ is algorithmically
  equivalent to itself then every free variable of $M_0$ appears in
  $\Delta$ with an appropriate type.  We say that such an object
  \isa{M\isactrlisub {\isadigit{0}}} is an \emph{applied variable}, defined formally
  as follows:
  
  \begin{center}
  \isa{M\isactrlisub {\isadigit{0}}} $::=$ \isa{x} $\mid$ \isa{M\isactrlisub {\isadigit{0}}\ x}
  \end{center}  

  \noindent 
  that is, it is a variable applied to a sequence of variables.  
  Clearly, applied variables are weak head normal forms:
  
  \begin{lemma}\labelLem{app-var-normal-forms}
    If \isa{M\isactrlisub {\isadigit{0}}} is an applied variable then \isa{M\isactrlisub {\isadigit{0}}} is in weak head normal form.
  \end{lemma}
  
  \noindent
  We then introduce a weak well-formedness relation
  \mbox{\isa{{\isasymDelta}\ {\isasymturnstile}\isactrlisub {\isadigit{0}}\ M\isactrlisub {\isadigit{0}}\ {\isacharcolon}\ {\isasymtau}}} for applied variables, defined as follows:

  \begin{center}
  \isa{\mbox{}\inferrule{\mbox{{\isacharparenleft}x{\isacharcomma}\ {\isasymtau}{\isacharparenright}\ {\isasymin}\ {\isasymDelta}}}{\mbox{{\isasymDelta}\ {\isasymturnstile}\isactrlisub {\isadigit{0}}\ x\ {\isacharcolon}\ {\isasymtau}}}} \hspace{6mm}
  \isa{\mbox{}\inferrule{\mbox{{\isasymDelta}\ {\isasymturnstile}\isactrlisub {\isadigit{0}}\ M\isactrlisub {\isadigit{0}}\ {\isacharcolon}\ {\isasymtau}\isactrlisub {\isadigit{1}}\ {\isasymrightarrow}\ {\isasymtau}\isactrlisub {\isadigit{2}}}\\\ \mbox{{\isacharparenleft}y{\isacharcomma}\ {\isasymtau}\isactrlisub {\isadigit{1}}{\isacharparenright}\ {\isasymin}\ {\isasymDelta}}}{\mbox{{\isasymDelta}\ {\isasymturnstile}\isactrlisub {\isadigit{0}}\ M\isactrlisub {\isadigit{0}}\ y\ {\isacharcolon}\ {\isasymtau}\isactrlisub {\isadigit{2}}}}}
  \end{center}
  
  \noindent
  It is easy to show that that \isa{{\isasymturnstile}\isactrlisub {\isadigit{0}}} satisfies strengthening:

  \begin{lemma}\labelLem{weak-ok-strengthening}
  \isa{{\normalsize{}If\,}\ \mbox{{\isacharparenleft}y{\isacharcomma}\ {\isasymtau}{\isacharprime}{\isacharparenright}{\isacharcolon}{\isacharcolon}{\isasymDelta}\ {\isasymturnstile}\isactrlisub {\isadigit{0}}\ M\isactrlisub {\isadigit{0}}\ {\isacharcolon}\ {\isasymtau}}\ {\normalsize \,and\,}\ \mbox{y\ {\isasymsharp}\ M\isactrlisub {\isadigit{0}}}\ {\normalsize\linebreak[0] \,then\,\linebreak[0]}\ \mbox{{\isasymDelta}\ {\isasymturnstile}\isactrlisub {\isadigit{0}}\ M\isactrlisub {\isadigit{0}}\ {\isacharcolon}\ {\isasymtau}{\isachardot}}}
  \end{lemma}

  Furthermore, if an applied variable is algorithmically or
  structurally equivalent to itself, then it is weakly well-formed:

  \begin{lemma}
  Suppose \isa{M\isactrlisub {\isadigit{0}}} is an applied variable and  \mbox{\isa{{\isasymturnstile}\ {\isasymDelta}\ sctx}}.
  \begin{compactenum}
  \item If \isa{{\isasymDelta}\ {\isasymturnstile}\isactrlbsub {\isasymSigma}\isactrlesub \ M\isactrlisub {\isadigit{0}}\ {\isasymLeftrightarrow}\ M\isactrlisub {\isadigit{0}}\ {\isacharcolon}\ {\isasymtau}} 
        then \isa{{\isasymDelta}\ {\isasymturnstile}\isactrlisub {\isadigit{0}}\ M\isactrlisub {\isadigit{0}}\ {\isacharcolon}\ {\isasymtau}}.
  \item If \isa{{\isasymDelta}\ {\isasymturnstile}\isactrlbsub {\isasymSigma}\isactrlesub \ M\isactrlisub {\isadigit{0}}\ {\isasymleftrightarrow}\ M\isactrlisub {\isadigit{0}}\ {\isacharcolon}\ {\isasymtau}} 
        then \isa{{\isasymDelta}\ {\isasymturnstile}\isactrlisub {\isadigit{0}}\ M\isactrlisub {\isadigit{0}}\ {\isacharcolon}\ {\isasymtau}}.
  \end{compactenum}
  \end{lemma}

  \begin{proof}
    Induction on derivations.  \refLem{app-var-normal-forms} is needed
    to show that the cases involving weak head reduction are vacuous.
    The only other interesting case is the case for an extensionality rule
    
    \begin{center}
    \isa{\mbox{}\inferrule{\mbox{{\isacharparenleft}x{\isacharcomma}\ {\isasymtau}\isactrlisub {\isadigit{1}}{\isacharparenright}{\isacharcolon}{\isacharcolon}{\isasymDelta}\ {\isasymturnstile}\isactrlbsub {\isasymSigma}\isactrlesub \ M\isactrlisub {\isadigit{0}}\ x\ {\isasymLeftrightarrow}\ M\isactrlisub {\isadigit{0}}\ x\ {\isacharcolon}\ {\isasymtau}\isactrlisub {\isadigit{2}}}\\\ \mbox{x\ {\isasymsharp}\ {\isacharparenleft}{\isasymDelta}{\isacharcomma}\ M\isactrlisub {\isadigit{0}}{\isacharcomma}\ M\isactrlisub {\isadigit{0}}{\isacharparenright}}}{\mbox{{\isasymDelta}\ {\isasymturnstile}\isactrlbsub {\isasymSigma}\isactrlesub \ M\isactrlisub {\isadigit{0}}\ {\isasymLeftrightarrow}\ M\isactrlisub {\isadigit{0}}\ {\isacharcolon}\ {\isasymtau}\isactrlisub {\isadigit{1}}\ {\isasymrightarrow}\ {\isasymtau}\isactrlisub {\isadigit{2}}}}}
    \end{center}     

    \noindent
    By induction, we have that \isa{{\isacharparenleft}x{\isacharcomma}\ {\isasymtau}\isactrlisub {\isadigit{1}}{\isacharparenright}{\isacharcolon}{\isacharcolon}{\isasymDelta}\ {\isasymturnstile}\isactrlisub {\isadigit{0}}\ M\isactrlisub {\isadigit{0}}\ x\ {\isacharcolon}\ {\isasymtau}\isactrlisub {\isadigit{2}}}.  By inversion, we can show that \isa{{\isacharparenleft}x{\isacharcomma}\ {\isasymtau}\isactrlisub {\isadigit{1}}{\isacharparenright}{\isacharcolon}{\isacharcolon}{\isasymDelta}\ {\isasymturnstile}\isactrlisub {\isadigit{0}}\ M\isactrlisub {\isadigit{0}}\ {\isacharcolon}\ {\isasymtau}\isactrlisub {\isadigit{1}}\ {\isasymrightarrow}\ {\isasymtau}\isactrlisub {\isadigit{2}}}.  To complete the
    proof, we use \refLem{weak-ok-strengthening} to show that \isa{{\isasymDelta}\ {\isasymturnstile}\isactrlisub {\isadigit{0}}\ M\isactrlisub {\isadigit{0}}\ {\isacharcolon}\ {\isasymtau}\isactrlisub {\isadigit{1}}\ {\isasymrightarrow}\ {\isasymtau}\isactrlisub {\isadigit{2}}}, which follows since \isa{x\ {\isasymsharp}\ M\isactrlisub {\isadigit{0}}}.
  \end{proof}

  \begin{corollary}\labelCor{var-alg-eq-unique}
    \isa{{\normalsize{}If\,}\ {\isasymDelta}\ {\isasymturnstile}\isactrlbsub {\isasymSigma}\isactrlesub \ x\ {\isasymLeftrightarrow}\ x\ {\isacharcolon}\ {\isasymtau}\ {\normalsize \,and\,}\ {\isasymturnstile}\ {\isasymDelta}\ sctx\ {\normalsize \,then\,}\ {\isacharparenleft}x{\isacharcomma}\ {\isasymtau}{\isacharparenright}\ {\isasymin}\ {\isasymDelta}{\isachardot}}
  \end{corollary}

  \noindent
  We also need to establish strengthening for weak algorithmic type
  equivalence:

  \begin{lemma}[(Strengthening of weak type equivalence)]
    \labelLem{weak-strengthening}\mbox{}\\  
  If \isa{{\isasymDelta}{\isacharprime}\ {\isacharat}\ {\isacharbrackleft}{\isacharparenleft}x{\isacharcomma}\ {\isasymtau}{\isacharparenright}{\isacharbrackright}\ {\isacharat}\ {\isasymDelta}\ {\isasymturnstile}\isactrlbsub {\isasymSigma}\isactrlesub \ A\ {\isasymrightleftharpoons}\ B\ {\isacharcolon}\ {\isasymkappa}} and
  \isa{x\ {\isasymsharp}\ {\isacharparenleft}{\isasymDelta}{\isacharprime}{\isacharcomma}\ A{\isacharcomma}\ B{\isacharparenright}} then\\
  \isa{{\isasymDelta}{\isacharprime}\ {\isacharat}\ {\isasymDelta}\ {\isasymturnstile}\isactrlbsub {\isasymSigma}\isactrlesub \ A\ {\isasymrightleftharpoons}\ B\ {\isacharcolon}\ {\isasymkappa}}.
  \end{lemma}

  \begin{proof}
    Straightforward induction on derivations.  Note that we need
    \refLem{algorithmic-strengthening} here in the case for
    structural equivalence of type applications.
  \end{proof}

  \noindent We now establish the admissibility of extensionality for
  weak type equivalence: 

  \begin{lemma}[(Extensionality of weak type equivalence)]
    \labelLem{weak-extensionality}\mbox{}\\    
  If \isa{{\isacharparenleft}x{\isacharcomma}\ {\isasymtau}{\isacharparenright}{\isacharcolon}{\isacharcolon}{\isasymDelta}\ {\isasymturnstile}\isactrlbsub {\isasymSigma}\isactrlesub \ A\ x\ {\isasymrightleftharpoons}\ B\ x\ {\isacharcolon}\ {\isasymkappa}} and
  \isa{x\ {\isasymsharp}\ {\isacharparenleft}{\isasymDelta}{\isacharcomma}\ A{\isacharcomma}\ B{\isacharparenright}} and
  \isa{{\isasymturnstile}\ {\isasymDelta}\ sctx} then\\
  \isa{{\isasymDelta}\ {\isasymturnstile}\isactrlbsub {\isasymSigma}\isactrlesub \ A\ {\isasymrightleftharpoons}\ B\ {\isacharcolon}\ {\isasymtau}\ {\isasymrightarrow}\ {\isasymkappa}}.
  \end{lemma} 

  \begin{proof}
    By inversion, we have subderivations \isa{{\isacharparenleft}x{\isacharcomma}\ {\isasymtau}{\isacharparenright}{\isacharcolon}{\isacharcolon}{\isasymDelta}\ {\isasymturnstile}\isactrlbsub {\isasymSigma}\isactrlesub \ A\ {\isasymrightleftharpoons}\ B\ {\isacharcolon}\ {\isasymtau}{\isacharprime}\ {\isasymrightarrow}\ {\isasymkappa}} and
    \isa{{\isacharparenleft}x{\isacharcomma}\ {\isasymtau}{\isacharparenright}{\isacharcolon}{\isacharcolon}{\isasymDelta}\ {\isasymturnstile}\isactrlbsub {\isasymSigma}\isactrlesub \ x\ {\isasymLeftrightarrow}\ x\ {\isacharcolon}\ {\isasymtau}{\isacharprime}} for some \isa{{\isasymtau}{\isacharprime}}.   Using \refCor{var-alg-eq-unique} on the second
    subderivation we have that \isa{{\isacharparenleft}x{\isacharcomma}\ {\isasymtau}{\isacharprime}{\isacharparenright}\ {\isasymin}\ {\isacharparenleft}x{\isacharcomma}\ {\isasymtau}{\isacharparenright}{\isacharcolon}{\isacharcolon}{\isasymDelta}} and using the validity of \isa{{\isacharparenleft}x{\isacharcomma}\ {\isasymtau}{\isacharparenright}{\isacharcolon}{\isacharcolon}{\isasymDelta}} we know that \isa{{\isasymtau}\ {\isacharequal}\ {\isasymtau}{\isacharprime}}.
    Hence, \isa{{\isacharparenleft}x{\isacharcomma}\ {\isasymtau}{\isacharparenright}{\isacharcolon}{\isacharcolon}{\isasymDelta}\ {\isasymturnstile}\isactrlbsub {\isasymSigma}\isactrlesub \ A\ {\isasymrightleftharpoons}\ B\ {\isacharcolon}\ {\isasymtau}\ {\isasymrightarrow}\ {\isasymkappa}}.
    Using \refLem{weak-strengthening} we conclude \isa{{\isasymDelta}\ {\isasymturnstile}\isactrlbsub {\isasymSigma}\isactrlesub \ A\ {\isasymrightleftharpoons}\ B\ {\isacharcolon}\ {\isasymtau}\ {\isasymrightarrow}\ {\isasymkappa}}.
  \end{proof}

  \begin{lemma}\labelLem{alg-str-implies-weak}
    Suppose \isa{{\isasymturnstile}\ {\isasymDelta}\ sctx}.  Then:
    \begin{compactenum}  
    \item If \isa{{\isasymDelta}\ {\isasymturnstile}\isactrlbsub {\isasymSigma}\isactrlesub \ A\ {\isasymLeftrightarrow}\ B\ {\isacharcolon}\ {\isasymkappa}} 
        then \isa{{\isasymDelta}\ {\isasymturnstile}\isactrlbsub {\isasymSigma}\isactrlesub \ A\ {\isasymrightleftharpoons}\ B\ {\isacharcolon}\ {\isasymkappa}}.
    \item If \isa{{\isasymDelta}\ {\isasymturnstile}\isactrlbsub {\isasymSigma}\isactrlesub \ A\ {\isasymleftrightarrow}\ B\ {\isacharcolon}\ {\isasymkappa}} 
        then \isa{{\isasymDelta}\ {\isasymturnstile}\isactrlbsub {\isasymSigma}\isactrlesub \ A\ {\isasymrightleftharpoons}\ B\ {\isacharcolon}\ {\isasymkappa}}.  
    \end{compactenum}
  \end{lemma}
   
  \begin{proof}
    By induction on the structure of derivations.  The case for the
    algorithmic type extensionality rule requires
    Lem.~\ref{lem:weak-extensionality}.
  \end{proof}

  \noindent
  The proof of \refThm{soundness} is completed as follows.

  \begin{lemma}[(Soundness of algorithmic type equivalence)]
    \labelLem{alg-type-equiv-soundness}
    ~
    \begin{compactenum}
    \item If \isa{{\isasymGamma}\isactrlisup {\isacharminus}\ {\isasymturnstile}\isactrlbsub {\isasymSigma}\isactrlisup {\isacharminus}\isactrlesub \ A\ {\isasymLeftrightarrow}\ B\ {\isacharcolon}\ K\isactrlisup {\isacharminus}} and
             \isa{{\isasymGamma}\ {\isasymturnstile}\isactrlisub {\isasymSigma}\ A\ {\isacharcolon}\ K} and
             \mbox{\isa{{\isasymGamma}\ {\isasymturnstile}\isactrlisub {\isasymSigma}\ B\ {\isacharcolon}\ K}}
          then \\\isa{{\isasymGamma}\ {\isasymturnstile}\isactrlisub {\isasymSigma}\ A\ {\isacharequal}\ B\ {\isacharcolon}\ K}.
    \item If \isa{{\isasymGamma}\isactrlisup {\isacharminus}\ {\isasymturnstile}\isactrlbsub {\isasymSigma}\isactrlisup {\isacharminus}\isactrlesub \ A\ {\isasymleftrightarrow}\ B\ {\isacharcolon}\ {\isasymkappa}} and
             \isa{{\isasymGamma}\ {\isasymturnstile}\isactrlisub {\isasymSigma}\ A\ {\isacharcolon}\ K} and
             \mbox{\isa{{\isasymGamma}\ {\isasymturnstile}\isactrlisub {\isasymSigma}\ B\ {\isacharcolon}\ L}}
          then \\
          \isa{{\isasymGamma}\ {\isasymturnstile}\isactrlisub {\isasymSigma}\ A\ {\isacharequal}\ B\ {\isacharcolon}\ K},
          \isa{{\isasymGamma}\ {\isasymturnstile}\isactrlisub {\isasymSigma}\ K\ {\isacharequal}\ L\ {\isacharcolon}\ kind},
          \isa{K\isactrlisup {\isacharminus}\ {\isacharequal}\ {\isasymkappa}} and
          \isa{L\isactrlisup {\isacharminus}\ {\isacharequal}\ {\isasymkappa}}.
      \end{compactenum}
  \end{lemma}

  \begin{proof}
  Immediate using Lem.~\ref{lem:alg-str-implies-weak} and
  \ref{lem:weak-soundness}.
  \end{proof}

   \begin{lemma}[(Soundness of algorithmic kind equivalence)]
     \labelLem{alg-kind-equiv-soundness}\mbox{}

   \isa{{\normalsize{}If\,}\ {\isasymGamma}\isactrlisup {\isacharminus}\ {\isasymturnstile}\isactrlbsub {\isasymSigma}\isactrlisup {\isacharminus}\isactrlesub \ K\ {\isasymLeftrightarrow}\ L\ {\isacharcolon}\ kind\isactrlisup {\isacharminus}\ {\normalsize \,and\,}\ {\isasymGamma}\ {\isasymturnstile}\isactrlisub {\isasymSigma}\ K\ {\isacharcolon}\ kind\ {\normalsize \,and\,}\ {\isasymGamma}\ {\isasymturnstile}\isactrlisub {\isasymSigma}\ L\ {\isacharcolon}\ kind\ {\normalsize \,then\,}\ {\isasymGamma}\ {\isasymturnstile}\isactrlisub {\isasymSigma}\ K\ {\isacharequal}\ L\ {\isacharcolon}\ kind{\isachardot}}
   \end{lemma}
  
  \begin{proof}
    As in \HP, using \refLem{alg-type-equiv-soundness} as necessary.
  \end{proof}

  \noindent
   \refThm{soundness} follows immediately from
   Lem.~\ref{lem:alg-object-equiv-soundness},
   \ref{lem:alg-type-equiv-soundness} and~\ref{lem:alg-kind-equiv-soundness}.%
\end{isamarkuptext}%
\isamarkuptrue%
\isamarkupsubsection{Algorithmic typechecking%
}
\isamarkuptrue%
\begin{isamarkuptext}%
\labelSec{typechecking}%
\end{isamarkuptext}%
\isamarkuptrue%
\begin{isamarkuptext}%
\noindent
  After the soundness and completeness proof, \HP introduces an
  algorithmic version of the typechecking judgment, proves additional
  syntactic properties of definitional equivalence, sketches proofs of
  decidability, and discusses quasicanonical forms and adequacy of LF
  encodings of object languages. We formalized many of these results
  and we will discuss them in the next few sections.

  %%%%%%%%%%%%%%%%%%%%%%%%%%%%%%%%%%%%%%%%%%%%%%%%%%%%%%%%%%%%%%%%%%
  \begin{figure}[tb]
    %\small

\fbox{\isa{{\isasymturnstile}\ {\isasymSigma}\ {\isasymRightarrow}\ sig}}
\begin{center}
  \isa{\mbox{}\inferrule{\mbox{}}{\mbox{{\isasymturnstile}\ {\isacharbrackleft}{\isacharbrackright}\ {\isasymRightarrow}\ sig}}}
  \qquad 
  \isa{\mbox{}\inferrule{\mbox{{\isasymturnstile}\ {\isasymSigma}\ {\isasymRightarrow}\ sig}\\\ \mbox{{\isacharbrackleft}{\isacharbrackright}\ {\isasymturnstile}\isactrlisub {\isasymSigma}\ A\ {\isasymRightarrow}\ type}\\\ \mbox{c\ {\isasymsharp}\ {\isasymSigma}}}{\mbox{{\isasymturnstile}\ {\isacharparenleft}c{\isacharcomma}\ A{\isacharparenright}{\isacharcolon}{\isacharcolon}{\isasymSigma}\ {\isasymRightarrow}\ sig}}}
  \smallskip\\
  \isa{\mbox{}\inferrule{\mbox{{\isasymturnstile}\ {\isasymSigma}\ {\isasymRightarrow}\ sig}\\\ \mbox{{\isacharbrackleft}{\isacharbrackright}\ {\isasymturnstile}\isactrlisub {\isasymSigma}\ K\ {\isasymRightarrow}\ kind}\\\ \mbox{a\ {\isasymsharp}\ {\isasymSigma}}}{\mbox{{\isasymturnstile}\ {\isacharparenleft}a{\isacharcomma}\ K{\isacharparenright}{\isacharcolon}{\isacharcolon}{\isasymSigma}\ {\isasymRightarrow}\ sig}}}
\end{center}

 \fbox{\isa{{\isasymturnstile}\isactrlisub {\isasymSigma}\ {\isasymGamma}\ {\isasymRightarrow}\ ctx}}

 \begin{center}
   \isa{\mbox{}\inferrule{\mbox{{\isasymturnstile}\ {\isasymSigma}\ {\isasymRightarrow}\ sig}}{\mbox{{\isasymturnstile}\isactrlisub {\isasymSigma}\ {\isacharbrackleft}{\isacharbrackright}\ {\isasymRightarrow}\ ctx}}} \qquad \isa{\mbox{}\inferrule{\mbox{{\isasymturnstile}\isactrlisub {\isasymSigma}\ {\isasymGamma}\ {\isasymRightarrow}\ ctx}\\\ \mbox{{\isasymGamma}\ {\isasymturnstile}\isactrlisub {\isasymSigma}\ A\ {\isasymRightarrow}\ type}\\\ \mbox{x\ {\isasymsharp}\ {\isasymGamma}}}{\mbox{{\isasymturnstile}\isactrlisub {\isasymSigma}\ {\isacharparenleft}x{\isacharcomma}\ A{\isacharparenright}{\isacharcolon}{\isacharcolon}{\isasymGamma}\ {\isasymRightarrow}\ ctx}}}

 \end{center} 

\fbox{\isa{{\isasymGamma}\ {\isasymturnstile}\isactrlisub {\isasymSigma}\ M\ {\isasymRightarrow}\ A}}
  
\begin{center}
  \isa{\mbox{}\inferrule{\mbox{{\isasymturnstile}\isactrlisub {\isasymSigma}\ {\isasymGamma}\ {\isasymRightarrow}\ ctx}\\\ \mbox{{\isacharparenleft}x{\isacharcomma}\ A{\isacharparenright}\ {\isasymin}\ {\isasymGamma}}}{\mbox{{\isasymGamma}\ {\isasymturnstile}\isactrlisub {\isasymSigma}\ x\ {\isasymRightarrow}\ A}}} 
  \smallskip\\ 
  \isa{\mbox{}\inferrule{\mbox{{\isasymturnstile}\isactrlisub {\isasymSigma}\ {\isasymGamma}\ {\isasymRightarrow}\ ctx}\\\ \mbox{{\isacharparenleft}c{\isacharcomma}\ A{\isacharparenright}\ {\isasymin}\ {\isasymSigma}}}{\mbox{{\isasymGamma}\ {\isasymturnstile}\isactrlisub {\isasymSigma}\ c\ {\isasymRightarrow}\ A}}} 
  \smallskip\\
  
  \isa{\mbox{}\inferrule{\mbox{{\isasymGamma}\ {\isasymturnstile}\isactrlisub {\isasymSigma}\ M\isactrlisub {\isadigit{1}}\ {\isasymRightarrow}\ {\isasymPi}x{\isacharcolon}A\isactrlisub {\isadigit{2}}{\isacharprime}{\isachardot}\ A\isactrlisub {\isadigit{1}}}\\\ \mbox{{\isasymGamma}\ {\isasymturnstile}\isactrlisub {\isasymSigma}\ M\isactrlisub {\isadigit{2}}\ {\isasymRightarrow}\ A\isactrlisub {\isadigit{2}}}\\\ \mbox{{\isasymGamma}\isactrlisup {\isacharminus}\ {\isasymturnstile}\isactrlbsub {\isasymSigma}\isactrlisup {\isacharminus}\isactrlesub \ A\isactrlisub {\isadigit{2}}\ {\isasymLeftrightarrow}\ A\isactrlisub {\isadigit{2}}{\isacharprime}\ {\isacharcolon}\ type\isactrlisup {\isacharminus}}\\\ \mbox{x\ {\isasymsharp}\ {\isasymGamma}}}{\mbox{{\isasymGamma}\ {\isasymturnstile}\isactrlisub {\isasymSigma}\ M\isactrlisub {\isadigit{1}}\ M\isactrlisub {\isadigit{2}}\ {\isasymRightarrow}\ A\isactrlisub {\isadigit{1}}{\isacharbrackleft}x{\isacharcolon}{\isacharequal}M\isactrlisub {\isadigit{2}}{\isacharbrackright}}}}
  \smallskip\\
  \isa{\mbox{}\inferrule{\mbox{{\isasymGamma}\ {\isasymturnstile}\isactrlisub {\isasymSigma}\ A\isactrlisub {\isadigit{1}}\ {\isasymRightarrow}\ type}\\\ \mbox{{\isacharparenleft}x{\isacharcomma}\ A\isactrlisub {\isadigit{1}}{\isacharparenright}{\isacharcolon}{\isacharcolon}{\isasymGamma}\ {\isasymturnstile}\isactrlisub {\isasymSigma}\ M\isactrlisub {\isadigit{2}}\ {\isasymRightarrow}\ A\isactrlisub {\isadigit{2}}}\\\ \mbox{x\ {\isasymsharp}\ {\isacharparenleft}{\isasymGamma}{\isacharcomma}\ A\isactrlisub {\isadigit{1}}{\isacharparenright}}}{\mbox{{\isasymGamma}\ {\isasymturnstile}\isactrlisub {\isasymSigma}\ {\isasymlambda}x{\isacharcolon}A\isactrlisub {\isadigit{1}}{\isachardot}\ M\isactrlisub {\isadigit{2}}\ {\isasymRightarrow}\ {\isasymPi}x{\isacharcolon}A\isactrlisub {\isadigit{1}}{\isachardot}\ A\isactrlisub {\isadigit{2}}}}}
\end{center}

\fbox{\isa{{\isasymGamma}\ {\isasymturnstile}\isactrlisub {\isasymSigma}\ A\ {\isasymRightarrow}\ K}}
  
\begin{center}
  \isa{\mbox{}\inferrule{\mbox{{\isasymturnstile}\isactrlisub {\isasymSigma}\ {\isasymGamma}\ {\isasymRightarrow}\ ctx}\\\ \mbox{{\isacharparenleft}a{\isacharcomma}\ K{\isacharparenright}\ {\isasymin}\ {\isasymSigma}}}{\mbox{{\isasymGamma}\ {\isasymturnstile}\isactrlisub {\isasymSigma}\ a\ {\isasymRightarrow}\ K}}} 
  \smallskip\\ 
  \isa{\mbox{}\inferrule{\mbox{{\isasymGamma}\ {\isasymturnstile}\isactrlisub {\isasymSigma}\ A\ {\isasymRightarrow}\ {\isasymPi}x{\isacharcolon}A\isactrlisub {\isadigit{2}}{\isacharprime}{\isachardot}\ K\isactrlisub {\isadigit{1}}}\\\ \mbox{{\isasymGamma}\ {\isasymturnstile}\isactrlisub {\isasymSigma}\ M\ {\isasymRightarrow}\ A\isactrlisub {\isadigit{2}}}\\\ \mbox{{\isasymGamma}\isactrlisup {\isacharminus}\ {\isasymturnstile}\isactrlbsub {\isasymSigma}\isactrlisup {\isacharminus}\isactrlesub \ A\isactrlisub {\isadigit{2}}\ {\isasymLeftrightarrow}\ A\isactrlisub {\isadigit{2}}{\isacharprime}\ {\isacharcolon}\ type\isactrlisup {\isacharminus}}\\\ \mbox{x\ {\isasymsharp}\ {\isasymGamma}}}{\mbox{{\isasymGamma}\ {\isasymturnstile}\isactrlisub {\isasymSigma}\ A\ M\ {\isasymRightarrow}\ K\isactrlisub {\isadigit{1}}{\isacharbrackleft}x{\isacharcolon}{\isacharequal}M{\isacharbrackright}}}}
   
  \isa{\mbox{}\inferrule{\mbox{{\isasymGamma}\ {\isasymturnstile}\isactrlisub {\isasymSigma}\ A\isactrlisub {\isadigit{1}}\ {\isasymRightarrow}\ type}\\\ \mbox{{\isacharparenleft}x{\isacharcomma}\ A\isactrlisub {\isadigit{1}}{\isacharparenright}{\isacharcolon}{\isacharcolon}{\isasymGamma}\ {\isasymturnstile}\isactrlisub {\isasymSigma}\ A\isactrlisub {\isadigit{2}}\ {\isasymRightarrow}\ type}\\\ \mbox{x\ {\isasymsharp}\ {\isacharparenleft}{\isasymGamma}{\isacharcomma}\ A\isactrlisub {\isadigit{1}}{\isacharparenright}}}{\mbox{{\isasymGamma}\ {\isasymturnstile}\isactrlisub {\isasymSigma}\ {\isasymPi}x{\isacharcolon}A\isactrlisub {\isadigit{1}}{\isachardot}\ A\isactrlisub {\isadigit{2}}\ {\isasymRightarrow}\ type}}}
\end{center}

\fbox{\isa{{\isasymGamma}\ {\isasymturnstile}\isactrlisub {\isasymSigma}\ K\ {\isasymRightarrow}\ kind}}

\begin{center}
  \isa{\mbox{}\inferrule{\mbox{{\isasymturnstile}\isactrlisub {\isasymSigma}\ {\isasymGamma}\ {\isasymRightarrow}\ ctx}}{\mbox{{\isasymGamma}\ {\isasymturnstile}\isactrlisub {\isasymSigma}\ type\ {\isasymRightarrow}\ kind}}} \quad \isa{\mbox{}\inferrule{\mbox{{\isasymGamma}\ {\isasymturnstile}\isactrlisub {\isasymSigma}\ A\ {\isasymRightarrow}\ type}\\\ \mbox{{\isacharparenleft}x{\isacharcomma}\ A{\isacharparenright}{\isacharcolon}{\isacharcolon}{\isasymGamma}\ {\isasymturnstile}\isactrlisub {\isasymSigma}\ K\ {\isasymRightarrow}\ kind}\\\ \mbox{x\ {\isasymsharp}\ {\isacharparenleft}{\isasymGamma}{\isacharcomma}\ A{\isacharparenright}}}{\mbox{{\isasymGamma}\ {\isasymturnstile}\isactrlisub {\isasymSigma}\ {\isasymPi}x{\isacharcolon}A{\isachardot}\ K\ {\isasymRightarrow}\ kind}}}
\end{center}
  \caption{Algorithmic typechecking rules}\labelFig{lf-alg-tc}
  \end{figure}
  %%%%%%%%%%%%%%%%%%%%%%%%%%%%%%%%%%%%%%%%%%%%%%%%%%%%%%%%%%%%%%%%%%

  The typechecking algorithm in \HP traverses terms, types and kinds
  in a syntax-directed manner, using the algorithmic equivalence
  judgment in certain places.  The definition of algorithmic
  typechecking in \HP omitted explicit definitions of algorithmic
  signature and context validity.  In our formalization, we added
  these (obvious) rules, as shown in \refFig{lf-alg-tc}.  The
  remaining rules are the same as in \HP except for a trivial
  typographical error in the rule for type constants.  Proving the
  soundness and completeness of algorithmic typechecking is a (mostly)
  straightforward exercise using soundness and completeness of
  algorithmic equivalence and various syntactic properties:

  \begin{theorem}[(Soundness of algorithmic typechecking)]\labelThm{algorithmic-tc-soundness}
  ~
  \begin{compactenum}
  \item \isa{{\normalsize{}If\,}\ {\isasymturnstile}\ {\isasymSigma}\ {\isasymRightarrow}\ sig\ {\normalsize \,then\,}\ {\isasymturnstile}\ {\isasymSigma}\ sig{\isachardot}}
  \item \isa{{\normalsize{}If\,}\ {\isasymturnstile}\isactrlisub {\isasymSigma}\ {\isasymGamma}\ {\isasymRightarrow}\ ctx\ {\normalsize \,then\,}\ {\isasymturnstile}\isactrlisub {\isasymSigma}\ {\isasymGamma}\ ctx{\isachardot}}
  \item \isa{{\normalsize{}If\,}\ {\isasymGamma}\ {\isasymturnstile}\isactrlisub {\isasymSigma}\ M\ {\isasymRightarrow}\ A\ {\normalsize \,then\,}\ {\isasymGamma}\ {\isasymturnstile}\isactrlisub {\isasymSigma}\ M\ {\isacharcolon}\ A{\isachardot}}
  \item \isa{{\normalsize{}If\,}\ {\isasymGamma}\ {\isasymturnstile}\isactrlisub {\isasymSigma}\ A\ {\isasymRightarrow}\ K\ {\normalsize \,then\,}\ {\isasymGamma}\ {\isasymturnstile}\isactrlisub {\isasymSigma}\ A\ {\isacharcolon}\ K{\isachardot}}
  \item \isa{{\normalsize{}If\,}\ {\isasymGamma}\ {\isasymturnstile}\isactrlisub {\isasymSigma}\ K\ {\isasymRightarrow}\ kind\ {\normalsize \,then\,}\ {\isasymGamma}\ {\isasymturnstile}\isactrlisub {\isasymSigma}\ K\ {\isacharcolon}\ kind{\isachardot}}
  \end{compactenum}
  \end{theorem} 
  \begin{theorem}[(Completeness of algorithmic typechecking)]\labelThm{algorithmic-tc-completeness}
  ~
  \begin{compactenum}
  \item \isa{{\normalsize{}If\,}\ {\isasymturnstile}\ {\isasymSigma}\ sig\ {\normalsize \,then\,}\ {\isasymturnstile}\ {\isasymSigma}\ {\isasymRightarrow}\ sig{\isachardot}}
  \item \isa{{\normalsize{}If\,}\ {\isasymturnstile}\isactrlisub {\isasymSigma}\ {\isasymGamma}\ ctx\ {\normalsize \,then\,}\ {\isasymturnstile}\isactrlisub {\isasymSigma}\ {\isasymGamma}\ {\isasymRightarrow}\ ctx{\isachardot}}
  \item \isa{{\normalsize{}If\,}\ {\isasymGamma}\ {\isasymturnstile}\isactrlisub {\isasymSigma}\ M\ {\isacharcolon}\ A\ {\normalsize \,then\,}\ {\isasymexists}A{\isacharprime}{\isachardot}\ {\isasymGamma}\ {\isasymturnstile}\isactrlisub {\isasymSigma}\ M\ {\isasymRightarrow}\ A{\isacharprime}\ \textrm{and\linebreak[1]}\ {\isasymGamma}\ {\isasymturnstile}\isactrlisub {\isasymSigma}\ A\ {\isacharequal}\ A{\isacharprime}\ {\isacharcolon}\ type{\isachardot}}
  \item \isa{{\normalsize{}If\,}\ {\isasymGamma}\ {\isasymturnstile}\isactrlisub {\isasymSigma}\ A\ {\isacharcolon}\ K\ {\normalsize \,then\,}\ {\isasymexists}K{\isacharprime}{\isachardot}\ {\isasymGamma}\ {\isasymturnstile}\isactrlisub {\isasymSigma}\ A\ {\isasymRightarrow}\ K{\isacharprime}\ \textrm{and\linebreak[1]}\ {\isasymGamma}\ {\isasymturnstile}\isactrlisub {\isasymSigma}\ K\ {\isacharequal}\ K{\isacharprime}\ {\isacharcolon}\ kind{\isachardot}}
  \item \isa{{\normalsize{}If\,}\ {\isasymGamma}\ {\isasymturnstile}\isactrlisub {\isasymSigma}\ K\ {\isacharcolon}\ kind\ {\normalsize \,then\,}\ {\isasymGamma}\ {\isasymturnstile}\isactrlisub {\isasymSigma}\ K\ {\isasymRightarrow}\ kind{\isachardot}}
  \end{compactenum}
  \end{theorem}%
\end{isamarkuptext}%
\isamarkuptrue%
\isamarkupsubsection{Strengthening and strong extensionality%
}
\isamarkuptrue%
\begin{isamarkuptext}%
\labelSec{strengthening}%
\end{isamarkuptext}%
\isamarkuptrue%
\begin{isamarkuptext}%
The strengthening property states that all of the definitional
  judgments are preserved by removing an unused variable from the
  context.  We already established strengthening for the algorithmic
  equivalence judgments (\refLem{algorithmic-strengthening}).  In
  order to establish strengthening for the algorithmic typechecking
  judgments, we need a stronger freshness lemma for algorithmic
  typechecking, which was not discussed in \HP:

  \begin{lemma}[(Strong algorithmic freshness)]\labelLem{strong-alg-freshness}
  Let \mbox{\isa{{\isasymGamma}\ {\isacharequal}\ {\isasymGamma}\isactrlisub {\isadigit{1}}\ {\isacharat}\ {\isacharbrackleft}{\isacharparenleft}x{\isacharcomma}\ B{\isacharparenright}{\isacharbrackright}\ {\isacharat}\ {\isasymGamma}\isactrlisub {\isadigit{2}}}}.
  \begin{compactenum}
  \item If \isa{{\isasymGamma}\ {\isasymturnstile}\isactrlisub {\isasymSigma}\ M\ {\isasymRightarrow}\ A} and \isa{x\ {\isasymsharp}\ {\isacharparenleft}{\isasymGamma}\isactrlisub {\isadigit{1}}{\isacharcomma}\ M{\isacharparenright}} then 
   \isa{x\ {\isasymsharp}\ A}.
  \item If \isa{{\isasymGamma}\ {\isasymturnstile}\isactrlisub {\isasymSigma}\ A\ {\isasymRightarrow}\ K} and \isa{x\ {\isasymsharp}\ {\isacharparenleft}{\isasymGamma}\isactrlisub {\isadigit{1}}{\isacharcomma}\ A{\isacharparenright}} then 
   \isa{x\ {\isasymsharp}\ K}.
  \end{compactenum}
  \end{lemma}

  We can now prove strengthening for algorithmic typechecking by
  induction on derivations:

  \begin{theorem}[(Strengthening of algorithmic typechecking)]
\mbox{}\\
    Let \mbox{\isa{{\isasymGamma}\ {\isacharequal}\ {\isasymGamma}\isactrlisub {\isadigit{1}}\ {\isacharat}\ {\isacharbrackleft}{\isacharparenleft}x{\isacharcomma}\ B{\isacharparenright}{\isacharbrackright}\ {\isacharat}\ {\isasymGamma}\isactrlisub {\isadigit{2}}}}.
  \begin{compactenum}
  \item If \isa{{\isasymturnstile}\isactrlisub {\isasymSigma}\ {\isasymGamma}\ {\isasymRightarrow}\ ctx} and \isa{x\ {\isasymsharp}\ {\isasymGamma}\isactrlisub {\isadigit{1}}} then \isa{{\isasymturnstile}\isactrlisub {\isasymSigma}\ {\isasymGamma}\isactrlisub {\isadigit{1}}\ {\isacharat}\ {\isasymGamma}\isactrlisub {\isadigit{2}}\ {\isasymRightarrow}\ ctx}.
  \item If \isa{{\isasymGamma}\ {\isasymturnstile}\isactrlisub {\isasymSigma}\ K\ {\isasymRightarrow}\ kind} and \isa{x\ {\isasymsharp}\ {\isacharparenleft}{\isasymGamma}\isactrlisub {\isadigit{1}}{\isacharcomma}\ K{\isacharparenright}}
  then \isa{{\isasymGamma}\isactrlisub {\isadigit{1}}\ {\isacharat}\ {\isasymGamma}\isactrlisub {\isadigit{2}}\ {\isasymturnstile}\isactrlisub {\isasymSigma}\ K\ {\isasymRightarrow}\ kind}.
  \item If \isa{{\isasymGamma}\ {\isasymturnstile}\isactrlisub {\isasymSigma}\ A\ {\isasymRightarrow}\ K} and \isa{x\ {\isasymsharp}\ {\isacharparenleft}{\isasymGamma}\isactrlisub {\isadigit{1}}{\isacharcomma}\ A{\isacharparenright}}
  then \isa{{\isasymGamma}\isactrlisub {\isadigit{1}}\ {\isacharat}\ {\isasymGamma}\isactrlisub {\isadigit{2}}\ {\isasymturnstile}\isactrlisub {\isasymSigma}\ A\ {\isasymRightarrow}\ K}.
  \item If \isa{{\isasymGamma}\ {\isasymturnstile}\isactrlisub {\isasymSigma}\ M\ {\isasymRightarrow}\ A} and \isa{x\ {\isasymsharp}\ {\isacharparenleft}{\isasymGamma}\isactrlisub {\isadigit{1}}{\isacharcomma}\ M{\isacharparenright}}
  then \isa{{\isasymGamma}\isactrlisub {\isadigit{1}}\ {\isacharat}\ {\isasymGamma}\isactrlisub {\isadigit{2}}\ {\isasymturnstile}\isactrlisub {\isasymSigma}\ M\ {\isasymRightarrow}\ A}.
  \end{compactenum}
  \end{theorem}
  \begin{proof}
    The proof is straightforward, using strengthening for algorithmic
    equivalence; parts (1--4) need to be proved in the order stated
    above since we need strengthening for contexts everywhere, we need
    strengthening for kinds to prove strengthening for types, and so
    on.  \refLem{strong-alg-freshness} is needed in the cases for
    object and type application.
  \end{proof}

  Finally, we can prove strengthening for the definitional system.

  \begin{theorem}[(Strengthening)]\labelThm{strengthening}
  Let \mbox{\isa{{\isasymGamma}\ {\isacharequal}\ {\isasymGamma}\isactrlisub {\isadigit{1}}\ {\isacharat}\ {\isacharbrackleft}{\isacharparenleft}x{\isacharcomma}\ B{\isacharparenright}{\isacharbrackright}\ {\isacharat}\ {\isasymGamma}\isactrlisub {\isadigit{2}}}}.
  \begin{compactenum}
  \item If \isa{{\isasymturnstile}\isactrlisub {\isasymSigma}\ {\isasymGamma}\ ctx} and \isa{x\ {\isasymsharp}\ {\isasymGamma}\isactrlisub {\isadigit{1}}} then \isa{{\isasymturnstile}\isactrlisub {\isasymSigma}\ {\isasymGamma}\isactrlisub {\isadigit{1}}\ {\isacharat}\ {\isasymGamma}\isactrlisub {\isadigit{2}}\ ctx}.
  \item If \isa{{\isasymGamma}\ {\isasymturnstile}\isactrlisub {\isasymSigma}\ K\ {\isacharcolon}\ kind} and \isa{x\ {\isasymsharp}\ {\isacharparenleft}{\isasymGamma}\isactrlisub {\isadigit{1}}{\isacharcomma}\ K{\isacharparenright}}
  then \isa{{\isasymGamma}\isactrlisub {\isadigit{1}}\ {\isacharat}\ {\isasymGamma}\isactrlisub {\isadigit{2}}\ {\isasymturnstile}\isactrlisub {\isasymSigma}\ K\ {\isacharcolon}\ kind}.
  \item If \isa{{\isasymGamma}\ {\isasymturnstile}\isactrlisub {\isasymSigma}\ K\ {\isacharequal}\ L\ {\isacharcolon}\ kind} and \isa{x\ {\isasymsharp}\ {\isacharparenleft}{\isasymGamma}\isactrlisub {\isadigit{1}}{\isacharcomma}\ K{\isacharcomma}\ L{\isacharparenright}}
  then \isa{{\isasymGamma}\isactrlisub {\isadigit{1}}\ {\isacharat}\ {\isasymGamma}\isactrlisub {\isadigit{2}}\ {\isasymturnstile}\isactrlisub {\isasymSigma}\ K\ {\isacharequal}\ L\ {\isacharcolon}\ kind}.
  \item If \isa{{\isasymGamma}\ {\isasymturnstile}\isactrlisub {\isasymSigma}\ A\ {\isacharcolon}\ K} and \isa{x\ {\isasymsharp}\ {\isacharparenleft}{\isasymGamma}\isactrlisub {\isadigit{1}}{\isacharcomma}\ A{\isacharparenright}}
  then \isa{{\isasymGamma}\isactrlisub {\isadigit{1}}\ {\isacharat}\ {\isasymGamma}\isactrlisub {\isadigit{2}}\ {\isasymturnstile}\isactrlisub {\isasymSigma}\ A\ {\isacharcolon}\ K}. \label{part:strengthening-type-valid}
  \item If \isa{{\isasymGamma}\ {\isasymturnstile}\isactrlisub {\isasymSigma}\ A\ {\isacharequal}\ B\ {\isacharcolon}\ K} and \isa{x\ {\isasymsharp}\ {\isacharparenleft}{\isasymGamma}\isactrlisub {\isadigit{1}}{\isacharcomma}\ A{\isacharcomma}\ B{\isacharparenright}}
  then \isa{{\isasymGamma}\isactrlisub {\isadigit{1}}\ {\isacharat}\ {\isasymGamma}\isactrlisub {\isadigit{2}}\ {\isasymturnstile}\isactrlisub {\isasymSigma}\ A\ {\isacharequal}\ B\ {\isacharcolon}\ K}.
  \item If \isa{{\isasymGamma}\ {\isasymturnstile}\isactrlisub {\isasymSigma}\ M\ {\isacharcolon}\ A} and \isa{x\ {\isasymsharp}\ {\isacharparenleft}{\isasymGamma}\isactrlisub {\isadigit{1}}{\isacharcomma}\ M{\isacharparenright}}
  then \isa{{\isasymGamma}\isactrlisub {\isadigit{1}}\ {\isacharat}\ {\isasymGamma}\isactrlisub {\isadigit{2}}\ {\isasymturnstile}\isactrlisub {\isasymSigma}\ M\ {\isacharcolon}\ A}. \label{part:strengthening-object-valid}
  \item If \isa{{\isasymGamma}\ {\isasymturnstile}\isactrlisub {\isasymSigma}\ M\ {\isacharequal}\ N\ {\isacharcolon}\ A} and \isa{x\ {\isasymsharp}\ {\isacharparenleft}{\isasymGamma}\isactrlisub {\isadigit{1}}{\isacharcomma}\ M{\isacharcomma}\ N{\isacharparenright}}
  then \isa{{\isasymGamma}\isactrlisub {\isadigit{1}}\ {\isacharat}\ {\isasymGamma}\isactrlisub {\isadigit{2}}\ {\isasymturnstile}\isactrlisub {\isasymSigma}\ M\ {\isacharequal}\ N\ {\isacharcolon}\ A}. 
  \end{compactenum}
  \end{theorem}

  \begin{proof}
    The proof follows the sketch in the article, using algorithmic
    strengthening and soundness and completeness of the algorithmic
    judgments, but some care is needed.  Part 1 is straightforward,
    but we must prove the remaining cases in the specific order
    listed: first kind validity, then kind equivalence, then type
    validity, etc.  The reason is that to prove strengthening for the
    equivalence judgments, we need strengthening for the corresponding
    validity judgments because of the validity side-conditions on
    \refThm{soundness}. In turn, to prove strengthening for the object
    and type validity judgments, we need strengthening for type and
    kind equivalence respectively, because of the respective type and
    kind equivalence judgments in the conclusions of
    \refThm{algorithmic-tc-completeness}.
    \refLem{strong-alg-freshness} is needed in parts
    (\ref{part:strengthening-type-valid}) and
    (\ref{part:strengthening-object-valid}).
  \end{proof}

  \HP also sketched a proof of admissibility of a stronger version
  of the extensionality rule which omits the well-formedness checks:
  \begin{center}
    \isa{\mbox{}\inferrule{\mbox{{\isacharparenleft}x{\isacharcomma}\ A\isactrlisub {\isadigit{1}}{\isacharparenright}{\isacharcolon}{\isacharcolon}{\isasymGamma}\ {\isasymturnstile}\isactrlisub {\isasymSigma}\ M\ x\ {\isacharequal}\ N\ x\ {\isacharcolon}\ A\isactrlisub {\isadigit{2}}}\\\ \mbox{x\ {\isasymsharp}\ {\isacharparenleft}M{\isacharcomma}\ N{\isacharparenright}}}{\mbox{{\isasymGamma}\ {\isasymturnstile}\isactrlisub {\isasymSigma}\ M\ {\isacharequal}\ N\ {\isacharcolon}\ {\isasymPi}x{\isacharcolon}A\isactrlisub {\isadigit{1}}{\isachardot}\ A\isactrlisub {\isadigit{2}}}}}
  \end{center}
  However, the short proof
  sketched in the article actually requires a substantial amount of
  work to formalize. The first two steps of their informal proof were as follows:
  \begin{compactenum}
  \item By validity, we have \isa{{\isacharparenleft}x{\isacharcomma}\ A\isactrlisub {\isadigit{1}}{\isacharparenright}{\isacharcolon}{\isacharcolon}{\isasymGamma}\ {\isasymturnstile}\isactrlisub {\isasymSigma}\ M\ x\ {\isacharcolon}\ A\isactrlisub {\isadigit{2}}}.
  \item By inversion, we have \isa{{\isacharparenleft}x{\isacharcomma}\ A\isactrlisub {\isadigit{1}}{\isacharparenright}{\isacharcolon}{\isacharcolon}{\isasymGamma}\ {\isasymturnstile}\isactrlisub {\isasymSigma}\ M\ {\isacharcolon}\ {\isasymPi}x{\isacharcolon}B\isactrlisub {\isadigit{1}}{\isachardot}\ B\isactrlisub {\isadigit{2}}} 
    and 
    \isa{{\isacharparenleft}x{\isacharcomma}\ A\isactrlisub {\isadigit{1}}{\isacharparenright}{\isacharcolon}{\isacharcolon}{\isasymGamma}\ {\isasymturnstile}\isactrlisub {\isasymSigma}\ x\ {\isacharcolon}\ B\isactrlisub {\isadigit{1}}}.
  \end{compactenum}
  However, step (2) above does not follow immediately from the
  inversion lemmas proved earlier.  In particular, we only know that
  $M$ will have a type of the form
  \isa{{\isasymPi}y{\isacharcolon}B\isactrlisub {\isadigit{1}}{\isachardot}\ B\isactrlisub {\isadigit{2}}} for some \isa{y}, \isa{B\isactrlisub {\isadigit{1}}}
  and 
  \isa{B\isactrlisub {\isadigit{2}}}
  such that
  \isa{{\isacharparenleft}x{\isacharcomma}\ A\isactrlisub {\isadigit{1}}{\isacharparenright}{\isacharcolon}{\isacharcolon}{\isasymGamma}\ {\isasymturnstile}\isactrlisub {\isasymSigma}\ M\ {\isacharcolon}\ {\isasymPi}y{\isacharcolon}B\isactrlisub {\isadigit{1}}{\isachardot}\ B\isactrlisub {\isadigit{2}}}
  and 
  \isa{{\isacharparenleft}x{\isacharcomma}\ A\isactrlisub {\isadigit{1}}{\isacharparenright}{\isacharcolon}{\isacharcolon}{\isasymGamma}\ {\isasymturnstile}\isactrlisub {\isasymSigma}\ y\ {\isacharcolon}\ B\isactrlisub {\isadigit{1}}}
  and
  \isa{{\isacharparenleft}x{\isacharcomma}\ A\isactrlisub {\isadigit{1}}{\isacharparenright}{\isacharcolon}{\isacharcolon}{\isasymGamma}\ {\isasymturnstile}\isactrlisub {\isasymSigma}\ A\isactrlisub {\isadigit{2}}\ {\isacharequal}\ B\isactrlisub {\isadigit{2}}{\isacharbrackleft}y{\isacharcolon}{\isacharequal}x{\isacharbrackright}\ {\isacharcolon}\ type}.  Moreover, in this case we
  cannot use the strong version of the inversion lemma to avoid this
  problem, because \isa{x} is already in use in the context.

  Although their proof looks rigorous and detailed, here Harper and
  Pfenning appear to employ implicit ``without loss of generality''
  reasoning about inversion and renaming that is not easy to formalize
  directly.  Instead we needed to show carefully that:

  \begin{lemma}\labelLem{extensional-inversion}
  \isa{{\normalsize{}If\,}\ \mbox{{\isacharparenleft}x{\isacharcomma}\ A\isactrlisub {\isadigit{1}}{\isacharparenright}{\isacharcolon}{\isacharcolon}{\isasymGamma}\ {\isasymturnstile}\isactrlisub {\isasymSigma}\ M\ x\ {\isacharcolon}\ A\isactrlisub {\isadigit{2}}}\ {\normalsize \,and\,}\ \mbox{x\ {\isasymsharp}\ M}\ {\normalsize\linebreak[0] \,then\,\linebreak[0]}\ \mbox{{\isasymGamma}\ {\isasymturnstile}\isactrlisub {\isasymSigma}\ M\ {\isacharcolon}\ {\isasymPi}x{\isacharcolon}A\isactrlisub {\isadigit{1}}{\isachardot}\ A\isactrlisub {\isadigit{2}}{\isachardot}}}
  \end{lemma} 
  \begin{proof}
    The proof proceeds by applying validity and inversion principles,
    as discussed above.  One subtle freshness side-condition is the
    fact that \isa{x} is fresh for \isa{{\isasymPi}y{\isacharcolon}B\isactrlisub {\isadigit{1}}{\isachardot}\ B\isactrlisub {\isadigit{2}}}, and this is proved by
    translating to the algorithmic typechecking system and using
    \refLem{strong-alg-freshness}.
  \end{proof}

  \noindent 
  Strong extensionality then follows essentially as in \HP, using
  \refLem{extensional-inversion} to fill the gap identified above:

  \begin{theorem}[(Strong extensionality)]\mbox{}\\
  If \isa{{\isacharparenleft}x{\isacharcomma}\ A\isactrlisub {\isadigit{1}}{\isacharparenright}{\isacharcolon}{\isacharcolon}{\isasymGamma}\ {\isasymturnstile}\isactrlisub {\isasymSigma}\ M\ x\ {\isacharequal}\ N\ x\ {\isacharcolon}\ A\isactrlisub {\isadigit{2}}}
  and  \isa{x\ {\isasymsharp}\ {\isacharparenleft}M{\isacharcomma}\ N{\isacharparenright}}
  then\\
   \isa{{\isasymGamma}\ {\isasymturnstile}\isactrlisub {\isasymSigma}\ M\ {\isacharequal}\ N\ {\isacharcolon}\ {\isasymPi}x{\isacharcolon}A\isactrlisub {\isadigit{1}}{\isachardot}\ A\isactrlisub {\isadigit{2}}}.
  \end{theorem}%
\end{isamarkuptext}%
\isamarkuptrue%
\isamarkupsubsection{Decidability%
}
\isamarkuptrue%
\begin{isamarkuptext}%
\labelSec{decidability}%
\end{isamarkuptext}%
\isamarkuptrue%
\begin{isamarkuptext}%
\HP also sketches proofs of the decidability of the algorithmic
  judgments (and hence also the definitional system).  Reasoning about
  decidability within Isabelle/HOL is not straightforward because
  Isabelle/HOL is based on classical logic.  Thus, unlike constructive
  logics or type theories, we cannot infer decidability of \isa{P}
  simply by proving \isa{P\ {\isasymor}\ {\isasymnot}\ P}.  Furthermore, given a
  relation $R$ definable in Isabelle/HOL, it is not clear how best to
  formalize the informal statement ``$R$ is decidable''.

  As a sanity check, we have shown that weak head reduction is
  strongly normalizing for well-formed terms.  We write \isa{M{\isasymDown}}
  to indicate that \isa{M} is strongly normalizing under weak head
  reduction.  This proof uses techniques and definitions from the
  example formalization of strong normalization for the simply-typed
  lambda calculus in the Nominal Datatype Package.

  \begin{theorem}\labelLem{whn-head} 
    \isa{{\normalsize{}If\,}\ {\isasymGamma}\ {\isasymturnstile}\isactrlisub {\isasymSigma}\ M\ {\isacharcolon}\ A\ {\normalsize \,then\,}\ M{\isasymDown}{\isachardot}}
  \end{theorem}
  
  \begin{proof}
    We first show the (standard) property that if \isa{M\ N{\isasymDown}}
    then \isa{M{\isasymDown}}.  We then show that if \isa{{\isasymDelta}\ {\isasymturnstile}\isactrlbsub {\isasymSigma}\isactrlesub \ M\ {\isasymLeftrightarrow}\ N\ {\isacharcolon}\ {\isasymtau}} then \isa{M{\isasymDown}} by induction on derivations.  The
    main result follows by reflexivity and \refThm{completeness}.
  \end{proof}

  Turning now to the issue of formalizing decidability properties in
  Isabelle/HOL, we considered the following options:

  \paragraph*{Formalizing computability theory} It should be possible
  to define Turing machines (or some other universal model of
  computation) within Isabelle/HOL and derive enough of the theory of
  computation to be able to prove that the algorithmic equivalence and
  typechecking relations are decidable.  It appears to be an open
  question how to formalize proofs of decidability in Isabelle/HOL,
  especially for algorithms over complex data structures such as
  nominal datatypes.  Although this would probably be the most
  satisfying solution, it would also require a major additional
  formalization effort, including a great deal of work that is
  orthogonal to the issues addressed here. Another possibility would be to restrict Isabelle/HOL to a 
  constructive fragment, but this seems even more difficult and
  time-consuming since Isabelle/HOL makes extensive use of choice
  principles and the law of excluded middle.
  We therefore view fully
  formalizing decidability in this way as beyond the scope of this
  article. Instead, we consider other techniques that stop short of
  full formalization while providing some convincing evidence for
  decidability.

  \paragraph*{Bounded-height derivations}
  We could define height-bounded versions of the algorithmic
  typechecking relations and prove that there is a computable bound
  on the height needed to derive any derivable judgment in the
  system.  That is, there exists a computable $h$ such that for any
  inputs $x_1,\ldots,x_n$, there is a derivation of
  $J(x_1,\ldots,x_n)$ if and only if there is a derivation of height
  at most $h(x_1,\ldots,x_n)$.  
  
  This seems reasonable intuitively, but there are several problems.
  First, it is not obvious how to obtain a closed-form, recursively
  defined height bound for the number of steps needed for algorithmic
  equivalence for the same reason it is difficult to give an explicit
  termination measure for weak head normalization.  Second, even if we
  could find such an $h$, this approach begs the question of how to
  prove that $h$ is computable.  It is clearly not enough to simply
  require that some $h$ exists, because the Axiom of Choice can be
  used to define $h$ nonconstructively.  Finally, inductively defined
  judgments in Isabelle/HOL may themselves involve nonconstructive
  features, including equality at or quantification over infinite
  types, negation of undecidable properties, and choice operators.
  Although the definitions we have in mind do not use these
  facilities, there is no easy way to certify this within
  Isabelle/HOL.

  \paragraph*{Inductive definability}
  We have formalized what we believe is the essence of the
  decidability proof using the following methodology. For each
  inductively defined relation \isa{R} we wish to prove decidable,
  possibly under some constraints \isa{P}:
  
  \begin{compactenum}
  \item Inductively define a complement relation \isa{R{\isacharprime}}.
\item (Exclusion) Prove that \isa{{\isasymnot}\ {\isacharparenleft}R\ \textrm{and\linebreak[1]}\ R{\isacharprime}{\isacharparenright}}.
  \item (Exhaustion) Prove that \isa{P} implies \isa{R\ {\isasymor}\ R{\isacharprime}}.
  \item Observe (informally) that \isa{R} and \isa{R{\isacharprime}} are
    recursively enumerable since they are defined inductively by rules
    without recourse to nonconstructive features such as negation or
    universal quantification in the hypotheses.  Conclude (informally)
    that \isa{P} implies \isa{R} is both r.e. and co-r.e.,
    hence decidable. \label{part:informal-step}
  \end{compactenum}
  
  This approach exploits an intuitive connection between inductively
  definable predicates and recursively enumerable sets in step
  (\ref{part:informal-step}).  It is important to note that this
  intuition is not rigorously formalized.  We argue that this approach
  does force us to perform all of the case analysis that would be
  necessary in a proper decidability proof, but the only way to be
  certain of this is to fully formalize a substantial amount of
  computability theory in Isabelle/HOL, which as we have discussed
  above would be a major undertaking in its own
  right. 

  We call a formula $R$ \emph{quasidecidable} if both $R$ and its
  negation are equivalent to inductively defined relations, as
  described above.  This is an informal (and intensional) property; we
  have \emph{not} defined quasidecidability explicitly in
  Isabelle/HOL.  We have the following lemma, analogous to \HP's Lemma
  6.1:

  \begin{theorem}[(Quasidecidability of algorithmic equivalence)]\labelThm{alg-eq-quasidecidable}
  ~  
  \begin{compactenum}  
  \item If \isa{{\isasymDelta}\ {\isasymturnstile}\isactrlbsub {\isasymSigma}\isactrlesub \ M\ {\isasymLeftrightarrow}\ M{\isacharprime}\ {\isacharcolon}\ {\isasymtau}} and \isa{{\isasymDelta}\ {\isasymturnstile}\isactrlbsub {\isasymSigma}\isactrlesub \ N\ {\isasymLeftrightarrow}\ N{\isacharprime}\ {\isacharcolon}\ {\isasymtau}} 
  then \isa{{\isasymDelta}\ {\isasymturnstile}\isactrlbsub {\isasymSigma}\isactrlesub \ M\ {\isasymLeftrightarrow}\ N\ {\isacharcolon}\ {\isasymtau}} is quasidecidable.
  \item If \isa{{\isasymDelta}\ {\isasymturnstile}\isactrlbsub {\isasymSigma}\isactrlesub \ M\ {\isasymleftrightarrow}\ M{\isacharprime}\ {\isacharcolon}\ {\isasymtau}\isactrlisub {\isadigit{1}}} and \isa{{\isasymDelta}\ {\isasymturnstile}\isactrlbsub {\isasymSigma}\isactrlesub \ N\ {\isasymleftrightarrow}\ N{\isacharprime}\ {\isacharcolon}\ {\isasymtau}\isactrlisub {\isadigit{2}}} 
  then \isa{{\isasymexists}{\isasymtau}\isactrlisub {\isadigit{3}}{\isachardot}\ {\isasymDelta}\ {\isasymturnstile}\isactrlbsub {\isasymSigma}\isactrlesub \ M\ {\isasymleftrightarrow}\ N\ {\isacharcolon}\ {\isasymtau}\isactrlisub {\isadigit{3}}} is quasidecidable.
  \item If \isa{{\isasymDelta}\ {\isasymturnstile}\isactrlbsub {\isasymSigma}\isactrlesub \ A\ {\isasymLeftrightarrow}\ A{\isacharprime}\ {\isacharcolon}\ {\isasymkappa}} and \isa{{\isasymDelta}\ {\isasymturnstile}\isactrlbsub {\isasymSigma}\isactrlesub \ B\ {\isasymLeftrightarrow}\ B{\isacharprime}\ {\isacharcolon}\ {\isasymkappa}} 
  then \isa{{\isasymDelta}\ {\isasymturnstile}\isactrlbsub {\isasymSigma}\isactrlesub \ A\ {\isasymLeftrightarrow}\ B\ {\isacharcolon}\ {\isasymkappa}} is quasidecidable.
  \item If \isa{{\isasymDelta}\ {\isasymturnstile}\isactrlbsub {\isasymSigma}\isactrlesub \ A\ {\isasymleftrightarrow}\ A{\isacharprime}\ {\isacharcolon}\ {\isasymkappa}\isactrlisub {\isadigit{1}}} and \isa{{\isasymDelta}\ {\isasymturnstile}\isactrlbsub {\isasymSigma}\isactrlesub \ B\ {\isasymleftrightarrow}\ B{\isacharprime}\ {\isacharcolon}\ {\isasymkappa}\isactrlisub {\isadigit{2}}} 
  then \isa{{\isasymexists}{\isasymkappa}\isactrlisub {\isadigit{3}}{\isachardot}\ {\isasymDelta}\ {\isasymturnstile}\isactrlbsub {\isasymSigma}\isactrlesub \ A\ {\isasymleftrightarrow}\ B\ {\isacharcolon}\ {\isasymkappa}\isactrlisub {\isadigit{3}}} is quasidecidable.
  \item If \isa{{\isasymDelta}\ {\isasymturnstile}\isactrlbsub {\isasymSigma}\isactrlesub \ K\ {\isasymLeftrightarrow}\ K{\isacharprime}\ {\isacharcolon}\ kind\isactrlisup {\isacharminus}}  and \isa{{\isasymDelta}\ {\isasymturnstile}\isactrlbsub {\isasymSigma}\isactrlesub \ L\ {\isasymLeftrightarrow}\ L{\isacharprime}\ {\isacharcolon}\ kind\isactrlisup {\isacharminus}}
  then \isa{{\isasymDelta}\ {\isasymturnstile}\isactrlbsub {\isasymSigma}\isactrlesub \ K\ {\isasymLeftrightarrow}\ L\ {\isacharcolon}\ kind\isactrlisup {\isacharminus}} is quasidecidable.
  \end{compactenum}
  \end{theorem}

  We further proved that the algorithmic typechecking
  judgments are quasidecidable, which is the key step in \HP's Theorem
  6.5.  Proving exclusivity required establishing uniqueness of
  algorithmic typechecking.

  \begin{lemma}[(Uniqueness of algorithmic types)]
  ~  
  \begin{compactenum}  
  \item \isa{{\normalsize{}If\,}\ \mbox{{\isasymGamma}\ {\isasymturnstile}\isactrlisub {\isasymSigma}\ M\ {\isasymRightarrow}\ A}\ {\normalsize \,and\,}\ \mbox{{\isasymGamma}\ {\isasymturnstile}\isactrlisub {\isasymSigma}\ M\ {\isasymRightarrow}\ A{\isacharprime}}\ {\normalsize\linebreak[0] \,then\,\linebreak[0]}\ \mbox{A\ {\isacharequal}\ A{\isacharprime}{\isachardot}}}
  \item \isa{{\normalsize{}If\,}\ \mbox{{\isasymGamma}\ {\isasymturnstile}\isactrlisub {\isasymSigma}\ A\ {\isasymRightarrow}\ K}\ {\normalsize \,and\,}\ \mbox{{\isasymGamma}\ {\isasymturnstile}\isactrlisub {\isasymSigma}\ A\ {\isasymRightarrow}\ K{\isacharprime}}\ {\normalsize\linebreak[0] \,then\,\linebreak[0]}\ \mbox{K\ {\isacharequal}\ K{\isacharprime}{\isachardot}}}
  \end{compactenum}
  \end{lemma}

  Equipped with \refThm{alg-eq-quasidecidable} and the uniqueness
  lemma above, we can show a form of \HP's Theorem 6.2.  Note that
  uses of \refThm{alg-eq-quasidecidable} are safe because we always
  call the algorithmic equivalence judgments on terms that are
  well-formed, and hence (by \refThm{soundness}) algorithmically
  equivalent to themselves.

  \begin{theorem}[(Quasidecidability of algorithmic typechecking)]
  ~  
  \begin{compactenum}  
    \item For any \isa{{\isasymSigma}}, \isa{{\isasymturnstile}\ {\isasymSigma}\ {\isasymRightarrow}\ sig} is quasidecidable.
    \item For any \isa{{\isasymSigma}{\isacharcomma}{\isasymGamma}}, if \isa{{\isasymturnstile}\ {\isasymSigma}\ {\isasymRightarrow}\ sig} holds then
      \isa{{\isasymturnstile}\isactrlisub {\isasymSigma}\ {\isasymGamma}\ {\isasymRightarrow}\ ctx} is quasidecidable.
    \item For any \isa{{\isasymSigma}{\isacharcomma}{\isasymGamma}{\isacharcomma}M}, if \isa{{\isasymturnstile}\isactrlisub {\isasymSigma}\ {\isasymGamma}\ {\isasymRightarrow}\ ctx} holds 
      then \isa{{\isasymexists}A{\isachardot}\ {\isasymGamma}\ {\isasymturnstile}\isactrlisub {\isasymSigma}\ M\ {\isasymRightarrow}\ A} is quasidecidable.
    \item For any \isa{{\isasymSigma}{\isacharcomma}{\isasymGamma}{\isacharcomma}A}, if \isa{{\isasymturnstile}\isactrlisub {\isasymSigma}\ {\isasymGamma}\ {\isasymRightarrow}\ ctx} holds
      then \isa{{\isasymexists}K{\isachardot}\ {\isasymGamma}\ {\isasymturnstile}\isactrlisub {\isasymSigma}\ A\ {\isasymRightarrow}\ K} is quasidecidable.
    \item For any \isa{{\isasymSigma}{\isacharcomma}{\isasymGamma}{\isacharcomma}K}, if \isa{{\isasymturnstile}\isactrlisub {\isasymSigma}\ {\isasymGamma}\ {\isasymRightarrow}\ ctx} holds 
      then \isa{{\isasymGamma}\ {\isasymturnstile}\isactrlisub {\isasymSigma}\ K\ {\isasymRightarrow}\ kind} is quasidecidable.
  \end{compactenum}
  \end{theorem}%
\end{isamarkuptext}%
\isamarkuptrue%
\isamarkupsubsection{Quasicanonical forms%
}
\isamarkuptrue%
\begin{isamarkuptext}%
\labelSec{quasicanonical}%
\end{isamarkuptext}%
\isamarkuptrue%
\begin{isamarkuptext}%
Section 7 of \HP discusses \emph{quasicanonical
    forms} which can be used to study the \emph{adequacy}, or
  correctness, of LF encodings. Quasicanonical forms are untyped
  $\lambda$-terms that correspond to the $\beta$-normal, $\eta$-long
  forms of well-typed LF terms.  Quasicanonical forms \isa{$\bar{\bar{O}}$} and quasiatomic forms \isa{$\bar{O}$} are given by
  the grammar rules:

  \begin{center}
  \isa{$\bar{\bar{O}}$} ::= \isa{$\bar{O}$} $\mid$ \isa{{\isasymlambda}x{\isachardot}$\bar{\bar{O}}$} $\quad$
  \isa{$\bar{O}$} ::= \isa{x} $\mid$ \isa{c} 
                        $\mid$ \isa{$\bar{O}$\ $\bar{\bar{O}}$}
  \end{center}
  
  \noindent 
  \HP introduces instrumented algorithmic equivalence judgments that
  construct quasicanonical forms for algorithmically and structurally
  equivalent terms, respectively.  The rules are shown in
  \refFig{quasi-alg-equiv}.

  %%%%%%%%%%%%%%%%%%%%%%%%%%%%%%%%%%%%%%%%%%%%%%%%%%%%%%%%%%%%%%%%%%
  \begin{figure}[tb]
      \fbox{\isa{{\isasymDelta}\ {\isasymturnstile}\isactrlbsub {\isasymSigma}\isactrlesub \ M\ {\isasymLeftrightarrow}\ N\ {\isacharcolon}\ {\isasymtau}\ {\isasymUp}\ $\bar{\bar{O}}$}}
      \begin{center}
        \begin{tabular}{@ {}c@ {}}
          \isa{\mbox{}\inferrule{\mbox{M\ $\stackrel{\mathrm{whr}}{\longrightarrow}$\ M{\isacharprime}}\\\ \mbox{{\isasymDelta}\ {\isasymturnstile}\isactrlbsub {\isasymSigma}\isactrlesub \ M{\isacharprime}\ {\isasymLeftrightarrow}\ N\ {\isacharcolon}\ a\isactrlisup {\isacharminus}\ {\isasymUp}\ $\bar{\bar{O}}$}}{\mbox{{\isasymDelta}\ {\isasymturnstile}\isactrlbsub {\isasymSigma}\isactrlesub \ M\ {\isasymLeftrightarrow}\ N\ {\isacharcolon}\ a\isactrlisup {\isacharminus}\ {\isasymUp}\ $\bar{\bar{O}}$}}}
          \quad 
          \isa{\mbox{}\inferrule{\mbox{N\ $\stackrel{\mathrm{whr}}{\longrightarrow}$\ N{\isacharprime}}\\\ \mbox{{\isasymDelta}\ {\isasymturnstile}\isactrlbsub {\isasymSigma}\isactrlesub \ M\ {\isasymLeftrightarrow}\ N{\isacharprime}\ {\isacharcolon}\ a\isactrlisup {\isacharminus}\ {\isasymUp}\ $\bar{\bar{O}}$}}{\mbox{{\isasymDelta}\ {\isasymturnstile}\isactrlbsub {\isasymSigma}\isactrlesub \ M\ {\isasymLeftrightarrow}\ N\ {\isacharcolon}\ a\isactrlisup {\isacharminus}\ {\isasymUp}\ $\bar{\bar{O}}$}}}\smallskip\\  
      
          \isa{\mbox{}\inferrule{\mbox{{\isasymDelta}\ {\isasymturnstile}\isactrlbsub {\isasymSigma}\isactrlesub \ M\ {\isasymleftrightarrow}\ N\ {\isacharcolon}\ a\isactrlisup {\isacharminus}\ {\isasymdown}\ $\bar{O}$}}{\mbox{{\isasymDelta}\ {\isasymturnstile}\isactrlbsub {\isasymSigma}\isactrlesub \ M\ {\isasymLeftrightarrow}\ N\ {\isacharcolon}\ a\isactrlisup {\isacharminus}\ {\isasymUp}\ $\bar{O}$}}}
          \quad 
          \isa{\mbox{}\inferrule{\mbox{{\isacharparenleft}x{\isacharcomma}\ {\isasymtau}{\isacharparenright}{\isacharcolon}{\isacharcolon}{\isasymDelta}\ {\isasymturnstile}\isactrlbsub {\isasymSigma}\isactrlesub \ M\ x\ {\isasymLeftrightarrow}\ N\ x\ {\isacharcolon}\ {\isasymtau}{\isacharprime}\ {\isasymUp}\ $\bar{\bar{O}}$}\\\ \mbox{x\ {\isasymsharp}\ {\isacharparenleft}{\isasymDelta}{\isacharcomma}\ M{\isacharcomma}\ N{\isacharparenright}}}{\mbox{{\isasymDelta}\ {\isasymturnstile}\isactrlbsub {\isasymSigma}\isactrlesub \ M\ {\isasymLeftrightarrow}\ N\ {\isacharcolon}\ {\isasymtau}\ {\isasymrightarrow}\ {\isasymtau}{\isacharprime}\ {\isasymUp}\ {\isasymlambda}x{\isachardot}$\bar{\bar{O}}$}}}
        \end{tabular}

      \end{center} 
      \fbox{\isa{{\isasymDelta}\ {\isasymturnstile}\isactrlbsub {\isasymSigma}\isactrlesub \ M\ {\isasymleftrightarrow}\ N\ {\isacharcolon}\ {\isasymtau}\ {\isasymdown}\ $\bar{O}$}}
      \begin{center}
        \begin{tabular}{@ {}c@ {}}
          \isa{\mbox{}\inferrule{\mbox{{\isacharparenleft}x{\isacharcomma}\ {\isasymtau}{\isacharparenright}\ {\isasymin}\ {\isasymDelta}}\\\ \mbox{{\isasymturnstile}\ {\isasymSigma}\ ssig}\\\ \mbox{{\isasymturnstile}\ {\isasymDelta}\ sctx}}{\mbox{{\isasymDelta}\ {\isasymturnstile}\isactrlbsub {\isasymSigma}\isactrlesub \ x\ {\isasymleftrightarrow}\ x\ {\isacharcolon}\ {\isasymtau}\ {\isasymdown}\ x}}}
          \quad 
          \isa{\mbox{}\inferrule{\mbox{{\isacharparenleft}c{\isacharcomma}\ {\isasymkappa}{\isacharparenright}\ {\isasymin}\ {\isasymSigma}}\\\ \mbox{{\isasymturnstile}\ {\isasymSigma}\ ssig}\\\ \mbox{{\isasymturnstile}\ {\isasymDelta}\ sctx}}{\mbox{{\isasymDelta}\ {\isasymturnstile}\isactrlbsub {\isasymSigma}\isactrlesub \ c\ {\isasymleftrightarrow}\ c\ {\isacharcolon}\ {\isasymkappa}\ {\isasymdown}\ c}}}\smallskip\\  
          \isa{\mbox{}\inferrule{\mbox{{\isasymDelta}\ {\isasymturnstile}\isactrlbsub {\isasymSigma}\isactrlesub \ M\isactrlisub {\isadigit{1}}\ {\isasymleftrightarrow}\ N\isactrlisub {\isadigit{1}}\ {\isacharcolon}\ {\isasymtau}\isactrlisub {\isadigit{2}}\ {\isasymrightarrow}\ {\isasymtau}\isactrlisub {\isadigit{1}}\ {\isasymdown}\ $\bar{O}_1$}\\\ \mbox{{\isasymDelta}\ {\isasymturnstile}\isactrlbsub {\isasymSigma}\isactrlesub \ M\isactrlisub {\isadigit{2}}\ {\isasymLeftrightarrow}\ N\isactrlisub {\isadigit{2}}\ {\isacharcolon}\ {\isasymtau}\isactrlisub {\isadigit{2}}\ {\isasymUp}\ $\bar{\bar{O}}_2$}}{\mbox{{\isasymDelta}\ {\isasymturnstile}\isactrlbsub {\isasymSigma}\isactrlesub \ M\isactrlisub {\isadigit{1}}\ M\isactrlisub {\isadigit{2}}\ {\isasymleftrightarrow}\ N\isactrlisub {\isadigit{1}}\ N\isactrlisub {\isadigit{2}}\ {\isacharcolon}\ {\isasymtau}\isactrlisub {\isadigit{1}}\ {\isasymdown}\ $\bar{O}_1$\ $\bar{\bar{O}}_2$}}}
        \end{tabular}
      \end{center}
      \caption{Algorithmic equivalence rules instrumented to produce
        quasicanonical forms.}\labelFig{quasi-alg-equiv}
  \end{figure}
  %%%%%%%%%%%%%%%%%%%%%%%%%%%%%%%%%%%%%%%%%%%%%%%%%%%%%%%%%%%%%%%%%%	

  It is straightforward to show that quasi-canonical and quasi-atomic
  forms exist and are unique (provided that \isa{{\isasymSigma}} and
  \isa{{\isasymDelta}} are valid).  
  \begin{lemma}[(Properties of quasicanonical forms)]\labelLem{qcan-basic-properties}
    \mbox{}
    \begin{compactenum}
    \item \isa{{\normalsize{}If\,}\ {\isasymDelta}\ {\isasymturnstile}\isactrlbsub {\isasymSigma}\isactrlesub \ M\ {\isasymLeftrightarrow}\ N\ {\isacharcolon}\ {\isasymtau}\ {\normalsize \,then\,}\ {\isasymexists}QC{\isachardot}\ {\isasymDelta}\ {\isasymturnstile}\isactrlbsub {\isasymSigma}\isactrlesub \ M\ {\isasymLeftrightarrow}\ N\ {\isacharcolon}\ {\isasymtau}\ {\isasymUp}\ QC{\isachardot}}
    \item \isa{{\normalsize{}If\,}\ {\isasymDelta}\ {\isasymturnstile}\isactrlbsub {\isasymSigma}\isactrlesub \ M\ {\isasymleftrightarrow}\ N\ {\isacharcolon}\ {\isasymtau}\ {\normalsize \,then\,}\ {\isasymexists}QA{\isachardot}\ {\isasymDelta}\ {\isasymturnstile}\isactrlbsub {\isasymSigma}\isactrlesub \ M\ {\isasymleftrightarrow}\ N\ {\isacharcolon}\ {\isasymtau}\ {\isasymdown}\ QA{\isachardot}}
    \item \isa{{\normalsize{}If\,}\ {\isasymDelta}\ {\isasymturnstile}\isactrlbsub {\isasymSigma}\isactrlesub \ M\ {\isasymLeftrightarrow}\ N\ {\isacharcolon}\ {\isasymtau}\ {\isasymUp}\ $\bar{\bar{O}}$\ {\normalsize \,then\,}\ {\isasymDelta}\ {\isasymturnstile}\isactrlbsub {\isasymSigma}\isactrlesub \ M\ {\isasymLeftrightarrow}\ N\ {\isacharcolon}\ {\isasymtau}{\isachardot}}
    \item \isa{{\normalsize{}If\,}\ {\isasymDelta}\ {\isasymturnstile}\isactrlbsub {\isasymSigma}\isactrlesub \ M\ {\isasymleftrightarrow}\ N\ {\isacharcolon}\ {\isasymtau}\ {\isasymdown}\ $\bar{O}$\ {\normalsize \,then\,}\ {\isasymDelta}\ {\isasymturnstile}\isactrlbsub {\isasymSigma}\isactrlesub \ M\ {\isasymleftrightarrow}\ N\ {\isacharcolon}\ {\isasymtau}{\isachardot}}
    \item \isa{{\normalsize{}If\,}\ {\isasymDelta}\ {\isasymturnstile}\isactrlbsub {\isasymSigma}\isactrlesub \ M\ {\isasymLeftrightarrow}\ N\ {\isacharcolon}\ {\isasymtau}\ {\isasymUp}\ $\bar{\bar{O}}$\ {\normalsize \,and\,}\ M\ $\stackrel{\mathrm{whr}}{\longrightarrow}$\ M{\isacharprime}\ {\normalsize \,then\,}\ {\isasymDelta}\ {\isasymturnstile}\isactrlbsub {\isasymSigma}\isactrlesub \ M{\isacharprime}\ {\isasymLeftrightarrow}\ N\ {\isacharcolon}\ {\isasymtau}\ {\isasymUp}\ $\bar{\bar{O}}${\isachardot}}
    \item \isa{{\normalsize{}If\,}\ {\isasymDelta}\ {\isasymturnstile}\isactrlbsub {\isasymSigma}\isactrlesub \ M\ {\isasymLeftrightarrow}\ N\ {\isacharcolon}\ {\isasymtau}\ {\isasymUp}\ $\bar{\bar{O}}$\ {\normalsize \,and\,}\ N\ $\stackrel{\mathrm{whr}}{\longrightarrow}$\ N{\isacharprime}\ {\normalsize \,then\,}\ {\isasymDelta}\ {\isasymturnstile}\isactrlbsub {\isasymSigma}\isactrlesub \ M\ {\isasymLeftrightarrow}\ N{\isacharprime}\ {\isacharcolon}\ {\isasymtau}\ {\isasymUp}\ $\bar{\bar{O}}${\isachardot}}
    \end{compactenum}
  \end{lemma}

  \begin{theorem}[(Uniqueness of quasicanonical forms)]
    \mbox{}
    \begin{compactenum}
    \item \isa{{\normalsize{}If\,}\ {\isasymturnstile}\ {\isasymDelta}\ sctx\ {\normalsize \,and\,}\ {\isasymturnstile}\ {\isasymSigma}\ ssig\ {\normalsize \,and\,}\ {\isasymDelta}\ {\isasymturnstile}\isactrlbsub {\isasymSigma}\isactrlesub \ M\ {\isasymLeftrightarrow}\ N\ {\isacharcolon}\ {\isasymtau}\ {\isasymUp}\ $\bar{\bar{O}}_1$\ {\normalsize \,and\,}\ {\isasymDelta}\ {\isasymturnstile}\isactrlbsub {\isasymSigma}\isactrlesub \ M\ {\isasymLeftrightarrow}\ N\ {\isacharcolon}\ {\isasymtau}\ {\isasymUp}\ $\bar{\bar{O}}_2$\ {\normalsize \,then\,}\ $\bar{\bar{O}}_1$\ {\isacharequal}\ $\bar{\bar{O}}_2${\isachardot}}
    \item \isa{{\normalsize{}If\,}\ {\isasymturnstile}\ {\isasymDelta}\ sctx\ {\normalsize \,and\,}\ {\isasymturnstile}\ {\isasymSigma}\ ssig\ {\normalsize \,and\,}\ {\isasymDelta}\ {\isasymturnstile}\isactrlbsub {\isasymSigma}\isactrlesub \ M\ {\isasymleftrightarrow}\ N\ {\isacharcolon}\ {\isasymtau}\ {\isasymdown}\ $\bar{O}_1$\ {\normalsize \,and\,}\ {\isasymDelta}\ {\isasymturnstile}\isactrlbsub {\isasymSigma}\isactrlesub \ M\ {\isasymleftrightarrow}\ N\ {\isacharcolon}\ {\isasymtau}{\isacharprime}\ {\isasymdown}\ $\bar{O}_2$\ {\normalsize \,then\,}\ {\isasymtau}\ {\isacharequal}\ {\isasymtau}{\isacharprime}\ \textrm{and\linebreak[1]}\ $\bar{O}_1$\ {\isacharequal}\ $\bar{O}_2${\isachardot}}
    \end{compactenum}
  \end{theorem}
  \begin{proof}
    By induction on derivations, using \refLem{qcan-basic-properties}(5,6) in the
    cases involving weak head reduction.
  \end{proof}
  The main result about these forms in
  \HP is that well-formed LF terms can be recovered from
  quasicanonical forms and type information.  To show this, we write
  \isa{N\ {\isasymUp}\ $\bar{\bar{O}}$} or \isa{N\ {\isasymdown}\ $\bar{O}$} for
  the relations that relate objects \isa{N} with their
  quasicanonical forms \isa{$\bar{\bar{O}}$} or quasiatomic forms \isa{$\bar{O}$}, respectively, where the type-labels have been erased.
  (\HP defined this notion as a partial function, which would be
  difficult to define with the
  Nominal Datatype Package at the time of writing.)
  These relations are defined as follows:
  \begin{center}
    \begin{tabular}{@ {}c@ {}}
      \isa{\mbox{}\inferrule{\mbox{}}{\mbox{x\ {\isasymdown}\ x}}}$\qquad$
      \isa{\mbox{}\inferrule{\mbox{}}{\mbox{c\ {\isasymdown}\ c}}}$\qquad$
      \isa{\mbox{}\inferrule{\mbox{M\ {\isasymdown}\ $\bar{O}$}\\\ \mbox{N\ {\isasymUp}\ $\bar{\bar{O}}$}}{\mbox{{\isacharparenleft}M\ N{\isacharparenright}\ {\isasymdown}\ $\bar{O}$\ $\bar{\bar{O}}$}}}
      $\qquad$
      \isa{\mbox{}\inferrule{\mbox{M\ {\isasymUp}\ $\bar{\bar{O}}$}}{\mbox{{\isacharparenleft}{\isasymlambda}x{\isacharcolon}A{\isachardot}\ M{\isacharparenright}\ {\isasymUp}\ {\isasymlambda}x{\isachardot}$\bar{\bar{O}}$}}}$\qquad$
      \isa{\mbox{}\inferrule{\mbox{M\ {\isasymdown}\ $\bar{O}$}}{\mbox{M\ {\isasymUp}\ $\bar{O}$}}}
    \end{tabular}
  \end{center}

  In the proof of the Quasicanonical Forms theorem (Theorem 7.1 of
  \HP) we found it necessary to prove several nontrivial auxiliary
  lemmas such as the admissibility of $\eta$-equivalence (which was
  not discussed in \HP):

  \begin{lemma}[(Eta-equivalence)]
    \isa{{\normalsize{}If\,}\ x\ {\isasymsharp}\ {\isasymGamma}\ {\normalsize \,and\,}\ {\isasymGamma}\ {\isasymturnstile}\isactrlisub {\isasymSigma}\ M\ {\isacharcolon}\ {\isasymPi}x{\isacharcolon}A\isactrlisub {\isadigit{1}}{\isachardot}\ A\isactrlisub {\isadigit{2}}\ {\normalsize \,then\,}\ {\isasymGamma}\ {\isasymturnstile}\isactrlisub {\isasymSigma}\ M\ {\isacharequal}\ {\isasymlambda}x{\isacharcolon}A\isactrlisub {\isadigit{1}}{\isachardot}\ M\ x\ {\isacharcolon}\ {\isasymPi}x{\isacharcolon}A\isactrlisub {\isadigit{1}}{\isachardot}\ A\isactrlisub {\isadigit{2}}{\isachardot}}
  \end{lemma}

  The following theorem is stated slightly differently than the
  corresponding theorem in \HP (Theorem 7.1), but their version
  follows immediately from this version.

  \begin{theorem}[(Quasicanonical forms)]\labelThm{quasicanonical-forms-exist-witness}
    ~
    \begin{compactenum}
    \item \isa{{\normalsize{}If\,}\ \mbox{{\isasymGamma}\isactrlisup {\isacharminus}\ {\isasymturnstile}\isactrlbsub {\isasymSigma}\isactrlisup {\isacharminus}\isactrlesub \ M\isactrlisub {\isadigit{1}}\ {\isasymLeftrightarrow}\ M\isactrlisub {\isadigit{2}}\ {\isacharcolon}\ A\isactrlisup {\isacharminus}\ {\isasymUp}\ $\bar{\bar{O}}$}\ {\normalsize \,and\,}\ \mbox{{\isasymGamma}\ {\isasymturnstile}\isactrlisub {\isasymSigma}\ M\isactrlisub {\isadigit{1}}\ {\isacharcolon}\ A}\ {\normalsize \,and\,}\ \mbox{{\isasymGamma}\ {\isasymturnstile}\isactrlisub {\isasymSigma}\ M\isactrlisub {\isadigit{2}}\ {\isacharcolon}\ A}\ {\normalsize\linebreak[0] \,then\,\linebreak[0]}\ \mbox{{\isasymexists}N{\isachardot}\ N\ {\isasymUp}\ $\bar{\bar{O}}$\ \textrm{and\linebreak[1]}\ {\isasymGamma}\ {\isasymturnstile}\isactrlisub {\isasymSigma}\ N\ {\isacharcolon}\ A\ \textrm{and\linebreak[1]}\ {\isasymGamma}\ {\isasymturnstile}\isactrlisub {\isasymSigma}\ M\isactrlisub {\isadigit{1}}\ {\isacharequal}\ N\ {\isacharcolon}\ A\ \textrm{and\linebreak[1]}\ {\isasymGamma}\ {\isasymturnstile}\isactrlisub {\isasymSigma}\ M\isactrlisub {\isadigit{2}}\ {\isacharequal}\ N\ {\isacharcolon}\ A{\isachardot}}}
    \item \isa{{\normalsize{}If\,}\ {\isasymGamma}\isactrlisup {\isacharminus}\ {\isasymturnstile}\isactrlbsub {\isasymSigma}\isactrlisup {\isacharminus}\isactrlesub \ M\isactrlisub {\isadigit{1}}\ {\isasymleftrightarrow}\ M\isactrlisub {\isadigit{2}}\ {\isacharcolon}\ {\isasymtau}\ {\isasymdown}\ $\bar{O}$\ {\normalsize \,and\,}\ {\isasymGamma}\ {\isasymturnstile}\isactrlisub {\isasymSigma}\ M\isactrlisub {\isadigit{1}}\ {\isacharcolon}\ A\isactrlisub {\isadigit{1}}\ {\normalsize \,and\,}\ {\isasymGamma}\ {\isasymturnstile}\isactrlisub {\isasymSigma}\ M\isactrlisub {\isadigit{2}}\ {\isacharcolon}\ A\isactrlisub {\isadigit{2}}\ {\normalsize \,then\,}\ {\isasymGamma}\ {\isasymturnstile}\isactrlisub {\isasymSigma}\ A\isactrlisub {\isadigit{1}}\ {\isacharequal}\ A\isactrlisub {\isadigit{2}}\ {\isacharcolon}\ type\ \textrm{and\linebreak[1]}\ A\isactrlisub {\isadigit{1}}\isactrlisup {\isacharminus}\ {\isacharequal}\ {\isasymtau}\ \textrm{and\linebreak[1]}\ A\isactrlisub {\isadigit{2}}\isactrlisup {\isacharminus}\ {\isacharequal}\ {\isasymtau}\ \textrm{and\linebreak[1]}\ {\isacharparenleft}{\isasymexists}N{\isachardot}\ N\ {\isasymdown}\ $\bar{O}$\ \textrm{and\linebreak[1]}\ {\isasymGamma}\ {\isasymturnstile}\isactrlisub {\isasymSigma}\ N\ {\isacharcolon}\ A\isactrlisub {\isadigit{1}}\ \textrm{and\linebreak[1]}\ {\isasymGamma}\ {\isasymturnstile}\isactrlisub {\isasymSigma}\ M\isactrlisub {\isadigit{1}}\ {\isacharequal}\ N\ {\isacharcolon}\ A\isactrlisub {\isadigit{1}}\ \textrm{and\linebreak[1]}\ {\isasymGamma}\ {\isasymturnstile}\isactrlisub {\isasymSigma}\ M\isactrlisub {\isadigit{2}}\ {\isacharequal}\ N\ {\isacharcolon}\ A\isactrlisub {\isadigit{2}}{\isacharparenright}{\isachardot}}
    \end{compactenum}
  \end{theorem}%
\end{isamarkuptext}%
\isamarkuptrue%
\isamarkupsubsection{Adequacy%
}
\isamarkuptrue%
\begin{isamarkuptext}%
\labelSec{adequacy}%
\end{isamarkuptext}%
\isamarkuptrue%
\begin{isamarkuptext}%
Conventionally, adequacy is the property that the terms of the
  object language are in a bijective correspondence with the
  well-formed LF terms of a given type, modulo LF equality.  Moreover,
  the bijection should be \emph{compositional}\footnote{This term is
    used in \HP without being defined, but this is the definition used
    in other articles which discuss
    adequacy, for example~\cite{HarperHonsellPlotkin93,Pfenning01}.} in the sense
  that substitution for the object language is preserved and reflected
  by substitution in LF.  The exact statement of the adequacy theorem
  for a given language depends on the language and its definition of
  substitution.  To illustrate how quasicanonical forms could be used
  for reasoning about adequacy, \HP introduces a small example
  language of first-order terms \isa{t} and formulas \isa{{\isasymphi}}, similar to the following:

  \begin{center}
  \isa{t{\isacharcomma}u} ::= \isa{x} $\mid$ \isa{f{\isacharparenleft}t{\isacharcomma}u{\isacharparenright}} $\quad$
  \isa{{\isasymphi}{\isacharcomma}{\isasympsi}} ::= \isa{t\ {\isacharequal}\ u} $\mid$ \isa{{\isasymphi}\ {\isasymand}\ {\isasympsi}} $\mid$ \isa{{\isasymforall}x{\isachardot}{\isasymphi}}
  \end{center}

  \noindent
  along with an appropriate LF signature \isa{{\isasymSigma}\isactrlisub F\isactrlisub O} with types \isa{{\isasymiota}} for
  first-order terms, \isa{o} for first-order formulas, and
  constants

  \begin{center}
  \begin{tabular}{rclcrcl}
  \isa{c\isactrlisub f} &:& \isa{{\isasymiota}\ {\isasymrightarrow}\ {\isasymiota}\ {\isasymrightarrow}\ {\isasymiota}} &&
  \isa{c\isactrlisub {\isacharequal}}  &:& \isa{{\isasymiota}\ {\isasymrightarrow}\ {\isasymiota}\ {\isasymrightarrow}\ o} \\
  \isa{c\isactrlisub {\isasymand}} &:& \isa{o\ {\isasymrightarrow}\ o\ {\isasymrightarrow}\ o} &&
  \isa{c\isactrlisub {\isasymforall}}  &:& \isa{{\isacharparenleft}{\isasymiota}\ {\isasymrightarrow}\ o{\isacharparenright}\ {\isasymrightarrow}\ o}\;.
  \end{tabular}
  \end{center}  

  \noindent
  \HP then defines translation judgments \isa{{\isasymGamma}\ {\isasymturnstile}\ t\ {\isasymtranslates}\ M\ {\isacharcolon}\ {\isasymiota}} and \isa{{\isasymGamma}\ {\isasymturnstile}\ {\isasymphi}\ {\isasymtranslates}\ M\ {\isacharcolon}\ o}
  relating LF terms \isa{M} with first-order terms and formulas
  \isa{t\ {\isacharcolon}\ {\isasymiota}} and \isa{{\isasymphi}\ {\isacharcolon}\ o}.  Note that unlike
  most other judgments in this article, the translations are
  \emph{not} implicitly parametrized by a signature \isa{{\isasymSigma}}
  since they only refer to constants from the fixed signature \isa{{\isasymSigma}\isactrlisub F\isactrlisub O}.  The rules for the translation are
  shown in \refFig{adequacy}.

  %%%%%%%%%%%%%%%%%%%%%%%%%%%%%%%%%%%%%%%%%%%%%%%%%%%%%%%%%%%%%%%%%%
  \begin{figure}[tb]
      \fbox{\isa{{\isasymGamma}\ {\isasymturnstile}\ t\ {\isasymtranslates}\ $\bar{\bar{M}}$\ {\isacharcolon}\ {\isasymiota}}}
       \begin{center}
         \begin{tabular}{@ {}c@ {}}
           \isa{\mbox{}\inferrule{\mbox{{\isacharparenleft}x{\isacharcomma}\ {\isasymiota}{\isacharparenright}\ {\isasymin}\ {\isasymGamma}}}{\mbox{{\isasymGamma}\ {\isasymturnstile}\ x\ {\isasymtranslates}\ x\ {\isacharcolon}\ {\isasymiota}}}} \quad \isa{\mbox{}\inferrule{\mbox{{\isasymGamma}\ {\isasymturnstile}\ t\isactrlisub {\isadigit{1}}\ {\isasymtranslates}\ $\bar{\bar{M}}_1$\ {\isacharcolon}\ {\isasymiota}}\\\ \mbox{{\isasymGamma}\ {\isasymturnstile}\ t\isactrlisub {\isadigit{2}}\ {\isasymtranslates}\ $\bar{\bar{M}}_2$\ {\isacharcolon}\ {\isasymiota}}}{\mbox{{\isasymGamma}\ {\isasymturnstile}\ f{\isacharparenleft}t\isactrlisub {\isadigit{1}}{\isacharcomma}t\isactrlisub {\isadigit{2}}{\isacharparenright}\ {\isasymtranslates}\ c\isactrlisub f\ $\bar{\bar{M}}_1$\ $\bar{\bar{M}}_2$\ {\isacharcolon}\ {\isasymiota}}}}
         \end{tabular}
       \end{center}
       \fbox{\isa{{\isasymGamma}\ {\isasymturnstile}\ {\isasymphi}\ {\isasymtranslates}\ $\bar{\bar{M}}$\ {\isacharcolon}\ o}}
       \begin{center}
        \begin{tabular}{@ {}c@ {}}
          \isa{\mbox{}\inferrule{\mbox{{\isasymGamma}\ {\isasymturnstile}\ t\isactrlisub {\isadigit{1}}\ {\isasymtranslates}\ $\bar{\bar{M}}_1$\ {\isacharcolon}\ {\isasymiota}}\\\ \mbox{{\isasymGamma}\ {\isasymturnstile}\ t\isactrlisub {\isadigit{2}}\ {\isasymtranslates}\ $\bar{\bar{M}}_2$\ {\isacharcolon}\ {\isasymiota}}}{\mbox{{\isasymGamma}\ {\isasymturnstile}\ t\isactrlisub {\isadigit{1}}\ {\isacharequal}\ t\isactrlisub {\isadigit{2}}\ {\isasymtranslates}\ c\isactrlisub {\isacharequal}\ $\bar{\bar{M}}_1$\ $\bar{\bar{M}}_2$\ {\isacharcolon}\ o}}} 
          \quad 
          \isa{\mbox{}\inferrule{\mbox{{\isasymGamma}\ {\isasymturnstile}\ {\isasymphi}\isactrlisub {\isadigit{1}}\ {\isasymtranslates}\ $\bar{\bar{M}}_1$\ {\isacharcolon}\ o}\\\ \mbox{{\isasymGamma}\ {\isasymturnstile}\ {\isasymphi}\isactrlisub {\isadigit{2}}\ {\isasymtranslates}\ $\bar{\bar{M}}_2$\ {\isacharcolon}\ o}}{\mbox{{\isasymGamma}\ {\isasymturnstile}\ {\isasymphi}\isactrlisub {\isadigit{1}}\ {\isasymand}\ {\isasymphi}\isactrlisub {\isadigit{2}}\ {\isasymtranslates}\ c\isactrlisub {\isasymand}\ $\bar{\bar{M}}_1$\ $\bar{\bar{M}}_2$\ {\isacharcolon}\ o}}}
          \smallskip\\
          \isa{\mbox{}\inferrule{\mbox{{\isacharparenleft}x{\isacharcomma}\ {\isasymiota}{\isacharparenright}{\isacharcolon}{\isacharcolon}{\isasymGamma}\ {\isasymturnstile}\ {\isasymphi}\ {\isasymtranslates}\ $\bar{\bar{M}}$\ {\isacharcolon}\ o}\\\ \mbox{x\ {\isasymsharp}\ {\isasymGamma}}}{\mbox{{\isasymGamma}\ {\isasymturnstile}\ {\isasymforall}x{\isachardot}{\isasymphi}\ {\isasymtranslates}\ c\isactrlisub {\isasymforall}\ {\isasymlambda}x{\isachardot}$\bar{\bar{M}}$\ {\isacharcolon}\ o}}}
        \end{tabular}
      \end{center}
   \caption{Adequacy translation}\labelFig{adequacy}
   \end{figure}
   %%%%%%%%%%%%%%%%%%%%%%%%%%%%%%%%%%%%%%%%%%%%%%%%%%%%%%%%%%%%%%%%%%

  Harper and Pfenning then formulate the adequacy property for this
  language in their Theorem 7.2 as follows:

  \begin{theorem}
    [(Adequacy for syntax of first-order logic)] Let $\Gamma$ be a
    context of the form $x_1:\iota,\ldots,x_n:\iota$ for some $n \geq
    0$.
    \begin{compactenum}
    \item The relation \isa{{\isasymGamma}\ {\isasymturnstile}\ t\ {\isasymtranslates}\ $\bar{\bar{M}}$\ {\isacharcolon}\ {\isasymiota}} is a compositional
      bijection between terms $t$ of first-order logic over variables
      $x_1,\ldots,x_n$ and quasi-canonical forms \isa{$\bar{\bar{M}}$} of type $\iota$
      relative to $\Gamma$.
    \item The relation \isa{{\isasymGamma}\ {\isasymturnstile}\ {\isasymphi}\ {\isasymtranslates}\ $\bar{\bar{M}}$\ {\isacharcolon}\ o} is a compositional bijection between
      formulas $t$ of first-order logic over variables $x_1,\ldots,x_n$
      and quasi-canonical forms \isa{$\bar{\bar{M}}$} of type $o$ relative to $\Gamma$.
    \end{compactenum}
  \end{theorem}

  Their proof sketch involves first showing that (for all appropriate
  $\Gamma$) the translations are bijections, and then proving
  compositionality by induction over the structure of terms and
  formulas.  

  Unfortunately, the statement of this theorem is ambiguous or at
  least incomplete.  The reason is that Harper and Pfenning do not
  explicitly define what it means for a bijection to be
  compositional.  Even assuming the standard definition of
  compositionality as substitution preservation, \HP did not provide a
  definition of substitution for quasicanonical forms.

  If we wish to substitute a quasicanonical form for a variable $y$ in
  another quasicanonical form, the result is not always
  quasicanonical.  For example, if we substitute $\lambda x.M$ for $y$
  in $y~N$, we get $(\lambda x.M)~N$, which is not quasicanonical.
  This illustrates that quasicanonical forms are not closed under
  substitution of quasicanonical forms for variables, because
  variables are quasiatomic forms and substituting a
  $\lambda$-expression for a variable may introduce $\beta$-redexes.

  It has been observed elsewhere (apparently first
  by~\citeN{watkins03}) that substitution can be defined for
  well-formed quasicanonical expressions in a \emph{hereditary} way
  that recursively renormalizes any $\beta$-redexes introduced by
  substitution.  \citeN{HarperLicata07} have shown
  how this idea can be used as the basis for a variant of LF called
  \emph{Canonical LF} in which all expressions are maintained in
  canonical form.

  In our initial formalization (reported
  in~\cite{UrbanCheneyBerghofer08}) we misinterpreted the definition
  of the translation slightly by defining the adequacy translations to
  relate first-order terms and formulas to \emph{quasiatomic} forms.
  It is easy to define substitution of quasiatomic forms for variables
  since no reduction can be introduced in doing so.  Consequently, we
  proved a variant of \HP's Theorem 7.2 with the word ``quasicanonical''
  replaced by ``quasiatomic''.  However, even with this modification,
  the formal proof is not as easy as the sketch in \HP suggests; for
  example, we needed to prove weakening, exchange, and substitution
  lemmas for the translation judgment in order to establish
  compositionality.

  After we discovered and corrected the mismatch between our
  definition and the original translation, we were still able to prove
  that the translations are bijections.  To establish
  compositionality, we also formalized hereditary substitution
  (using a simple form of Harper and Licata's definition) and showed
  that the translation maps object-language substitution to hereditary
  substitution.

  Formalizing \HP's Theorem 7.2 thus appears to require either
  changing their translation or introducing hereditary substitution, a
  nontrivial concept that was not mentioned in \HP.  The Canonical LF
  approach now appears to be the preferred starting point for research
  on extensions to LF.  Developing a full and satisfying formalization
  of hereditary substitutions and adequacy properties (and relating
  \HP's version of LF to Harper and Licata's development of Canonical
  LF~\citeyear{HarperLicata07}) would be a significant independent
  undertaking.  Therefore, we prefer to leave further study of
  adequacy based on hereditary substitution for future work.

\end{isamarkuptext}%
\isamarkuptrue%
\isamarkupsection{Code Generation%
}
\isamarkuptrue%
\begin{isamarkuptext}%
\labelSec{locally-nameless}
  Since type checking in LF can be part of the trusted code base of
  proof-carrying code, \citeN{appel03trustworthy} were very careful to
  implement it as cleanly as possible and in as few lines of code as
  possible. Their motivation was that a small and clean implementation
  can be manually inspected and hence can be made robust against, for
  example, Thompson-style attacks \cite{Thompson84}. For this they
  explicitly set out to minimize the number of library functions they
  have to trust in order for their implementation to be correct.
  However, they relied upon the correctness of the type-checking
  algorithm in \HP.
 
  In this paper we have formally proved that both the equivalence
  checking and type-checking algorithms from \HP are sound and
  complete. Consequently, we can remove this aspect from our ``trusted
  code base''. In this section we show how to obtain a verified
  executable ML-implementation of the type-checking algorithm from our
  proof of correctness.

  Isabelle/HOL contains a code generator implemented by
  \citeN{BerghoferNipkow02} which can translate inductive definitions
  into executable pure ML-code automatically. To be able to use this
  code generator, however, we need to invest some further work. The
  present version of this code generator can only deal with rules
  involving datatypes, not \emph{nominal} datatypes.  To surmount this
  problem we translate our nominal representation of kinds, types and
  terms into a locally nameless representation
  \cite{McKinnaP99,AydemirEtAl08}, which can be implemented as an
  ordinary Isabelle/HOL datatype. For the LF-syntax this gives rise
  to the definition:

  \begin{center}
  \begin{tabular}{lrl}
    \textit{Locally Nameless Kinds} &::=& \isa{type} $\mid$ \isa{{\isasymPi}A{\isachardot}\ K}\\
    \textit{Locally Nameless Types} &::=& \isa{a} 
  $\mid$ \isa{{\isasymPi}A\isactrlisub {\isadigit{1}}{\isachardot}\ A\isactrlisub {\isadigit{2}}} $\mid$ \isa{A\ M}\\
    \textit{Locally Nameless Objects} &::=& \isa{c} $\mid$ \isa{x} $\mid$ \isa{n} $\mid$ 
                      \isa{{\isasymlambda}A{\isachardot}\ M} $\mid$ \isa{M\isactrlisub {\isadigit{1}}\ M\isactrlisub {\isadigit{2}}}
  \end{tabular}
  \end{center}

  \noindent
  where terms contain de Bruijn indices $n$ for bound
  variables~\cite{debruijn72im}.  In comparison with ``pure'' de
  Bruijn representations, in the locally nameless representation free
  variables still have names.  This means we can continue using our
  implementation of signatures and contexts in judgments. With a
  ``pure'' de Bruijn representation, contexts would need to be
  referenced by numbers and positions.

  While the locally nameless representation is straightforward to
  implement in Isabelle/HOL, the translations between the nominal and
  locally nameless representation involve quite a lot of formalisation
  work. First we have to define a well-formedness predicate that
  ensures that there are no loose de Bruijn indices. We also need
  three substitution operations, namely substituting (well-formed)
  terms for free variables, written $(-)$\isa{{\isacharbrackleft}x\ {\isacharcolon}{\isacharequal}\ M{\isacharbrackright}},
  substituting terms for de Bruijn indices, written $(-)$\isa{{\isacharbrackleft}n\ {\isacharcolon}{\isacharequal}\ M{\isacharbrackright}}, and substituting de Bruijn indices for variables, written
  $(-)$\isa{{\isacharbrackleft}x\ {\isacharcolon}{\isacharequal}\ n{\isacharbrackright}}. In the latter we have to increase the de
  Bruijn index whenever the substitution moves under a binder. Also
  the translation functions between the nominal and locally nameless
  representations are non-trivial to define. In one
  direction the translation is a partial function and only total over
  well-formed locally nameless terms. In the other direction we use a
  translation depending on an explicit list of variables. The idea is
  to push a variable onto the list whenever the translation goes under
  a \isa{{\isasymlambda}}- or a \isa{{\isasymPi}}-abstraction. Now the de Bruijn index
  for a variable occurrence is the position of the variable in this list. The
  translation, written \isa{{\isacharbar}$-${\isacharbar}\isactrlbsub xs\isactrlesub }, can be formally
  defined as

  \begin{center}
  \begin{tabular}{r@ {\hspace{2mm}}c@ {\hspace{2mm}}ll}
  \isa{{\isacharbar}type{\isacharbar}\isactrlbsub xs\isactrlesub } & $=$ & \isa{type}\\
  \isa{{\isacharbar}{\isasymPi}x{\isacharcolon}A{\isachardot}\ K{\isacharbar}\isactrlbsub xs\isactrlesub } & $=$ & \isa{{\isasymPi}{\isacharbar}A{\isacharbar}\isactrlbsub xs\isactrlesub {\isachardot}\ {\isacharbar}K{\isacharbar}\isactrlbsub {\isacharparenleft}x{\isacharcolon}{\isacharcolon}xs{\isacharparenright}\isactrlesub } & provided \isa{x\ {\isasymsharp}\ xs}\smallskip\\
  \isa{{\isacharbar}a{\isacharbar}\isactrlbsub xs\isactrlesub } & $=$ & \isa{a}\\
  \isa{{\isacharbar}A\ M{\isacharbar}\isactrlbsub xs\isactrlesub } & $=$ & \isa{{\isacharbar}A{\isacharbar}\isactrlbsub xs\isactrlesub \ {\isacharbar}M{\isacharbar}\isactrlbsub xs\isactrlesub }\\
  \isa{{\isacharbar}{\isasymPi}x{\isacharcolon}A\isactrlisub {\isadigit{1}}{\isachardot}\ A\isactrlisub {\isadigit{2}}{\isacharbar}\isactrlbsub xs\isactrlesub } 
     & $=$ & \isa{{\isasymPi}{\isacharbar}A\isactrlisub {\isadigit{1}}{\isacharbar}\isactrlbsub xs\isactrlesub {\isachardot}\ {\isacharbar}A\isactrlisub {\isadigit{2}}{\isacharbar}\isactrlbsub {\isacharparenleft}x{\isacharcolon}{\isacharcolon}xs{\isacharparenright}\isactrlesub } 
  & provided \isa{x\ {\isasymsharp}\ xs}\smallskip\\
  \isa{{\isacharbar}c{\isacharbar}\isactrlbsub xs\isactrlesub } & $=$ & \isa{c}\\
  \isa{{\isacharbar}x{\isacharbar}\isactrlbsub xs\isactrlesub } & $=$ & \isa{index\ x\ xs\ {\isadigit{0}}}\\
  \isa{{\isacharbar}M\ N{\isacharbar}\isactrlbsub xs\isactrlesub } & $=$ & \isa{{\isacharbar}M{\isacharbar}\isactrlbsub xs\isactrlesub \ {\isacharbar}N{\isacharbar}\isactrlbsub xs\isactrlesub }\\
  \isa{{\isacharbar}{\isasymlambda}x{\isacharcolon}A{\isachardot}\ M{\isacharbar}\isactrlbsub xs\isactrlesub } & $=$ & \isa{{\isasymlambda}{\isacharbar}A{\isacharbar}\isactrlbsub xs\isactrlesub {\isachardot}\ {\isacharbar}M{\isacharbar}\isactrlbsub {\isacharparenleft}x{\isacharcolon}{\isacharcolon}xs{\isacharparenright}\isactrlesub } & provided \isa{x\ {\isasymsharp}\ xs}\\
  \end{tabular}
  \end{center}
  
  \noindent
  where the variable case is defined in terms of the auxiliary function 
  \isa{index\ x\ xs\ n}:
  
  \begin{center}
  \begin{tabular}{@ {}l@ {}}
  \isa{index\ x\ {\isacharbrackleft}{\isacharbrackright}\ n\ {\isacharequal}\ x}\\
  \isa{index\ x\ {\isacharparenleft}y{\isacharcolon}{\isacharcolon}ys{\isacharparenright}\ n\ {\isacharequal}\ {\isacharparenleft}\textrm{if}\ x\ {\isacharequal}\ y\ \textrm{then}\ n\ \textrm{else}\ index\ x\ ys\ {\isacharparenleft}Suc\ n{\isacharparenright}{\isacharparenright}}\\
  \end{tabular}  
  \end{center}

  \noindent
  The problem with this definition arises from the fact that inductions need 
  to be appropriately generalised in order to take the potentially growing list 
  of variables into account. This is sometimes easy to do, but sometimes
  we needed a lot of ingenuity to find the right lemmas to get inductions
  through.

  Having translated all our terms into the locally nameless representation,
  we solved the technical problem with the code generator in Isabelle/HOL. 
  However, there is a further problem that
  needs to solved: the algorithms specified so far are not yet
  concrete enough to be translated directly into runnable ML-code. For
  this consider again the algorithmic equivalence rule

  \begin{center}
  \isa{\mbox{}\inferrule{\mbox{{\isacharparenleft}x{\isacharcomma}\ {\isasymtau}\isactrlisub {\isadigit{1}}{\isacharparenright}{\isacharcolon}{\isacharcolon}{\isasymDelta}\ {\isasymturnstile}\isactrlbsub {\isasymSigma}\isactrlesub \ M\ x\ {\isasymLeftrightarrow}\ N\ x\ {\isacharcolon}\ {\isasymtau}\isactrlisub {\isadigit{2}}}\\\ \mbox{x\ {\isasymsharp}\ {\isacharparenleft}{\isasymDelta}{\isacharcomma}\ M{\isacharcomma}\ N{\isacharparenright}}}{\mbox{{\isasymDelta}\ {\isasymturnstile}\isactrlbsub {\isasymSigma}\isactrlesub \ M\ {\isasymLeftrightarrow}\ N\ {\isacharcolon}\ {\isasymtau}\isactrlisub {\isadigit{1}}\ {\isasymrightarrow}\ {\isasymtau}\isactrlisub {\isadigit{2}}}}}
  \end{center}

  \noindent
  from \refFig{alg-equiv}. This rule decides the equivalence between
  the terms \isa{M} and \isa{N} having function type. When read
  bottom-up, it states that we need to introduce a variable \isa{x}
  (any will do) that is fresh for \isa{{\isasymDelta}}, \isa{M} and
  \isa{N}.  ML does not have any built-in facilities for choosing
  such a fresh name (unlike, for example, FreshML by
  \citeN{PittsAM:frepbm}). This means for an ML-implementation of type
  and equivalence checking that we need to make explicit which fresh
  name should be chosen. An obvious choice is to inspect all free
  variables occurring in \isa{{\isasymDelta}}, \isa{M} and \isa{N},
  and produce a variable with a higher index. In our case, it suffices
  to compute the maximum index of all variables in scope and increase
  by one to obtain a fresh variable index. We are able to compute this
  index because names in the Nominal Datatype Package have a natural
  number as index and thus can be ordered.  This allows us to formulate
  algorithmic equivalence rules as follows

  \begin{center}
  \isa{\mbox{}\inferrule{\mbox{{\isacharparenleft}x{\isacharcomma}\ {\isasymtau}\isactrlisub {\isadigit{1}}{\isacharparenright}{\isacharcolon}{\isacharcolon}{\isasymDelta}\ \isactrlbsub ln\isactrlesub {\isasymturnstile}\isactrlbsub {\isasymSigma}\isactrlesub \ M\ x\ {\isasymLeftrightarrow}\ N\ x\ {\isacharcolon}\ {\isasymtau}\isactrlisub {\isadigit{2}}}\\\ \mbox{x\ {\isacharequal}\ maxi\ {\isacharparenleft}fv\ {\isasymDelta}\ {\isacharat}\ fv\ M\ {\isacharat}\ fv\ N{\isacharparenright}}}{\mbox{{\isasymDelta}\ \isactrlbsub ln\isactrlesub {\isasymturnstile}\isactrlbsub {\isasymSigma}\isactrlesub \ M\ {\isasymLeftrightarrow}\ N\ {\isacharcolon}\ {\isasymtau}\isactrlisub {\isadigit{1}}\ {\isasymrightarrow}\ {\isasymtau}\isactrlisub {\isadigit{2}}}}}
  \end{center}

  \noindent
  where \isa{fv} is a polymorphic function producing a list of free
  variables of a term or context,  and the function \isa{maxi} scans
  through a list of variables and returns the highest variable increased by
  one.%
\end{isamarkuptext}%
\isamarkuptrue%
\begin{isamarkuptext}%
%%%%%%%%%%%%%%%%%%%%%%%%%%%%%%%%%%%%%%%%%%%%%%%%%%%%%%%%%%%%%%%%%%
  \begin{figure}[tb]
    %\small

\fbox{\isa{\isactrlbsub ln\isactrlesub {\isasymturnstile}\ {\isasymSigma}\ {\isasymRightarrow}\ sig}}
\begin{center}
  \isa{\mbox{}\inferrule{\mbox{}}{\mbox{\isactrlbsub ln\isactrlesub {\isasymturnstile}\ {\isacharbrackleft}{\isacharbrackright}\ {\isasymRightarrow}\ sig}}}
  \qquad 
  \isa{\mbox{}\inferrule{\mbox{\isactrlbsub ln\isactrlesub {\isasymturnstile}\ {\isasymSigma}\ {\isasymRightarrow}\ sig}\\\ \mbox{{\isacharbrackleft}{\isacharbrackright}\ \isactrlbsub ln\isactrlesub {\isasymturnstile}\isactrlbsub {\isasymSigma}\isactrlesub \ A\ {\isasymRightarrow}\ type}\\\ \mbox{c\ {\isasymnotin}\ fi\ {\isasymSigma}}}{\mbox{\isactrlbsub ln\isactrlesub {\isasymturnstile}\ {\isacharparenleft}c{\isacharcomma}\ A{\isacharparenright}{\isacharcolon}{\isacharcolon}{\isasymSigma}\ {\isasymRightarrow}\ sig}}}
  \smallskip\\
  \isa{\mbox{}\inferrule{\mbox{\isactrlbsub ln\isactrlesub {\isasymturnstile}\ {\isasymSigma}\ {\isasymRightarrow}\ sig}\\\ \mbox{{\isacharbrackleft}{\isacharbrackright}\ \isactrlbsub ln\isactrlesub {\isasymturnstile}\isactrlbsub {\isasymSigma}\isactrlesub \ K\ {\isasymRightarrow}\ kind}\\\ \mbox{a\ {\isasymnotin}\ fi\ {\isasymSigma}}}{\mbox{\isactrlbsub ln\isactrlesub {\isasymturnstile}\ {\isacharparenleft}a{\isacharcomma}\ K{\isacharparenright}{\isacharcolon}{\isacharcolon}{\isasymSigma}\ {\isasymRightarrow}\ sig}}}
\end{center}

 \fbox{\isa{\isactrlbsub ln\isactrlesub {\isasymturnstile}\isactrlbsub {\isasymSigma}\isactrlesub \ {\isasymGamma}\ {\isasymRightarrow}\ ctx}}

 \begin{center}
   \isa{\mbox{}\inferrule{\mbox{\isactrlbsub ln\isactrlesub {\isasymturnstile}\ {\isasymSigma}\ {\isasymRightarrow}\ sig}}{\mbox{\isactrlbsub ln\isactrlesub {\isasymturnstile}\isactrlbsub {\isasymSigma}\isactrlesub \ {\isacharbrackleft}{\isacharbrackright}\ {\isasymRightarrow}\ ctx}}} \qquad \isa{\mbox{}\inferrule{\mbox{\isactrlbsub ln\isactrlesub {\isasymturnstile}\isactrlbsub {\isasymSigma}\isactrlesub \ {\isasymGamma}\ {\isasymRightarrow}\ ctx}\\\ \mbox{{\isasymGamma}\ \isactrlbsub ln\isactrlesub {\isasymturnstile}\isactrlbsub {\isasymSigma}\isactrlesub \ A\ {\isasymRightarrow}\ type}\\\ \mbox{x\ {\isasymnotin}\ fv\ {\isasymGamma}}}{\mbox{\isactrlbsub ln\isactrlesub {\isasymturnstile}\isactrlbsub {\isasymSigma}\isactrlesub \ {\isacharparenleft}x{\isacharcomma}\ A{\isacharparenright}{\isacharcolon}{\isacharcolon}{\isasymGamma}\ {\isasymRightarrow}\ ctx}}}

 \end{center} 

\fbox{\isa{{\isasymGamma}\ \isactrlbsub ln\isactrlesub {\isasymturnstile}\isactrlbsub {\isasymSigma}\isactrlesub \ M\ {\isasymRightarrow}\ A}}
  
\begin{center}
  \isa{\mbox{}\inferrule{\mbox{\isactrlbsub ln\isactrlesub {\isasymturnstile}\isactrlbsub {\isasymSigma}\isactrlesub \ {\isasymGamma}\ {\isasymRightarrow}\ ctx}\\\ \mbox{{\isacharparenleft}x{\isacharcomma}\ A{\isacharparenright}\ {\isasymin}\ {\isasymGamma}}}{\mbox{{\isasymGamma}\ \isactrlbsub ln\isactrlesub {\isasymturnstile}\isactrlbsub {\isasymSigma}\isactrlesub \ x\ {\isasymRightarrow}\ A}}} \quad 
  \isa{\mbox{}\inferrule{\mbox{\isactrlbsub ln\isactrlesub {\isasymturnstile}\isactrlbsub {\isasymSigma}\isactrlesub \ {\isasymGamma}\ {\isasymRightarrow}\ ctx}\\\ \mbox{{\isacharparenleft}c{\isacharcomma}\ A{\isacharparenright}\ {\isasymin}\ {\isasymSigma}}}{\mbox{{\isasymGamma}\ \isactrlbsub ln\isactrlesub {\isasymturnstile}\isactrlbsub {\isasymSigma}\isactrlesub \ c\ {\isasymRightarrow}\ A}}} 
  \smallskip\\
  
  \isa{\mbox{}\inferrule{\mbox{{\isasymGamma}\ \isactrlbsub ln\isactrlesub {\isasymturnstile}\isactrlbsub {\isasymSigma}\isactrlesub \ M\isactrlisub {\isadigit{1}}\ {\isasymRightarrow}\ {\isasymPi}A\isactrlisub {\isadigit{2}}{\isacharprime}{\isachardot}\ A\isactrlisub {\isadigit{1}}}\\\ \mbox{{\isasymGamma}\ \isactrlbsub ln\isactrlesub {\isasymturnstile}\isactrlbsub {\isasymSigma}\isactrlesub \ M\isactrlisub {\isadigit{2}}\ {\isasymRightarrow}\ A\isactrlisub {\isadigit{2}}}\\\ \mbox{{\isasymGamma}\isactrlisup {\isacharminus}\ \isactrlbsub ln\isactrlesub {\isasymturnstile}\isactrlbsub {\isasymSigma}\isactrlisup {\isacharminus}\isactrlesub \ A\isactrlisub {\isadigit{2}}\ {\isasymLeftrightarrow}\ A\isactrlisub {\isadigit{2}}{\isacharprime}\ {\isacharcolon}\ type\isactrlisup {\isacharminus}}}{\mbox{{\isasymGamma}\ \isactrlbsub ln\isactrlesub {\isasymturnstile}\isactrlbsub {\isasymSigma}\isactrlesub \ M\isactrlisub {\isadigit{1}}\ M\isactrlisub {\isadigit{2}}\ {\isasymRightarrow}\ A\isactrlisub {\isadigit{1}}{\isacharbrackleft}{\isadigit{0}}\ {\isacharcolon}{\isacharequal}\ M\isactrlisub {\isadigit{2}}{\isacharbrackright}}}}
  \smallskip\\
  \isa{\mbox{}\inferrule{\mbox{{\isasymGamma}\ \isactrlbsub ln\isactrlesub {\isasymturnstile}\isactrlbsub {\isasymSigma}\isactrlesub \ A\isactrlisub {\isadigit{1}}\ {\isasymRightarrow}\ type}\\\ \mbox{{\isacharparenleft}x{\isacharcomma}\ A\isactrlisub {\isadigit{1}}{\isacharparenright}{\isacharcolon}{\isacharcolon}{\isasymGamma}\ \isactrlbsub ln\isactrlesub {\isasymturnstile}\isactrlbsub {\isasymSigma}\isactrlesub \ M\isactrlisub {\isadigit{2}}{\isacharbrackleft}{\isadigit{0}}\ {\isacharcolon}{\isacharequal}\ x{\isacharbrackright}\ {\isasymRightarrow}\ A\isactrlisub {\isadigit{2}}}\\\ \mbox{x\ {\isacharequal}\ maxi\ {\isacharparenleft}fv\ {\isasymGamma}\ {\isacharat}\ fv\ M\isactrlisub {\isadigit{2}}\ {\isacharat}\ fv\ A\isactrlisub {\isadigit{1}}{\isacharparenright}}\\\ \mbox{A\isactrlisub {\isadigit{2}}{\isacharprime}\ {\isacharequal}\ A\isactrlisub {\isadigit{2}}{\isacharbrackleft}x\ {\isacharcolon}{\isacharequal}\ {\isadigit{0}}{\isacharbrackright}}}{\mbox{{\isasymGamma}\ \isactrlbsub ln\isactrlesub {\isasymturnstile}\isactrlbsub {\isasymSigma}\isactrlesub \ {\isasymlambda}A\isactrlisub {\isadigit{1}}{\isachardot}\ M\isactrlisub {\isadigit{2}}\ {\isasymRightarrow}\ {\isasymPi}A\isactrlisub {\isadigit{1}}{\isachardot}\ A\isactrlisub {\isadigit{2}}{\isacharprime}}}}
\end{center}

\fbox{\isa{{\isasymGamma}\ \isactrlbsub ln\isactrlesub {\isasymturnstile}\isactrlbsub {\isasymSigma}\isactrlesub \ A\ {\isasymRightarrow}\ K}}
  
\begin{center}
  \isa{\mbox{}\inferrule{\mbox{\isactrlbsub ln\isactrlesub {\isasymturnstile}\isactrlbsub {\isasymSigma}\isactrlesub \ {\isasymGamma}\ {\isasymRightarrow}\ ctx}\\\ \mbox{{\isacharparenleft}a{\isacharcomma}\ K{\isacharparenright}\ {\isasymin}\ {\isasymSigma}}}{\mbox{{\isasymGamma}\ \isactrlbsub ln\isactrlesub {\isasymturnstile}\isactrlbsub {\isasymSigma}\isactrlesub \ a\ {\isasymRightarrow}\ K}}} 
  \smallskip\\ 
  \isa{\mbox{}\inferrule{\mbox{{\isasymGamma}\ \isactrlbsub ln\isactrlesub {\isasymturnstile}\isactrlbsub {\isasymSigma}\isactrlesub \ A\ {\isasymRightarrow}\ {\isasymPi}A\isactrlisub {\isadigit{2}}{\isacharprime}{\isachardot}\ K\isactrlisub {\isadigit{1}}}\\\ \mbox{{\isasymGamma}\ \isactrlbsub ln\isactrlesub {\isasymturnstile}\isactrlbsub {\isasymSigma}\isactrlesub \ M\ {\isasymRightarrow}\ A\isactrlisub {\isadigit{2}}}\\\ \mbox{{\isasymGamma}\isactrlisup {\isacharminus}\ \isactrlbsub ln\isactrlesub {\isasymturnstile}\isactrlbsub {\isasymSigma}\isactrlisup {\isacharminus}\isactrlesub \ A\isactrlisub {\isadigit{2}}\ {\isasymLeftrightarrow}\ A\isactrlisub {\isadigit{2}}{\isacharprime}\ {\isacharcolon}\ type\isactrlisup {\isacharminus}}}{\mbox{{\isasymGamma}\ \isactrlbsub ln\isactrlesub {\isasymturnstile}\isactrlbsub {\isasymSigma}\isactrlesub \ A\ M\ {\isasymRightarrow}\ K\isactrlisub {\isadigit{1}}{\isacharbrackleft}{\isadigit{0}}\ {\isacharcolon}{\isacharequal}\ M{\isacharbrackright}}}}\smallskip\\
   
  \isa{\mbox{}\inferrule{\mbox{{\isasymGamma}\ \isactrlbsub ln\isactrlesub {\isasymturnstile}\isactrlbsub {\isasymSigma}\isactrlesub \ A\isactrlisub {\isadigit{1}}\ {\isasymRightarrow}\ type}\\\ \mbox{{\isacharparenleft}x{\isacharcomma}\ A\isactrlisub {\isadigit{1}}{\isacharparenright}{\isacharcolon}{\isacharcolon}{\isasymGamma}\ \isactrlbsub ln\isactrlesub {\isasymturnstile}\isactrlbsub {\isasymSigma}\isactrlesub \ A\isactrlisub {\isadigit{2}}{\isacharbrackleft}{\isadigit{0}}\ {\isacharcolon}{\isacharequal}\ x{\isacharbrackright}\ {\isasymRightarrow}\ type}\\\ \mbox{x\ {\isacharequal}\ maxi\ {\isacharparenleft}fv\ {\isasymGamma}\ {\isacharat}\ fv\ A\isactrlisub {\isadigit{1}}\ {\isacharat}\ fv\ A\isactrlisub {\isadigit{2}}{\isacharparenright}}}{\mbox{{\isasymGamma}\ \isactrlbsub ln\isactrlesub {\isasymturnstile}\isactrlbsub {\isasymSigma}\isactrlesub \ {\isasymPi}A\isactrlisub {\isadigit{1}}{\isachardot}\ A\isactrlisub {\isadigit{2}}\ {\isasymRightarrow}\ type}}}
\end{center}

\fbox{\isa{{\isasymGamma}\ \isactrlbsub ln\isactrlesub {\isasymturnstile}\isactrlbsub {\isasymSigma}\isactrlesub \ K\ {\isasymRightarrow}\ kind}}

\begin{center}
  \isa{\mbox{}\inferrule{\mbox{\isactrlbsub ln\isactrlesub {\isasymturnstile}\isactrlbsub {\isasymSigma}\isactrlesub \ {\isasymGamma}\ {\isasymRightarrow}\ ctx}}{\mbox{{\isasymGamma}\ \isactrlbsub ln\isactrlesub {\isasymturnstile}\isactrlbsub {\isasymSigma}\isactrlesub \ type\ {\isasymRightarrow}\ kind}}} \smallskip\\

  \isa{\mbox{}\inferrule{\mbox{{\isasymGamma}\ \isactrlbsub ln\isactrlesub {\isasymturnstile}\isactrlbsub {\isasymSigma}\isactrlesub \ A\ {\isasymRightarrow}\ type}\\\ \mbox{{\isacharparenleft}x{\isacharcomma}\ A{\isacharparenright}{\isacharcolon}{\isacharcolon}{\isasymGamma}\ \isactrlbsub ln\isactrlesub {\isasymturnstile}\isactrlbsub {\isasymSigma}\isactrlesub \ K{\isacharbrackleft}{\isadigit{0}}\ {\isacharcolon}{\isacharequal}\ x{\isacharbrackright}\ {\isasymRightarrow}\ kind}\\\ \mbox{x\ {\isacharequal}\ maxi\ {\isacharparenleft}fv\ {\isasymGamma}\ {\isacharat}\ fv\ A\ {\isacharat}\ fv\ K{\isacharparenright}}}{\mbox{{\isasymGamma}\ \isactrlbsub ln\isactrlesub {\isasymturnstile}\isactrlbsub {\isasymSigma}\isactrlesub \ {\isasymPi}A{\isachardot}\ K\ {\isasymRightarrow}\ kind}}}
\end{center}
  \caption{Algorithmic typechecking rules used for generating executable code.}\labelFig{lf-lalg-tc}
  \end{figure}
  %%%%%%%%%%%%%%%%%%%%%%%%%%%%%%%%%%%%%%%%%%%%%%%%%%%%%%%%%%%%%%%%%%%
\end{isamarkuptext}%
\isamarkuptrue%
\begin{isamarkuptext}%
In \refFig{lf-lalg-tc} we show the 
  rules for type checking in the locally nameless representation and with the
  explicit choice of fresh variables.  The locally nameless variants
  of these judgments are marked by the subscript $ln$.  We omit the
  locally nameless versions of the algorithmic equivalence rules but
  they are similar.  The functions \mbox{\isa{fi\ {\isacharparenleft}$-${\isacharparenright}}} and \isa{fv\ {\isacharparenleft}$-${\isacharparenright}} calculate the free
  identifiers and free variables of their arguments, respectively.

  It is important to note that it would be extremely inconvenient to
  build the concrete choice for a fresh variable into the rules that
  are used in the soundness and completeness proofs described in the
  earlier sections.  The reason is that several of the proofs would
  \emph{not} go through as stated in \HP since the choice is not fresh
  enough for all entities considered in some lemmas (an example is the
  weakening property, where the variable \isa{x} is assumed to be
  not just fresh for \isa{{\isasymDelta}}, \isa{M} and \isa{N}, but
  also for a larger context \isa{{\isasymDelta}{\isacharprime}}). It is however
  relatively straightforward to show the equivalence (i.e., they
  derive the same judgments, modulo translation) between the original
  rules and the rules with the concrete choice for fresh variables. We
  can show:

  \begin{lemma}[(Equivalence)]\mbox{}
    \begin{compactenum}
    \item \isa{{\isasymturnstile}\ {\isasymSigma}\ {\isasymRightarrow}\ sig} if and only if \isa{\isactrlbsub ln\isactrlesub {\isasymturnstile}\ {\isacharbar}{\isasymSigma}{\isacharbar}\ensuremath{\mathrm{{}_{[]}}}\ {\isasymRightarrow}\ sig}.
    \item \isa{{\isasymturnstile}\isactrlisub {\isasymSigma}\ {\isasymGamma}\ {\isasymRightarrow}\ ctx} if and only if \isa{\isactrlbsub ln\isactrlesub {\isasymturnstile}\isactrlbsub {\isacharbar}{\isasymSigma}{\isacharbar}\ensuremath{\mathrm{{}_{[]}}}\isactrlesub \ {\isacharbar}{\isasymGamma}{\isacharbar}\isactrlbsub {\isacharbrackleft}{\isacharbrackright}\isactrlesub \ {\isasymRightarrow}\ ctx}.
    \item \isa{{\isasymGamma}\ {\isasymturnstile}\isactrlisub {\isasymSigma}\ M\ {\isasymRightarrow}\ A} if and only if \isa{{\isacharbar}{\isasymGamma}{\isacharbar}\isactrlbsub {\isacharbrackleft}{\isacharbrackright}\isactrlesub \ \isactrlbsub ln\isactrlesub {\isasymturnstile}\isactrlbsub {\isacharbar}{\isasymSigma}{\isacharbar}\ensuremath{\mathrm{{}_{[]}}}\isactrlesub \ {\isacharbar}M{\isacharbar}\isactrlbsub {\isacharbrackleft}{\isacharbrackright}\isactrlesub \ {\isasymRightarrow}\ {\isacharbar}A{\isacharbar}\isactrlbsub {\isacharbrackleft}{\isacharbrackright}\isactrlesub }.
    \item \isa{{\isasymGamma}\ {\isasymturnstile}\isactrlisub {\isasymSigma}\ A\ {\isasymRightarrow}\ K} if and only if \isa{{\isacharbar}{\isasymGamma}{\isacharbar}\isactrlbsub {\isacharbrackleft}{\isacharbrackright}\isactrlesub \ \isactrlbsub ln\isactrlesub {\isasymturnstile}\isactrlbsub {\isacharbar}{\isasymSigma}{\isacharbar}\ensuremath{\mathrm{{}_{[]}}}\isactrlesub \ {\isacharbar}A{\isacharbar}\isactrlbsub {\isacharbrackleft}{\isacharbrackright}\isactrlesub \ {\isasymRightarrow}\ {\isacharbar}K{\isacharbar}\isactrlbsub {\isacharbrackleft}{\isacharbrackright}\isactrlesub }.
    \item \isa{{\isasymGamma}\ {\isasymturnstile}\isactrlisub {\isasymSigma}\ K\ {\isasymRightarrow}\ kind} if and only if \isa{{\isacharbar}{\isasymGamma}{\isacharbar}\isactrlbsub {\isacharbrackleft}{\isacharbrackright}\isactrlesub \ \isactrlbsub ln\isactrlesub {\isasymturnstile}\isactrlbsub {\isacharbar}{\isasymSigma}{\isacharbar}\ensuremath{\mathrm{{}_{[]}}}\isactrlesub \ {\isacharbar}K{\isacharbar}\isactrlbsub {\isacharbrackleft}{\isacharbrackright}\isactrlesub \ {\isasymRightarrow}\ kind}.
    \end{compactenum}
  \end{lemma}

  \noindent
  From the rules in \refFig{lf-lalg-tc} the code generator of
  Isabelle/HOL can generate ML-code. Of course the correctness of this code
  depends on the correctness of the generator. However it is
  relatively easy to inspect the generated ML-code and we are
  confident that it implements correctly the inductive definitions
  that have been proved to be sound and complete with respect to their
  specification.
  We have used the extracted ML-code to type-check several 
  LF example signatures.%
\end{isamarkuptext}%
\isamarkuptrue%
\isamarkupsection{Discussion%
}
\isamarkuptrue%
\begin{isamarkuptext}%
\labelSec{discussion}

It is difficult to argue objectively about the efficacy or usability
of tools for mechanized metatheory about languages with name-binding,
since there are substantial differences among systems, there are few
experts in the use of more than one system, and each such
formalization is a major undertaking.  Nevertheless, we believe it is
worthwhile to make some subjective observations about our experience
formalizing LF using Nominal Isabelle/HOL, and identify aspects of the two
systems that aided or hindered formalization.

  \paragraph*{Methodological observations}
  The formalization was performed by two of the authors; one is a
  developer of the Nominal Datatype Package and expert Isabelle/HOL
  user and the other had roughly three months' experience with these
  tools prior to starting the formalization.  We estimate that the
  total effort involved in conducting the formalizations in
  \refSec{formalization} was at most three person-months.  We worked
  on the code generation part intermittently and therefore do not have
  detailed information about the time required.  Although there is
  still room for improvement in both Isabelle/HOL and the Nominal
  Datatype Package, our experience suggests that these tools can now
  be used to perform significant formalizations within reasonable
  time-frames, at least by experienced users.

  It took approximately six person-weeks to formalize everything up to
  the soundness proof (including pondering why the omitted case for
  type extensionality did not go through).  However, once Harper and
  Pfenning confirmed that this case was indeed not handled correctly
  in their proof, one of the authors was able to check within 2 hours
  that adding a type-extensionality rule solves the problem.
  Re-checking the proof on paper would have meant reviewing
  approximately 31 pages of proofs.  Subsequently we checked the
  validity of a solution suggested by Harper and found another
  solution for the problem.  As a practical matter, the ability
  to rapidly evaluate the effects of changes to the system was
  essential for finding these solutions and evaluating other
  possibilities.  In a similar formalization project, the first author
  showed that a central lemma in the informal proof in his PhD-thesis
  can be repaired~\cite{Urban08}.

  \paragraph*{Comparing the formalization and informal proof}
  In our formalization, we attempted to follow the syntax, definitions
  and proofs given in \HP as closely as we could, and resisted the
  temptation to change their rules to make our task easier.  We
  found that nominal techniques were usually able to
  state results almost exactly as they are presented on paper; the
  main differences tended to involve freshness or validity
  side-conditions that were left implicit in \HP.  To
  illustrate this point, we have prepared this paper using Isabelle's
  documentation facilities~\cite{IsabelleTutorial}.  Most lemmas,
  theorems, and definitions in this paper have been generated directly from the
  formalization (the main exceptions are the quasidecidability and
  adequacy properties, which are paraphrased).  

  In this article, we have focused on the high-level ideas of the
  formalization and, in the main, downplayed the low-level details of
  proofs using the Nominal Datatype Package.  This is not because
  these details are embarrassing, but because (with a few
  clearly-marked exceptions) they are prosaic.  For example, as one
  would expect, our formalization also required formally proving many
  properties of substitution, swapping, freshness, contexts, erasure
  and so on.  We have not discussed these because they are routine,
  and rely upon techniques already covered in previous work on nominal
  techniques~\cite{Urban08JAR}.  However, we would like to point out
  here that the capability to define functions such as substitution
  and erasure as (nominal) primitive recursive functions, and use
  Isabelle/HOL's built-in simplifier to rewrite formulas involving
  these functions, was absolutely essential.  This is not to say that
  these proofs were always easy, but that difficulties involving
  name-binding were usually not the dominant factor.

  We have not explicitly stated when features such as strong nominal
  induction or inversion
  principles~\cite{UrbanBerghoferNorrish07,BerghoferUrban08} have been
  used (nor have we given complete explanations of these techniques),
  but our formalization relies upon them extensively.  When these
  principles could be used, the paper proof was usually easy to
  translate to a formal proof step-by-step (although as is often the
  case with formalizations, an informal proof step often translated to
  many formal steps or necessitated additional lemmas).  On the other hand, in a few cases nominal induction principles could
  not be applied, often because of subtleties involving binding.  When
  this was the case, proof cases involving binding were often much
  more labor-intensive because they required explicit reasoning about
  choosing fresh names, alpha-equivalence, swapping and substitution
  (see, for example, the proof of transitivity of algorithmic
  equivalence).

  Many of our proofs have been be written to match corresponding
  informal proofs closely using the Isar proof-language, as in an example
  by \citeN[Sec. 6]{Urban08JAR} of a typical substitution
  property.  However, writing readable Isar proofs is labor-intensive;
  the proof-script tactic language of classic Isabelle/HOL tends to be
  much easier to write but harder to read.

  The interested, or skeptical, reader is welcome to consult the formalization for
  these details, replay the proofs of key properties, compare them
  with those in \HP, and form his or her own opinion.

\begin{table}[tb]
  \caption{Summary of the formalization}\label{tab:meaningless-metrics}
\begin{center}\small
  \begin{tabular}{|l|p{4cm}|r|r|r|}
    \hline
    Theory 			& Description 							& Size (bytes)	& Lines	& Lemmas\\
    \hline
    \texttt{LF} 		& Syntax and definitional judgments of LF			& 125,975	& 2,631	& 103	\\
    \texttt{Erasure} 		& Simple types and kinds,  erasure				& 14,860	& 463	& 35	\\
    \texttt{PairOrdering} 	& Pair ordering used for transitivity				& 962		& 29 	& 3	\\
    \texttt{EquivAlg}		& Algorithmic equivalence judgments and properties		& 47,480	& 1,015	& 46	\\
    \texttt{Completeness} 	& Logical relation, completeness proof				& 54,575	& 778	& 22	\\
    \texttt{WeakEquivAlg} 	& Weak algorithmic typechecking					& 9,373		& 219 	& 7	\\
    \texttt{Soundness} 		& Subject reduction, soundness proofs						& 31,235	& 562	& 8	\\
    \texttt{TypeAlg} & Algorithmic typechecking  					& 13,139 & 244 & 5 \\
    \texttt{Decidability} 	& Quasidecidability	& 104,939	& 2,087	& 50	\\
    \texttt{Strengthening} 	& Strengthening and strong extensionality	& 28,940	& 591	& 15	\\
    \texttt{Canonical} 		& Quasicanonical forms			& 27,702	& 556	& 13	\\
    \texttt{Adequacy} 		& Adequacy example			& 29,777	& 736	& 45	\\
    \texttt{LocallyN}           & Translation to locally nameless syntax & 179,148      & 4,674  & 223 \\
\hline
    Total 		& 								& 668,105	& 14,585	& 575   \\

    \hline
  \end{tabular}
\end{center}
\end{table}

\paragraph*{Metrics about the formalization}
In Table~\ref{tab:meaningless-metrics}, we report some simple metrics
about our formalization such as the sizes, number of lines of text,
and number of lemmas in each theory in the main formalization.  As
Table~\ref{tab:meaningless-metrics} shows, the core \texttt{LF} theory
accounts for about 20\% of the development.  These syntactic
properties are mostly straightforward, and their proofs merit only
cursory discussion in \HP, but some lemmas have many cases which must
each be handled individually.  The \texttt{Decidability} theory
accounts for another 15\%; the quasidecidability proofs are verbose
but largely straightforward.  The \texttt{LocallyN} theory proves that
the nominal datatypes version of LF is equivalent to a locally
nameless formulation; this accounts for about 25\% of the development.
The effort involved in this part was therefore quite substantial: it can
be explained by the lack of automatic infrastructure for the locally nameless 
representation of binders in Isabelle/HOL, but also by the inherent subtleties
when working with this representation. A number of lemmas need to be 
carefully stated, and in a few cases in rather non-intuitive ways.  
The remaining theories account for at most 5--10\% of the
formalization each; the \texttt{WeakAlgorithm} theory defines the weak
algorithmic equivalence judgment and proves the additional properties
needed for the third solution, and accounts for only around 2\% of the
total development.

The merit of metrics such as proof size or number of lemmas is
debatable.  We have not attempted to distinguish between
meaningful lines of proof vs.~blank or comment lines; nor have we
distinguished between significant and trivial lemmas.  Nevertheless,
this information should at least convey an idea of the \emph{relative}
effort involved in each part of the proof.

\paragraph*{Correctness of the representation}
The facilities for defining and reasoning about languages with binding
provided by the Nominal Datatype Package are convenient, but their use
may not be persuasive to readers unfamiliar with nominal logic and
abstract syntax.  Thus, a skeptical reader might ask whether these
representations, definitions and reasoning principles are really
\emph{correct}; that is, whether they are equivalent to the
definitions in \HP, as formalized using some more conventional
approach to binding syntax.  For higher-order abstract syntax
representations, this property is often called \emph{adequacy}; this
term appears to have been coined in the context of
LF~\cite{HarperHonsellPlotkin93}, due to the potential problems
involved in reasoning about higher-order terms modulo alpha, beta and
eta-equivalence.

Adequacy is also important for nominal techniques and deserves further
study.  We believe that the techniques explored in existing work on
the semantics of nominal abstract syntax and its implementation in the
Nominal Datatype
Package~\cite{GabbayPitts02,Pitts03,Cheney06,Pitts06,Urban08JAR}
suffices for informally judging the correctness of our formalization.
There has also been some prior work on formalizing adequacy results
for nominal datatypes via isomorphisms.  \citeN{Urban08JAR} proves a
bijective correspondence between nominal datatypes and a conventional
named implementation of the $\lambda$-calculus modulo
$\alpha$-equivalence. \citeN{norrish:tphols2007} have formalized
isomorphisms between nominal and de Bruijn representations, and they
provide further citations to several other isomorphism results.
Our proof of equivalence to a locally nameless representation
described in \refSec{locally-nameless} also gives evidence for the
correctness of the nominal datatype representation.

In any case, our formalization has exposed some subtle issues which
make sense in the context of LF, independently of whether or not
nominal datatypes in Isabelle/HOL really capture our informal
intuitions about abstract syntax with binding.

\paragraph*{Reflecting on formalizing LF}

It has been observed (as discussed, for example, by \citeN{Pientka07})
that the process of formalization can suggest changes that both ease
formalization and clarify the original system.  Likewise, our
formalization provides a basis for reflecting on how the LF metatheory
might be adapted to make it easier to formalize.  Most obviously, many
of the problems we encountered with soundness disappear if we simply
add the omitted extensionality rule or change the equivalence
algorithm.

A more subtle complication we encountered was that since the
algorithmic rules in \HP do not enforce well-formedness, it is not
even guaranteed that a variable appearing in one of the terms being
compared also appears in the context $\Delta$.  This necessitates
extra freshness conditions on many rules and induction hypotheses to
ensure that strong nominal induction principles can be used safely.
Building these constraints into the algorithmic rules might make
several of the proofs about the equivalence algorithm cleaner.

Another practical consideration was that the syntax and rules of LF in
\HP exhibit redundancy, which leads to additional (albeit
straightforward) formalization effort.  For example, constants,
dependent products, and applications each appear at more than one
level of the syntax, resulting in proofs with redundant cases.
Similarly, because objects, kinds and types are defined by mutual
recursion, each inductive proof about syntax needs to have three
inductive hypotheses and ten cases. Likewise, any proof concerning the
definitional judgments needs to state eight simultaneous induction
hypotheses and thirty-five cases.  Collapsing the three levels of LF
syntax into one level, and collapsing the many definitional judgments
into a smaller number could make the formalization much less verbose,
as in Pure Type Systems~\cite{McKinnaP99}, at the cost of increasing
the distance between the paper version and the formalization.  On the
other hand, such an approach could also make it easier to generalize
proofs about LF to richer type theories.%
\end{isamarkuptext}%
\isamarkuptrue%
\isamarkupsection{Related and Future Work%
}
\isamarkuptrue%
\begin{isamarkuptext}%
\labelSec{related}

  \noindent
  \citeN{McKinnaP99}'s LEGO formalization of Pure Type Systems is
  probably the most extensive formalization of a dependent type theory
  in a theorem prover.  Their formalization introduced the locally
  nameless variant of de Bruijn's name-free approach~\cite{debruijn72im} and
  considered primarily syntactic properties of pure type systems with
  $\beta$-equivalence, including a proof of strengthening.
  \citeN{pollack95} subsequently verified the partial correctness of
  typechecking algorithms for certain classes of Pure Type Systems
  including LF.

  Completely formalizing metatheoretic and syntactic proofs about
  languages and logics with name-binding has been a long-standing open
  problem in computational logic.  We will not give a detailed survey
  of all of these techniques here, but mention a few recent
  developments.  In the last five years, catalyzed by the POPLMark
  Challenge~\cite{poplmark}, there has been renewed interest in this
  area.  \citeN{AydemirEtAl08} have developed a methodology for
  formalizing metatheory in Coq using the locally nameless
  representation to manage binding, and using cofinite quantification
  to handle fresh names.  Chlipala's \emph{parametric higher-order
    abstract syntax} is another recently developed technique for
  reasoning about abstract syntax in Coq, and has been applied to good
  effect in reasoning about compiler
  transformations~\cite{DBLP:conf/icfp/Chlipala08}.
  \citeN{westbrook09lfmtp} are developing CINIC, a variant of Coq that
  provides built-in support for nominal abstract syntax (generalizing
  a simple nominal type theory developed by \citeN{cheney08lfmtp}).
  \citeN{abella} have developed Abella, a proof assistant for
  reasoning about higher-order abstract syntax, inductive definitions,
  and generic quantification (similar to nominal logic's fresh-name
  quantifier).  \citeN{Schuermann} have recently discovered techniques
  for performing logical relations proofs in
  Twelf~\cite{PfenningSchuermann99}.  Formalizing the results in this
  article using these or other emerging tools would provide a useful
  comparison of these approaches, particularly concerning decidability
  proofs, which ought to be easier in constructive logics.

  Algorithms for equivalence and canonicalization for dependent type
  theories have been studied by several authors.  Prior work on
  equivalence checking for LF has focused on first checking
  well-formedness with respect to simple types, then $\beta$- or
  $\beta\eta$-normalizing; these approaches are discussed in detail
  by~\citeN{HarperPfenning05}.  Coquand's algorithm
  \citeyear{Coquand91} is similar to Harper and Pfenning's but
  operates on untyped terms.  Goguen's approach
  \citeyear{Goguen05POPL} involves first type-directed
  $\eta$-expansion and then $\beta$-normalization, and relies on
  standard properties such as the Church-Rosser theorem, strong
  normalization of $\beta$-reduction and strengthening.
  \citeN{Goguen05FOSSACS} extends this proof technique to show
  termination of Coquand's and Harper and Pfenning's algorithms, and
  gives a terminating type-directed algorithm for checking
  $\beta\eta$-equivalence in System F. It may be interesting to formalize
  these algorithms and proofs and compare with Harper and Pfenning's
  proof.

  Our formalization provides a foundation for several possible future
  investigations.  We are interested in extending our formalization to
  include verifying Twelf-style meta-reasoning about LF
  specifications, following Harper and Licata's detailed informal
  development of Canonical LF~\citeyear{HarperLicata07}.  Doing so
  could make it possible to extract Isabelle/HOL theorems from Twelf
  proofs, but as discussed earlier, formalizing Canonical LF,
  hereditary substitutions, and the rest of Harper and Licata's work
  appears to be a substantial challenge.

  It would also be interesting to extend our formalization to
  accommodate extensions to LF involving (ordered) linear logic,
  concurrency, proof-irrelevance, or singleton kinds, as discussed by
  \citeN[Sec.  8]{HarperPfenning05}.  We hope that anyone who proposes
  an extension to LF will be able to use our formalization as a
  starting point for verifying its metatheory.%
\end{isamarkuptext}%
\isamarkuptrue%
\isamarkupsection{Conclusions%
}
\isamarkuptrue%
\begin{isamarkuptext}%
\labelSec{concl}

  \noindent
  LF is an extremely convenient tool for defining logics and other
  calculi involving binding syntax. It has many compelling
  applications and underlies the system Twelf, which has a proven
  record in formalizing many programming language calculi.  Hence, it
  is of intrinsic interest to verify key properties of LF's
  metatheory, such as the correctness and decidability of the
  typechecking algorithms.  We have done so, using the Nominal
  Datatype Package for Isabelle/HOL. The infrastructure provided by
  this package allowed us to follow the proof of Harper and Pfenning
  closely.

  For our formalization we had the advantage of working from Harper
  and Pfenning's carefully-written informal proof, which withstood
  rigorous mechanical formalization rather well. Still we found in
  this informal proof one gap and numerous minor complications.  We
  have shown that they can be repaired. We have also partially
  verified the decidability of the equivalence and typechecking
  algorithms, although some work remains to formally prove
  decidability per se. Formalizing decidability proofs of any kind in
  Isabelle/HOL appears to be an open problem, so we leave this for
  future work.

  While verifying correctness of proofs is a central motivation for
  doing formalizations, it is not the only one.  There is a
  second important benefit---they can be used to experiment with
  changes to the system rapidly.  By replaying a modified
  formalization in a theorem prover one can immediately focus on
  places where the proof fails and attempt to repair them rather than
  re-checking the many cases that are unchanged.  This capability was
  essential in fixing the soundness proof, and it illustrates one of
  the distinctive advantages of performing such a formalization.  Had
  we attempted to repair the gap using only the paper proof,
  experimenting with different solutions would have required
  manually re-checking the roughly 31 pages of paper proofs for each change.

  Our formalization is not an end in itself but also provides a
  foundation for further study in several directions.  Researchers
  developing extensions to LF may find our formalization useful as a
  starting point for verifying the metatheory of such extensions.  We
  plan to further investigate hereditary substitutions and adequacy
  proofs in LF and Canonical LF.  More ambitiously, we contemplate
  formalizing the meaning and correctness of metatheoretic reasoning
  about LF specifications (as provided by the Twelf system) inside
  Isabelle/HOL, and extracting Isabelle/HOL theorems from Twelf
  proofs.

\appendixhead{URLend}

 \begin{acks}
    We are extremely grateful to Bob Harper for discussions about LF
    and the proof.  Benjamin Pierce and Stephanie Weirich have also
    made helpful comments on drafts of this paper.
  \end{acks}%
\end{isamarkuptext}%
\isamarkuptrue%
\isadelimtheory
\endisadelimtheory
\isatagtheory
\endisatagtheory
{\isafoldtheory}%
\isadelimtheory
\endisadelimtheory
\isanewline
\isanewline
\end{isabellebody}%

%% end generated text

\bibliographystyle{acmtrans}

 \begin{received}
 Received October 2009;
 revised April 2010;
 accepted April 2010
 \end{received}

\elecappendix

%% begin electronic appendix

\setcounter{lemma}{0}
\begin{isabellebody}%
\def\isabellecontext{Paper}%
\isadelimtheory
\endisadelimtheory
\isatagtheory
\endisatagtheory
{\isafoldtheory}%
\isadelimtheory
\endisadelimtheory
\begin{isamarkuptext}%
%\section*{APPENDIX}
%\setcounter{section}{0}
%\renewcommand{\thesection}{\Alph{section}}

\section{ Full statements of syntactic results}

  \begin{lemma}[(Freshness)]
\mbox{}
    \begin{compactenum}
      \item \isa{{\normalsize{}If\,}\ {\isasymturnstile}\ {\isasymSigma}\ sig\ {\normalsize \,then\,}\ x\ {\isasymsharp}\ {\isasymSigma}{\isachardot}}
      \item \isa{{\normalsize{}If\,}\ {\isasymturnstile}\isactrlisub {\isasymSigma}\ {\isasymGamma}\ ctx\ {\normalsize \,then\,}\ x\ {\isasymsharp}\ {\isasymSigma}{\isachardot}}
      \item \isa{{\normalsize{}If\,}\ {\isasymGamma}\ {\isasymturnstile}\isactrlisub {\isasymSigma}\ M\ {\isacharcolon}\ A\ {\normalsize \,and\,}\ x\ {\isasymsharp}\ {\isasymGamma}\ {\normalsize \,then\,}\ x\ {\isasymsharp}\ M\ \textrm{and\linebreak[1]}\ x\ {\isasymsharp}\ A{\isachardot}}
      \item \isa{{\normalsize{}If\,}\ {\isasymGamma}\ {\isasymturnstile}\isactrlisub {\isasymSigma}\ A\ {\isacharcolon}\ K\ {\normalsize \,and\,}\ x\ {\isasymsharp}\ {\isasymGamma}\ {\normalsize \,then\,}\ x\ {\isasymsharp}\ A\ \textrm{and\linebreak[1]}\ x\ {\isasymsharp}\ K{\isachardot}}
      \item \isa{{\normalsize{}If\,}\ {\isasymGamma}\ {\isasymturnstile}\isactrlisub {\isasymSigma}\ K\ {\isacharcolon}\ kind\ {\normalsize \,and\,}\ x\ {\isasymsharp}\ {\isasymGamma}\ {\normalsize \,then\,}\ x\ {\isasymsharp}\ K{\isachardot}}
      \item \isa{{\normalsize{}If\,}\ {\isasymGamma}\ {\isasymturnstile}\isactrlisub {\isasymSigma}\ M\ {\isacharequal}\ N\ {\isacharcolon}\ A\ {\normalsize \,and\,}\ x\ {\isasymsharp}\ {\isasymGamma}\ {\normalsize \,then\,}\ x\ {\isasymsharp}\ M\ \textrm{and\linebreak[1]}\ x\ {\isasymsharp}\ N\ \textrm{and\linebreak[1]}\ x\ {\isasymsharp}\ A{\isachardot}}
      \item \isa{{\normalsize{}If\,}\ {\isasymGamma}\ {\isasymturnstile}\isactrlisub {\isasymSigma}\ A\ {\isacharequal}\ B\ {\isacharcolon}\ K\ {\normalsize \,and\,}\ x\ {\isasymsharp}\ {\isasymGamma}\ {\normalsize \,then\,}\ x\ {\isasymsharp}\ A\ \textrm{and\linebreak[1]}\ x\ {\isasymsharp}\ B\ \textrm{and\linebreak[1]}\ x\ {\isasymsharp}\ K{\isachardot}}
      \item \isa{{\normalsize{}If\,}\ {\isasymGamma}\ {\isasymturnstile}\isactrlisub {\isasymSigma}\ K\ {\isacharequal}\ L\ {\isacharcolon}\ kind\ {\normalsize \,and\,}\ x\ {\isasymsharp}\ {\isasymGamma}\ {\normalsize \,then\,}\ x\ {\isasymsharp}\ K\ \textrm{and\linebreak[1]}\ x\ {\isasymsharp}\ L{\isachardot}}
  \end{compactenum}
  \end{lemma}
  
  \begin{lemma}[(Implicit Validity)]
\mbox{}
    \begin{compactenum}
      \item \isa{{\normalsize{}If\,}\ {\isasymturnstile}\isactrlisub {\isasymSigma}\ {\isasymGamma}\ ctx\ {\normalsize \,then\,}\ {\isasymturnstile}\ {\isasymSigma}\ sig{\isachardot}}
      \item \isa{{\normalsize{}If\,}\ {\isasymGamma}\ {\isasymturnstile}\isactrlisub {\isasymSigma}\ M\ {\isacharcolon}\ A\ {\normalsize \,then\,}\ {\isasymturnstile}\isactrlisub {\isasymSigma}\ {\isasymGamma}\ ctx\ \textrm{and\linebreak[1]}\ {\isasymturnstile}\ {\isasymSigma}\ sig{\isachardot}}
      \item \isa{{\normalsize{}If\,}\ {\isasymGamma}\ {\isasymturnstile}\isactrlisub {\isasymSigma}\ A\ {\isacharcolon}\ K\ {\normalsize \,then\,}\ {\isasymturnstile}\isactrlisub {\isasymSigma}\ {\isasymGamma}\ ctx\ \textrm{and\linebreak[1]}\ {\isasymturnstile}\ {\isasymSigma}\ sig{\isachardot}}
      \item \isa{{\normalsize{}If\,}\ {\isasymGamma}\ {\isasymturnstile}\isactrlisub {\isasymSigma}\ K\ {\isacharcolon}\ kind\ {\normalsize \,then\,}\ {\isasymturnstile}\isactrlisub {\isasymSigma}\ {\isasymGamma}\ ctx\ \textrm{and\linebreak[1]}\ {\isasymturnstile}\ {\isasymSigma}\ sig{\isachardot}}
      \item \isa{{\normalsize{}If\,}\ {\isasymGamma}\ {\isasymturnstile}\isactrlisub {\isasymSigma}\ M\ {\isacharequal}\ N\ {\isacharcolon}\ A\ {\normalsize \,then\,}\ {\isasymturnstile}\isactrlisub {\isasymSigma}\ {\isasymGamma}\ ctx\ \textrm{and\linebreak[1]}\ {\isasymturnstile}\ {\isasymSigma}\ sig{\isachardot}}
      \item \isa{{\normalsize{}If\,}\ {\isasymGamma}\ {\isasymturnstile}\isactrlisub {\isasymSigma}\ A\ {\isacharequal}\ B\ {\isacharcolon}\ K\ {\normalsize \,then\,}\ {\isasymturnstile}\isactrlisub {\isasymSigma}\ {\isasymGamma}\ ctx\ \textrm{and\linebreak[1]}\ {\isasymturnstile}\ {\isasymSigma}\ sig{\isachardot}}
      \item \isa{{\normalsize{}If\,}\ {\isasymGamma}\ {\isasymturnstile}\isactrlisub {\isasymSigma}\ K\ {\isacharequal}\ L\ {\isacharcolon}\ kind\ {\normalsize \,then\,}\ {\isasymturnstile}\isactrlisub {\isasymSigma}\ {\isasymGamma}\ ctx\ \textrm{and\linebreak[1]}\ {\isasymturnstile}\ {\isasymSigma}\ sig{\isachardot}}
   \end{compactenum}
  \end{lemma}

  \begin{lemma}[(Implicit Validity)]
    If \isa{{\isasymGamma}\ {\isasymturnstile}\isactrlisub {\isasymSigma}\ M\ {\isacharcolon}\ A} then \isa{{\isasymturnstile}\ {\isasymSigma}\ sig} and \isa{{\isasymturnstile}\isactrlisub {\isasymSigma}\ {\isasymGamma}\ ctx}.
  \end{lemma}

  \begin{lemma}[(Weakening)] 
  Suppose \isa{{\isasymturnstile}\isactrlisub {\isasymSigma}\ {\isasymGamma}\isactrlisub {\isadigit{2}}\ ctx} and \isa{{\isasymGamma}\isactrlisub {\isadigit{1}}\ {\isasymsubseteq}\ {\isasymGamma}\isactrlisub {\isadigit{2}}}.
    ~
    \begin{compactenum} 
    \item If \isa{{\isasymGamma}\isactrlisub {\isadigit{1}}\ {\isasymturnstile}\isactrlisub {\isasymSigma}\ M\ {\isacharcolon}\ A} 
          then \isa{{\isasymGamma}\isactrlisub {\isadigit{2}}\ {\isasymturnstile}\isactrlisub {\isasymSigma}\ M\ {\isacharcolon}\ A}.
    \item If \isa{{\isasymGamma}\isactrlisub {\isadigit{1}}\ {\isasymturnstile}\isactrlisub {\isasymSigma}\ A\ {\isacharcolon}\ K}
          then \isa{{\isasymGamma}\isactrlisub {\isadigit{2}}\ {\isasymturnstile}\isactrlisub {\isasymSigma}\ A\ {\isacharcolon}\ K}.
    \item If \isa{{\isasymGamma}\isactrlisub {\isadigit{1}}\ {\isasymturnstile}\isactrlisub {\isasymSigma}\ K\ {\isacharcolon}\ kind} 
          then \isa{{\isasymGamma}\isactrlisub {\isadigit{2}}\ {\isasymturnstile}\isactrlisub {\isasymSigma}\ K\ {\isacharcolon}\ kind}.
    \item If \isa{{\isasymGamma}\isactrlisub {\isadigit{1}}\ {\isasymturnstile}\isactrlisub {\isasymSigma}\ M\ {\isacharequal}\ N\ {\isacharcolon}\ A}
          then \isa{{\isasymGamma}\isactrlisub {\isadigit{2}}\ {\isasymturnstile}\isactrlisub {\isasymSigma}\ M\ {\isacharequal}\ N\ {\isacharcolon}\ A}.
    \item If \isa{{\isasymGamma}\isactrlisub {\isadigit{1}}\ {\isasymturnstile}\isactrlisub {\isasymSigma}\ A\ {\isacharequal}\ B\ {\isacharcolon}\ K} 
          then \isa{{\isasymGamma}\isactrlisub {\isadigit{2}}\ {\isasymturnstile}\isactrlisub {\isasymSigma}\ A\ {\isacharequal}\ B\ {\isacharcolon}\ K}.
    \item If \isa{{\isasymGamma}\isactrlisub {\isadigit{1}}\ {\isasymturnstile}\isactrlisub {\isasymSigma}\ K\ {\isacharequal}\ L\ {\isacharcolon}\ kind}
          then \isa{{\isasymGamma}\isactrlisub {\isadigit{2}}\ {\isasymturnstile}\isactrlisub {\isasymSigma}\ K\ {\isacharequal}\ L\ {\isacharcolon}\ kind}.
    \end{compactenum}
  \end{lemma}
  
  \begin{lemma}[(Substitution)]  Suppose \isa{{\isasymGamma}\isactrlisub {\isadigit{2}}\ {\isasymturnstile}\isactrlisub {\isasymSigma}\ P\ {\isacharcolon}\ C} and let 
    \mbox{\isa{{\isasymGamma}\ {\isacharequal}\ {\isasymGamma}\isactrlisub {\isadigit{1}}\ {\isacharat}\ {\isacharbrackleft}{\isacharparenleft}y{\isacharcomma}\ C{\isacharparenright}{\isacharbrackright}\ {\isacharat}\ {\isasymGamma}\isactrlisub {\isadigit{2}}}}.
    \begin{compactenum} 
    \item If \isa{{\isasymturnstile}\isactrlisub {\isasymSigma}\ {\isasymGamma}\ ctx} 
          then \isa{{\isasymturnstile}\isactrlisub {\isasymSigma}\ {\isasymGamma}\isactrlisub {\isadigit{1}}{\isacharbrackleft}y{\isacharcolon}{\isacharequal}P{\isacharbrackright}\ {\isacharat}\ {\isasymGamma}\isactrlisub {\isadigit{2}}\ ctx}.
    \item If \isa{{\isasymGamma}\ {\isasymturnstile}\isactrlisub {\isasymSigma}\ M\ {\isacharcolon}\ B}
          then \isa{{\isasymGamma}\isactrlisub {\isadigit{1}}{\isacharbrackleft}y{\isacharcolon}{\isacharequal}P{\isacharbrackright}\ {\isacharat}\ {\isasymGamma}\isactrlisub {\isadigit{2}}\ {\isasymturnstile}\isactrlisub {\isasymSigma}\ M{\isacharbrackleft}y{\isacharcolon}{\isacharequal}P{\isacharbrackright}\ {\isacharcolon}\ B{\isacharbrackleft}y{\isacharcolon}{\isacharequal}P{\isacharbrackright}}.
    \item If \isa{{\isasymGamma}\ {\isasymturnstile}\isactrlisub {\isasymSigma}\ B\ {\isacharcolon}\ K}
          then \isa{{\isasymGamma}\isactrlisub {\isadigit{1}}{\isacharbrackleft}y{\isacharcolon}{\isacharequal}P{\isacharbrackright}\ {\isacharat}\ {\isasymGamma}\isactrlisub {\isadigit{2}}\ {\isasymturnstile}\isactrlisub {\isasymSigma}\ B{\isacharbrackleft}y{\isacharcolon}{\isacharequal}P{\isacharbrackright}\ {\isacharcolon}\ K{\isacharbrackleft}y{\isacharcolon}{\isacharequal}P{\isacharbrackright}}.
    \item If \isa{{\isasymGamma}\ {\isasymturnstile}\isactrlisub {\isasymSigma}\ K\ {\isacharcolon}\ kind}
          then \isa{{\isasymGamma}\isactrlisub {\isadigit{1}}{\isacharbrackleft}y{\isacharcolon}{\isacharequal}P{\isacharbrackright}\ {\isacharat}\ {\isasymGamma}\isactrlisub {\isadigit{2}}\ {\isasymturnstile}\isactrlisub {\isasymSigma}\ K{\isacharbrackleft}y{\isacharcolon}{\isacharequal}P{\isacharbrackright}\ {\isacharcolon}\ kind}.
    \item If \isa{{\isasymGamma}\ {\isasymturnstile}\isactrlisub {\isasymSigma}\ M\ {\isacharequal}\ N\ {\isacharcolon}\ A}
          then \isa{{\isasymGamma}\isactrlisub {\isadigit{1}}{\isacharbrackleft}y{\isacharcolon}{\isacharequal}P{\isacharbrackright}\ {\isacharat}\ {\isasymGamma}\isactrlisub {\isadigit{2}}\ {\isasymturnstile}\isactrlisub {\isasymSigma}\ M{\isacharbrackleft}y{\isacharcolon}{\isacharequal}P{\isacharbrackright}\ {\isacharequal}\ N{\isacharbrackleft}y{\isacharcolon}{\isacharequal}P{\isacharbrackright}\ {\isacharcolon}\ A{\isacharbrackleft}y{\isacharcolon}{\isacharequal}P{\isacharbrackright}}.
    \item If \isa{{\isasymGamma}\ {\isasymturnstile}\isactrlisub {\isasymSigma}\ A\ {\isacharequal}\ B\ {\isacharcolon}\ K}
          then \isa{{\isasymGamma}\isactrlisub {\isadigit{1}}{\isacharbrackleft}y{\isacharcolon}{\isacharequal}P{\isacharbrackright}\ {\isacharat}\ {\isasymGamma}\isactrlisub {\isadigit{2}}\ {\isasymturnstile}\isactrlisub {\isasymSigma}\ A{\isacharbrackleft}y{\isacharcolon}{\isacharequal}P{\isacharbrackright}\ {\isacharequal}\ B{\isacharbrackleft}y{\isacharcolon}{\isacharequal}P{\isacharbrackright}\ {\isacharcolon}\ K{\isacharbrackleft}y{\isacharcolon}{\isacharequal}P{\isacharbrackright}}.
    \item If \isa{{\isasymGamma}\ {\isasymturnstile}\isactrlisub {\isasymSigma}\ K\ {\isacharequal}\ L\ {\isacharcolon}\ kind}
          then \isa{{\isasymGamma}\isactrlisub {\isadigit{1}}{\isacharbrackleft}y{\isacharcolon}{\isacharequal}P{\isacharbrackright}\ {\isacharat}\ {\isasymGamma}\isactrlisub {\isadigit{2}}\ {\isasymturnstile}\isactrlisub {\isasymSigma}\ K{\isacharbrackleft}y{\isacharcolon}{\isacharequal}P{\isacharbrackright}\ {\isacharequal}\ L{\isacharbrackleft}y{\isacharcolon}{\isacharequal}P{\isacharbrackright}\ {\isacharcolon}\ kind}.
    \end{compactenum}
  \end{lemma}

  \begin{lemma}[(Context
    Conversion)]\raggedright  ~Assume that
    \isa{{\isasymGamma}\ {\isasymturnstile}\isactrlisub {\isasymSigma}\ B\ {\isacharcolon}\ type} and \isa{{\isasymGamma}\ {\isasymturnstile}\isactrlisub {\isasymSigma}\ A\ {\isacharequal}\ B\ {\isacharcolon}\ type}.  Then:
    \begin{compactenum}
    \item If \isa{{\isacharparenleft}x{\isacharcomma}\ A{\isacharparenright}{\isacharcolon}{\isacharcolon}{\isasymGamma}\ {\isasymturnstile}\isactrlisub {\isasymSigma}\ M\ {\isacharcolon}\ C} then \isa{{\isacharparenleft}x{\isacharcomma}\ B{\isacharparenright}{\isacharcolon}{\isacharcolon}{\isasymGamma}\ {\isasymturnstile}\isactrlisub {\isasymSigma}\ M\ {\isacharcolon}\ C}
    \item If \isa{{\isacharparenleft}x{\isacharcomma}\ A{\isacharparenright}{\isacharcolon}{\isacharcolon}{\isasymGamma}\ {\isasymturnstile}\isactrlisub {\isasymSigma}\ C\ {\isacharcolon}\ K} then \isa{{\isacharparenleft}x{\isacharcomma}\ B{\isacharparenright}{\isacharcolon}{\isacharcolon}{\isasymGamma}\ {\isasymturnstile}\isactrlisub {\isasymSigma}\ C\ {\isacharcolon}\ K}
    \item If \isa{{\isacharparenleft}x{\isacharcomma}\ A{\isacharparenright}{\isacharcolon}{\isacharcolon}{\isasymGamma}\ {\isasymturnstile}\isactrlisub {\isasymSigma}\ K\ {\isacharcolon}\ kind} then \isa{{\isacharparenleft}x{\isacharcomma}\ B{\isacharparenright}{\isacharcolon}{\isacharcolon}{\isasymGamma}\ {\isasymturnstile}\isactrlisub {\isasymSigma}\ K\ {\isacharcolon}\ kind}
    \item If \isa{{\isacharparenleft}x{\isacharcomma}\ A{\isacharparenright}{\isacharcolon}{\isacharcolon}{\isasymGamma}\ {\isasymturnstile}\isactrlisub {\isasymSigma}\ C\ {\isacharequal}\ D\ {\isacharcolon}\ K} then \isa{{\isacharparenleft}x{\isacharcomma}\ B{\isacharparenright}{\isacharcolon}{\isacharcolon}{\isasymGamma}\ {\isasymturnstile}\isactrlisub {\isasymSigma}\ C\ {\isacharequal}\ D\ {\isacharcolon}\ K}
    \item If \isa{{\isacharparenleft}x{\isacharcomma}\ A{\isacharparenright}{\isacharcolon}{\isacharcolon}{\isasymGamma}\ {\isasymturnstile}\isactrlisub {\isasymSigma}\ K\ {\isacharequal}\ L\ {\isacharcolon}\ kind} then \isa{{\isacharparenleft}x{\isacharcomma}\ B{\isacharparenright}{\isacharcolon}{\isacharcolon}{\isasymGamma}\ {\isasymturnstile}\isactrlisub {\isasymSigma}\ K\ {\isacharequal}\ L\ {\isacharcolon}\ kind}
    \end{compactenum}
  \end{lemma}

  \begin{lemma}[(Functionality for Typing)]\raggedright
    Assume that \isa{{\isasymGamma}\ {\isasymturnstile}\isactrlisub {\isasymSigma}\ M\ {\isacharcolon}\ C} and \isa{{\isasymGamma}\ {\isasymturnstile}\isactrlisub {\isasymSigma}\ N\ {\isacharcolon}\ C} and \isa{{\isasymGamma}\ {\isasymturnstile}\isactrlisub {\isasymSigma}\ M\ {\isacharequal}\ N\ {\isacharcolon}\ C}.  Then:
    \begin{compactenum}
    \item If \isa{{\isasymGamma}{\isacharprime}\ {\isacharat}\ {\isacharbrackleft}{\isacharparenleft}y{\isacharcomma}\ C{\isacharparenright}{\isacharbrackright}\ {\isacharat}\ {\isasymGamma}\ {\isasymturnstile}\isactrlisub {\isasymSigma}\ P\ {\isacharcolon}\ B} then \isa{{\isasymGamma}{\isacharprime}{\isacharbrackleft}y{\isacharcolon}{\isacharequal}M{\isacharbrackright}\ {\isacharat}\ {\isasymGamma}\ {\isasymturnstile}\isactrlisub {\isasymSigma}\ P{\isacharbrackleft}y{\isacharcolon}{\isacharequal}M{\isacharbrackright}\ {\isacharequal}\ P{\isacharbrackleft}y{\isacharcolon}{\isacharequal}N{\isacharbrackright}\ {\isacharcolon}\ B{\isacharbrackleft}y{\isacharcolon}{\isacharequal}M{\isacharbrackright}}
    \item If \isa{{\isasymGamma}{\isacharprime}\ {\isacharat}\ {\isacharbrackleft}{\isacharparenleft}y{\isacharcomma}\ C{\isacharparenright}{\isacharbrackright}\ {\isacharat}\ {\isasymGamma}\ {\isasymturnstile}\isactrlisub {\isasymSigma}\ B\ {\isacharcolon}\ K} then \isa{{\isasymGamma}{\isacharprime}{\isacharbrackleft}y{\isacharcolon}{\isacharequal}M{\isacharbrackright}\ {\isacharat}\ {\isasymGamma}\ {\isasymturnstile}\isactrlisub {\isasymSigma}\ B{\isacharbrackleft}y{\isacharcolon}{\isacharequal}M{\isacharbrackright}\ {\isacharequal}\ B{\isacharbrackleft}y{\isacharcolon}{\isacharequal}N{\isacharbrackright}\ {\isacharcolon}\ K{\isacharbrackleft}y{\isacharcolon}{\isacharequal}M{\isacharbrackright}}
    \item If \isa{{\isasymGamma}{\isacharprime}\ {\isacharat}\ {\isacharbrackleft}{\isacharparenleft}y{\isacharcomma}\ C{\isacharparenright}{\isacharbrackright}\ {\isacharat}\ {\isasymGamma}\ {\isasymturnstile}\isactrlisub {\isasymSigma}\ K\ {\isacharcolon}\ kind} then \isa{{\isasymGamma}{\isacharprime}{\isacharbrackleft}y{\isacharcolon}{\isacharequal}M{\isacharbrackright}\ {\isacharat}\ {\isasymGamma}\ {\isasymturnstile}\isactrlisub {\isasymSigma}\ K{\isacharbrackleft}y{\isacharcolon}{\isacharequal}M{\isacharbrackright}\ {\isacharequal}\ K{\isacharbrackleft}y{\isacharcolon}{\isacharequal}N{\isacharbrackright}\ {\isacharcolon}\ kind}
    \end{compactenum}
  \end{lemma}

  \begin{lemma}[(Validity)]
  Objects, types and kinds appearing in derivable judgments are valid, that is 
    \begin{compactenum}
    \item \isa{{\normalsize{}If\,}\ {\isasymGamma}\ {\isasymturnstile}\isactrlisub {\isasymSigma}\ M\ {\isacharcolon}\ A\ {\normalsize \,then\,}\ {\isasymGamma}\ {\isasymturnstile}\isactrlisub {\isasymSigma}\ A\ {\isacharcolon}\ type{\isachardot}}
    \item \isa{{\normalsize{}If\,}\ {\isasymGamma}\ {\isasymturnstile}\isactrlisub {\isasymSigma}\ A\ {\isacharcolon}\ K\ {\normalsize \,then\,}\ {\isasymGamma}\ {\isasymturnstile}\isactrlisub {\isasymSigma}\ K\ {\isacharcolon}\ kind{\isachardot}}
    %% Part 5 is trivial
    %% \item \isa{{\normalsize{}If\,}\ {\isasymGamma}\ {\isasymturnstile}\isactrlisub {\isasymSigma}\ K\ {\isacharcolon}\ kind\ {\normalsize \,then\,}\ True{\isachardot}}
    \item \isa{{\normalsize{}If\,}\ {\isasymGamma}\ {\isasymturnstile}\isactrlisub {\isasymSigma}\ M\ {\isacharequal}\ N\ {\isacharcolon}\ B\ {\normalsize \,then\,}\ {\isasymGamma}\ {\isasymturnstile}\isactrlisub {\isasymSigma}\ M\ {\isacharcolon}\ B\ \textrm{and\linebreak[1]}\ {\isasymGamma}\ {\isasymturnstile}\isactrlisub {\isasymSigma}\ N\ {\isacharcolon}\ B\ \textrm{and\linebreak[1]}\ {\isasymGamma}\ {\isasymturnstile}\isactrlisub {\isasymSigma}\ B\ {\isacharcolon}\ type{\isachardot}}
    \item \isa{{\normalsize{}If\,}\ {\isasymGamma}\ {\isasymturnstile}\isactrlisub {\isasymSigma}\ A\ {\isacharequal}\ B\ {\isacharcolon}\ K\ {\normalsize \,then\,}\ {\isasymGamma}\ {\isasymturnstile}\isactrlisub {\isasymSigma}\ A\ {\isacharcolon}\ K\ \textrm{and\linebreak[1]}\ {\isasymGamma}\ {\isasymturnstile}\isactrlisub {\isasymSigma}\ B\ {\isacharcolon}\ K\ \textrm{and\linebreak[1]}\ {\isasymGamma}\ {\isasymturnstile}\isactrlisub {\isasymSigma}\ K\ {\isacharcolon}\ kind{\isachardot}}
    \item \isa{{\normalsize{}If\,}\ {\isasymGamma}\ {\isasymturnstile}\isactrlisub {\isasymSigma}\ K\ {\isacharequal}\ L\ {\isacharcolon}\ kind\ {\normalsize \,then\,}\ {\isasymGamma}\ {\isasymturnstile}\isactrlisub {\isasymSigma}\ K\ {\isacharcolon}\ kind\ \textrm{and\linebreak[1]}\ {\isasymGamma}\ {\isasymturnstile}\isactrlisub {\isasymSigma}\ L\ {\isacharcolon}\ kind{\isachardot}}
    \end{compactenum}
  \end{lemma}

  \begin{lemma}[(Typing inversion)]
    The validity rules are invertible, up to conversion of types and
    kinds.  
    \begin{compactenum}
    \item \isa{{\normalsize{}If\,}\ {\isasymGamma}\ {\isasymturnstile}\isactrlisub {\isasymSigma}\ x\ {\isacharcolon}\ A\ {\normalsize \,then\,}\ {\isasymexists}B{\isachardot}\ {\isacharparenleft}x{\isacharcomma}\ B{\isacharparenright}\ {\isasymin}\ {\isasymGamma}\ \textrm{and\linebreak[1]}\ {\isasymGamma}\ {\isasymturnstile}\isactrlisub {\isasymSigma}\ A\ {\isacharequal}\ B\ {\isacharcolon}\ type{\isachardot}}
    \item \isa{{\normalsize{}If\,}\ {\isasymGamma}\ {\isasymturnstile}\isactrlisub {\isasymSigma}\ c\ {\isacharcolon}\ A\ {\normalsize \,then\,}\ {\isasymexists}B{\isachardot}\ {\isacharparenleft}c{\isacharcomma}\ B{\isacharparenright}\ {\isasymin}\ {\isasymSigma}\ \textrm{and\linebreak[1]}\ {\isasymGamma}\ {\isasymturnstile}\isactrlisub {\isasymSigma}\ A\ {\isacharequal}\ B\ {\isacharcolon}\ type{\isachardot}}
    \item \isa{{\normalsize{}If\,}\ {\isasymGamma}\ {\isasymturnstile}\isactrlisub {\isasymSigma}\ M\isactrlisub {\isadigit{1}}\ M\isactrlisub {\isadigit{2}}\ {\isacharcolon}\ A\ {\normalsize \,then\,}\ {\isasymexists}x\ A\isactrlisub {\isadigit{1}}\ A\isactrlisub {\isadigit{2}}{\isachardot}\ {\isasymGamma}\ {\isasymturnstile}\isactrlisub {\isasymSigma}\ M\isactrlisub {\isadigit{1}}\ {\isacharcolon}\ {\isasymPi}x{\isacharcolon}A\isactrlisub {\isadigit{2}}{\isachardot}\ A\isactrlisub {\isadigit{1}}\ \textrm{and\linebreak[1]}\ {\isasymGamma}\ {\isasymturnstile}\isactrlisub {\isasymSigma}\ M\isactrlisub {\isadigit{2}}\ {\isacharcolon}\ A\isactrlisub {\isadigit{2}}\ \textrm{and\linebreak[1]}\ {\isasymGamma}\ {\isasymturnstile}\isactrlisub {\isasymSigma}\ A\ {\isacharequal}\ A\isactrlisub {\isadigit{1}}{\isacharbrackleft}x{\isacharcolon}{\isacharequal}M\isactrlisub {\isadigit{2}}{\isacharbrackright}\ {\isacharcolon}\ type{\isachardot}}
    \item \isa{{\normalsize{}If\,}\ {\isasymGamma}\ {\isasymturnstile}\isactrlisub {\isasymSigma}\ {\isasymlambda}x{\isacharcolon}A{\isachardot}\ M\ {\isacharcolon}\ B\ {\normalsize \,and\,}\ x\ {\isasymsharp}\ {\isasymGamma}\ {\normalsize \,then\,}\ {\isasymexists}A{\isacharprime}{\isachardot}\ {\isasymGamma}\ {\isasymturnstile}\isactrlisub {\isasymSigma}\ B\ {\isacharequal}\ {\isasymPi}x{\isacharcolon}A{\isachardot}\ A{\isacharprime}\ {\isacharcolon}\ type\ \textrm{and\linebreak[1]}\ {\isasymGamma}\ {\isasymturnstile}\isactrlisub {\isasymSigma}\ A\ {\isacharcolon}\ type\ \textrm{and\linebreak[1]}\ {\isacharparenleft}x{\isacharcomma}\ A{\isacharparenright}{\isacharcolon}{\isacharcolon}{\isasymGamma}\ {\isasymturnstile}\isactrlisub {\isasymSigma}\ M\ {\isacharcolon}\ A{\isacharprime}{\isachardot}}
    \item \isa{{\normalsize{}If\,}\ {\isasymGamma}\ {\isasymturnstile}\isactrlisub {\isasymSigma}\ {\isasymPi}x{\isacharcolon}A\isactrlisub {\isadigit{1}}{\isachardot}\ A\isactrlisub {\isadigit{2}}\ {\isacharcolon}\ K\ {\normalsize \,and\,}\ x\ {\isasymsharp}\ {\isasymGamma}\ {\normalsize \,then\,}\ {\isasymGamma}\ {\isasymturnstile}\isactrlisub {\isasymSigma}\ K\ {\isacharequal}\ type\ {\isacharcolon}\ kind\ \textrm{and\linebreak[1]}\ {\isasymGamma}\ {\isasymturnstile}\isactrlisub {\isasymSigma}\ A\isactrlisub {\isadigit{1}}\ {\isacharcolon}\ type\ \textrm{and\linebreak[1]}\ {\isacharparenleft}x{\isacharcomma}\ A\isactrlisub {\isadigit{1}}{\isacharparenright}{\isacharcolon}{\isacharcolon}{\isasymGamma}\ {\isasymturnstile}\isactrlisub {\isasymSigma}\ A\isactrlisub {\isadigit{2}}\ {\isacharcolon}\ type{\isachardot}}
    \item \isa{{\normalsize{}If\,}\ {\isasymGamma}\ {\isasymturnstile}\isactrlisub {\isasymSigma}\ c\ {\isacharcolon}\ K\ {\normalsize \,then\,}\ {\isasymexists}L{\isachardot}\ {\isacharparenleft}c{\isacharcomma}\ L{\isacharparenright}\ {\isasymin}\ {\isasymSigma}\ \textrm{and\linebreak[1]}\ {\isasymGamma}\ {\isasymturnstile}\isactrlisub {\isasymSigma}\ K\ {\isacharequal}\ L\ {\isacharcolon}\ kind{\isachardot}}
    \item \isa{{\normalsize{}If\,}\ {\isasymGamma}\ {\isasymturnstile}\isactrlisub {\isasymSigma}\ A\ M\ {\isacharcolon}\ K\ {\normalsize \,then\,}\ {\isasymexists}x\ A{\isadigit{1}}\ K{\isadigit{2}}{\isachardot}\ {\isasymGamma}\ {\isasymturnstile}\isactrlisub {\isasymSigma}\ A\ {\isacharcolon}\ {\isasymPi}x{\isacharcolon}A{\isadigit{1}}{\isachardot}\ K{\isadigit{2}}\ \textrm{and\linebreak[1]}\ {\isasymGamma}\ {\isasymturnstile}\isactrlisub {\isasymSigma}\ M\ {\isacharcolon}\ A{\isadigit{1}}\ \textrm{and\linebreak[1]}\ {\isasymGamma}\ {\isasymturnstile}\isactrlisub {\isasymSigma}\ K\ {\isacharequal}\ K{\isadigit{2}}{\isacharbrackleft}x{\isacharcolon}{\isacharequal}M{\isacharbrackright}\ {\isacharcolon}\ kind{\isachardot}}
    \item \isa{{\normalsize{}If\,}\ {\isasymGamma}\ {\isasymturnstile}\isactrlisub {\isasymSigma}\ {\isasymPi}x{\isacharcolon}A\isactrlisub {\isadigit{1}}{\isachardot}\ K\isactrlisub {\isadigit{2}}\ {\isacharcolon}\ kind\ {\normalsize \,and\,}\ x\ {\isasymsharp}\ {\isasymGamma}\ {\normalsize \,then\,}\ {\isasymGamma}\ {\isasymturnstile}\isactrlisub {\isasymSigma}\ A\isactrlisub {\isadigit{1}}\ {\isacharcolon}\ type\ \textrm{and\linebreak[1]}\ {\isacharparenleft}x{\isacharcomma}\ A\isactrlisub {\isadigit{1}}{\isacharparenright}{\isacharcolon}{\isacharcolon}{\isasymGamma}\ {\isasymturnstile}\isactrlisub {\isasymSigma}\ K\isactrlisub {\isadigit{2}}\ {\isacharcolon}\ kind{\isachardot}}
    \end{compactenum}
  \end{lemma}

  \begin{lemma}[(Equality inversion)]
    ~
    \begin{compactenum}
    \item \isa{{\normalsize{}If\,}\ {\isasymGamma}\ {\isasymturnstile}\isactrlisub {\isasymSigma}\ type\ {\isacharequal}\ L\ {\isacharcolon}\ kind\ {\normalsize \,then\,}\ L\ {\isacharequal}\ type{\isachardot}}
    \item \isa{{\normalsize{}If\,}\ {\isasymGamma}\ {\isasymturnstile}\isactrlisub {\isasymSigma}\ L\ {\isacharequal}\ type\ {\isacharcolon}\ kind\ {\normalsize \,then\,}\ L\ {\isacharequal}\ type{\isachardot}}
    \item \isa{{\normalsize{}If\,}\ {\isasymGamma}\ {\isasymturnstile}\isactrlisub {\isasymSigma}\ A\ {\isacharequal}\ {\isasymPi}x{\isacharcolon}B\isactrlisub {\isadigit{1}}{\isachardot}\ B\isactrlisub {\isadigit{2}}\ {\isacharcolon}\ type\ {\normalsize \,and\,}\ x\ {\isasymsharp}\ {\isasymGamma}\ {\normalsize \,then\,}\ {\isasymexists}A\isactrlisub {\isadigit{1}}\ A\isactrlisub {\isadigit{2}}{\isachardot}\ A\ {\isacharequal}\ {\isasymPi}x{\isacharcolon}A\isactrlisub {\isadigit{1}}{\isachardot}\ A\isactrlisub {\isadigit{2}}\ \textrm{and\linebreak[1]}\ {\isasymGamma}\ {\isasymturnstile}\isactrlisub {\isasymSigma}\ A\isactrlisub {\isadigit{1}}\ {\isacharequal}\ B\isactrlisub {\isadigit{1}}\ {\isacharcolon}\ type\ \textrm{and\linebreak[1]}\ {\isacharparenleft}x{\isacharcomma}\ A\isactrlisub {\isadigit{1}}{\isacharparenright}{\isacharcolon}{\isacharcolon}{\isasymGamma}\ {\isasymturnstile}\isactrlisub {\isasymSigma}\ A\isactrlisub {\isadigit{2}}\ {\isacharequal}\ B\isactrlisub {\isadigit{2}}\ {\isacharcolon}\ type{\isachardot}}
    \item \isa{{\normalsize{}If\,}\ {\isasymGamma}\ {\isasymturnstile}\isactrlisub {\isasymSigma}\ {\isasymPi}x{\isacharcolon}B\isactrlisub {\isadigit{1}}{\isachardot}\ B\isactrlisub {\isadigit{2}}\ {\isacharequal}\ B\ {\isacharcolon}\ type\ {\normalsize \,and\,}\ x\ {\isasymsharp}\ {\isasymGamma}\ {\normalsize \,then\,}\ {\isasymexists}A\isactrlisub {\isadigit{1}}\ A\isactrlisub {\isadigit{2}}{\isachardot}\ B\ {\isacharequal}\ {\isasymPi}x{\isacharcolon}A\isactrlisub {\isadigit{1}}{\isachardot}\ A\isactrlisub {\isadigit{2}}\ \textrm{and\linebreak[1]}\ {\isasymGamma}\ {\isasymturnstile}\isactrlisub {\isasymSigma}\ A\isactrlisub {\isadigit{1}}\ {\isacharequal}\ B\isactrlisub {\isadigit{1}}\ {\isacharcolon}\ type\ \textrm{and\linebreak[1]}\ {\isacharparenleft}x{\isacharcomma}\ A\isactrlisub {\isadigit{1}}{\isacharparenright}{\isacharcolon}{\isacharcolon}{\isasymGamma}\ {\isasymturnstile}\isactrlisub {\isasymSigma}\ A\isactrlisub {\isadigit{2}}\ {\isacharequal}\ B\isactrlisub {\isadigit{2}}\ {\isacharcolon}\ type{\isachardot}}
    \item \isa{{\normalsize{}If\,}\ {\isasymGamma}\ {\isasymturnstile}\isactrlisub {\isasymSigma}\ K\ {\isacharequal}\ {\isasymPi}x{\isacharcolon}B\isactrlisub {\isadigit{1}}{\isachardot}\ L\isactrlisub {\isadigit{2}}\ {\isacharcolon}\ kind\ {\normalsize \,and\,}\ x\ {\isasymsharp}\ {\isasymGamma}\ {\normalsize \,then\,}\ {\isasymexists}A\isactrlisub {\isadigit{1}}\ K\isactrlisub {\isadigit{2}}{\isachardot}\ K\ {\isacharequal}\ {\isasymPi}x{\isacharcolon}A\isactrlisub {\isadigit{1}}{\isachardot}\ K\isactrlisub {\isadigit{2}}\ \textrm{and\linebreak[1]}\ {\isasymGamma}\ {\isasymturnstile}\isactrlisub {\isasymSigma}\ A\isactrlisub {\isadigit{1}}\ {\isacharequal}\ B\isactrlisub {\isadigit{1}}\ {\isacharcolon}\ type\ \textrm{and\linebreak[1]}\ {\isacharparenleft}x{\isacharcomma}\ A\isactrlisub {\isadigit{1}}{\isacharparenright}{\isacharcolon}{\isacharcolon}{\isasymGamma}\ {\isasymturnstile}\isactrlisub {\isasymSigma}\ K\isactrlisub {\isadigit{2}}\ {\isacharequal}\ L\isactrlisub {\isadigit{2}}\ {\isacharcolon}\ kind{\isachardot}}
    \item \isa{{\normalsize{}If\,}\ {\isasymGamma}\ {\isasymturnstile}\isactrlisub {\isasymSigma}\ {\isasymPi}x{\isacharcolon}B\isactrlisub {\isadigit{1}}{\isachardot}\ L\isactrlisub {\isadigit{2}}\ {\isacharequal}\ L\ {\isacharcolon}\ kind\ {\normalsize \,and\,}\ x\ {\isasymsharp}\ {\isasymGamma}\ {\normalsize \,then\,}\ {\isasymexists}A\isactrlisub {\isadigit{1}}\ K\isactrlisub {\isadigit{2}}{\isachardot}\ L\ {\isacharequal}\ {\isasymPi}x{\isacharcolon}A\isactrlisub {\isadigit{1}}{\isachardot}\ K\isactrlisub {\isadigit{2}}\ \textrm{and\linebreak[1]}\ {\isasymGamma}\ {\isasymturnstile}\isactrlisub {\isasymSigma}\ A\isactrlisub {\isadigit{1}}\ {\isacharequal}\ B\isactrlisub {\isadigit{1}}\ {\isacharcolon}\ type\ \textrm{and\linebreak[1]}\ {\isacharparenleft}x{\isacharcomma}\ A\isactrlisub {\isadigit{1}}{\isacharparenright}{\isacharcolon}{\isacharcolon}{\isasymGamma}\ {\isasymturnstile}\isactrlisub {\isasymSigma}\ K\isactrlisub {\isadigit{2}}\ {\isacharequal}\ L\isactrlisub {\isadigit{2}}\ {\isacharcolon}\ kind{\isachardot}}
    \end{compactenum}
  \end{lemma}
 
  \begin{lemma}[(Product injectivity)]
  Suppose \isa{x\ {\isasymsharp}\ {\isasymGamma}}.
  \begin{compactenum}
  \item If \isa{{\isasymGamma}\ {\isasymturnstile}\isactrlisub {\isasymSigma}\ {\isasymPi}x{\isacharcolon}A\isactrlisub {\isadigit{1}}{\isachardot}\ A\isactrlisub {\isadigit{2}}\ {\isacharequal}\ {\isasymPi}x{\isacharcolon}B\isactrlisub {\isadigit{1}}{\isachardot}\ B\isactrlisub {\isadigit{2}}\ {\isacharcolon}\ type}
        then \isa{{\isasymGamma}\ {\isasymturnstile}\isactrlisub {\isasymSigma}\ A\isactrlisub {\isadigit{1}}\ {\isacharequal}\ B\isactrlisub {\isadigit{1}}\ {\isacharcolon}\ type}
             and 
             \isa{{\isacharparenleft}x{\isacharcomma}\ A\isactrlisub {\isadigit{1}}{\isacharparenright}{\isacharcolon}{\isacharcolon}{\isasymGamma}\ {\isasymturnstile}\isactrlisub {\isasymSigma}\ A\isactrlisub {\isadigit{2}}\ {\isacharequal}\ B\isactrlisub {\isadigit{2}}\ {\isacharcolon}\ type}.
  \item If \isa{{\isasymGamma}\ {\isasymturnstile}\isactrlisub {\isasymSigma}\ {\isasymPi}x{\isacharcolon}A{\isachardot}\ K\ {\isacharequal}\ {\isasymPi}x{\isacharcolon}B{\isachardot}\ L\ {\isacharcolon}\ kind}
        then \isa{{\isasymGamma}\ {\isasymturnstile}\isactrlisub {\isasymSigma}\ A\ {\isacharequal}\ B\ {\isacharcolon}\ type}
             and 
             \isa{{\isacharparenleft}x{\isacharcomma}\ A{\isacharparenright}{\isacharcolon}{\isacharcolon}{\isasymGamma}\ {\isasymturnstile}\isactrlisub {\isasymSigma}\ K\ {\isacharequal}\ L\ {\isacharcolon}\ kind}.
  \end{compactenum}
  \end{lemma}

  \begin{lemma}[(Strong versions of rules)]
    ~
    The following rules are admissible:
    \begin{compactenum}
    \item \isa{\mbox{}\inferrule{\mbox{{\isasymGamma}\ {\isasymturnstile}\isactrlisub {\isasymSigma}\ M\isactrlisub {\isadigit{1}}\ {\isacharcolon}\ {\isasymPi}x{\isacharcolon}A\isactrlisub {\isadigit{2}}{\isachardot}\ A\isactrlisub {\isadigit{1}}}\\\ \mbox{{\isasymGamma}\ {\isasymturnstile}\isactrlisub {\isasymSigma}\ M\isactrlisub {\isadigit{2}}\ {\isacharcolon}\ A\isactrlisub {\isadigit{2}}}}{\mbox{{\isasymGamma}\ {\isasymturnstile}\isactrlisub {\isasymSigma}\ M\isactrlisub {\isadigit{1}}\ M\isactrlisub {\isadigit{2}}\ {\isacharcolon}\ A\isactrlisub {\isadigit{1}}{\isacharbrackleft}x{\isacharcolon}{\isacharequal}M\isactrlisub {\isadigit{2}}{\isacharbrackright}}}}
    \item \isa{\mbox{}\inferrule{\mbox{{\isasymGamma}\ {\isasymturnstile}\isactrlisub {\isasymSigma}\ A\ {\isacharcolon}\ {\isasymPi}x{\isacharcolon}B{\isachardot}\ K}\\\ \mbox{{\isasymGamma}\ {\isasymturnstile}\isactrlisub {\isasymSigma}\ M\ {\isacharcolon}\ B}}{\mbox{{\isasymGamma}\ {\isasymturnstile}\isactrlisub {\isasymSigma}\ A\ M\ {\isacharcolon}\ K{\isacharbrackleft}x{\isacharcolon}{\isacharequal}M{\isacharbrackright}}}}
    \item \isa{\mbox{}\inferrule{\mbox{{\isacharparenleft}x{\isacharcomma}\ A\isactrlisub {\isadigit{1}}{\isacharparenright}{\isacharcolon}{\isacharcolon}{\isasymGamma}\ {\isasymturnstile}\isactrlisub {\isasymSigma}\ M\isactrlisub {\isadigit{2}}\ {\isacharequal}\ N\isactrlisub {\isadigit{2}}\ {\isacharcolon}\ A\isactrlisub {\isadigit{2}}}\\\ \mbox{{\isasymGamma}\ {\isasymturnstile}\isactrlisub {\isasymSigma}\ M\isactrlisub {\isadigit{1}}\ {\isacharequal}\ N\isactrlisub {\isadigit{1}}\ {\isacharcolon}\ A\isactrlisub {\isadigit{1}}}\\\ \mbox{x\ {\isasymsharp}\ {\isasymGamma}}}{\mbox{{\isasymGamma}\ {\isasymturnstile}\isactrlisub {\isasymSigma}\ {\isacharparenleft}{\isasymlambda}x{\isacharcolon}A\isactrlisub {\isadigit{1}}{\isachardot}\ M\isactrlisub {\isadigit{2}}{\isacharparenright}\ M\isactrlisub {\isadigit{1}}\ {\isacharequal}\ N\isactrlisub {\isadigit{2}}{\isacharbrackleft}x{\isacharcolon}{\isacharequal}N\isactrlisub {\isadigit{1}}{\isacharbrackright}\ {\isacharcolon}\ A\isactrlisub {\isadigit{2}}{\isacharbrackleft}x{\isacharcolon}{\isacharequal}M\isactrlisub {\isadigit{1}}{\isacharbrackright}}}}
    \item \isa{\mbox{}\inferrule{\mbox{{\isasymGamma}\ {\isasymturnstile}\isactrlisub {\isasymSigma}\ A\isactrlisub {\isadigit{1}}\ {\isacharequal}\ B\isactrlisub {\isadigit{1}}\ {\isacharcolon}\ type}\\\ \mbox{{\isacharparenleft}x{\isacharcomma}\ A\isactrlisub {\isadigit{1}}{\isacharparenright}{\isacharcolon}{\isacharcolon}{\isasymGamma}\ {\isasymturnstile}\isactrlisub {\isasymSigma}\ A\isactrlisub {\isadigit{2}}\ {\isacharequal}\ B\isactrlisub {\isadigit{2}}\ {\isacharcolon}\ type}\\\ \mbox{x\ {\isasymsharp}\ {\isasymGamma}}}{\mbox{{\isasymGamma}\ {\isasymturnstile}\isactrlisub {\isasymSigma}\ {\isasymPi}x{\isacharcolon}A\isactrlisub {\isadigit{1}}{\isachardot}\ A\isactrlisub {\isadigit{2}}\ {\isacharequal}\ {\isasymPi}x{\isacharcolon}B\isactrlisub {\isadigit{1}}{\isachardot}\ B\isactrlisub {\isadigit{2}}\ {\isacharcolon}\ type}}}
    \item \isa{\mbox{}\inferrule{\mbox{{\isasymGamma}\ {\isasymturnstile}\isactrlisub {\isasymSigma}\ A\ {\isacharequal}\ B\ {\isacharcolon}\ type}\\\ \mbox{{\isacharparenleft}x{\isacharcomma}\ A{\isacharparenright}{\isacharcolon}{\isacharcolon}{\isasymGamma}\ {\isasymturnstile}\isactrlisub {\isasymSigma}\ K\ {\isacharequal}\ L\ {\isacharcolon}\ kind}\\\ \mbox{x\ {\isasymsharp}\ {\isasymGamma}}}{\mbox{{\isasymGamma}\ {\isasymturnstile}\isactrlisub {\isasymSigma}\ {\isasymPi}x{\isacharcolon}A{\isachardot}\ K\ {\isacharequal}\ {\isasymPi}x{\isacharcolon}B{\isachardot}\ L\ {\isacharcolon}\ kind}}}
    \end{compactenum}
  \end{lemma}%
\end{isamarkuptext}%
\isamarkuptrue%
\isadelimtheory
\endisadelimtheory
\isatagtheory
\endisatagtheory
{\isafoldtheory}%
\isadelimtheory
\endisadelimtheory
\isanewline
\end{isabellebody}%

%% end electronic appendix

\end{document}